\documentclass[fleqn,usenatbib]{mnras}

\usepackage{newtxtext,newtxmath}

\usepackage[T1]{fontenc}

\DeclareRobustCommand{\VAN}[3]{#2}
\let\VANthebibliography\thebibliography
\def\thebibliography{\DeclareRobustCommand{\VAN}[3]{##3}\VANthebibliography}

\usepackage{graphicx}	
\usepackage{amsmath}	

\usepackage{comment} 
\usepackage{ulem}
\usepackage{subcaption} 

\newcommand\lgal{\textsc{L-Galaxies}}
\newcommand\gadgetfour{\textsc{Gadget-4}}
\newcommand\secref[1]{Sect.~\ref{#1}}
\newcommand\figref[1]{Fig.~\ref{#1}}

\newcommand\equref[1]{Eq.~\eqref{#1}}

\newcommand\appref[1]{Appendix~\ref{#1}}
\defcitealias{Henriques2015}{H15}
\defcitealias{Henriques2020}{H20}
\defcitealias{Ayromlou2021}{A21}
\defcitealias{Barrera2023}{B23}

\title[\textsc{L-Galaxies} and high-$z$ galaxy evolution]{Galaxies in the first two billion years: Earlier formation and quenching with surface-density modulated star formation and feedback}

\author[Vani et al.]{%
Akash Vani$^{1,2}$ \thanks{E-mail: vani@mpa-garching.mpg.de},
Mohammadreza Ayromlou$^{3}$, Guinevere Kauffmann$^{1}$, and Volker Springel$^{1}$
\vspace*{0.1cm}\\%
$^{1}$Max Planck Institute for Astrophysics, Karl-Schwarzschild-Str. 1, 85741 Garching bei München, Germany\\%
$^{2}$Ludwig-Maximilians-Universität München, Geschwister-Scholl-Platz 1, 80539 München, Germany \\%
$^{3}$Argelander-Institut f\"ur Astronomie, Auf dem H\"ugel 71, D-53121 Bonn, Germany\\
}

\date{Accepted XXX. Received YYY; in original form ZZZ}

\pubyear{\the\year{}}

\begin{document}
\label{firstpage}
\pagerange{\pageref{firstpage}--\pageref{lastpage}}
\maketitle

\begin{abstract}
JWST has revealed massive, UV-bright, compact, and in some cases quenched galaxies in the first two billion years of cosmic history, challenging current galaxy formation models. We develop an extended \lgal{} semi-analytic model, run on the MillenniumTNG dark-matter-only simulation, and follow galaxy evolution over $0 \le z \lesssim 15$. The model ties the star formation efficiency and the coupling of stellar feedback energy to the local cold gas surface density, making star formation efficient and feedback ineffective in dense gas. We include dissipative gas-rich mergers and disc instabilities in stellar and gaseous discs. All parameters are calibrated simultaneously against stellar mass functions and quenched fractions at $z\simeq0-4$ and UV luminosity functions at $z\simeq11-12$ with a new parallelised Markov chain Monte Carlo framework. The model reproduces the stellar mass function up to $z\approx 11$ and increases the abundance of massive and UV-bright galaxies at $z\gtrsim9$ by up to two orders of magnitude over the legacy version. Although the AGN feedback prescription is unchanged, the enhanced bulge growth and black hole fuelling make massive quenched galaxies at $z\simeq3-8$ two orders of magnitude more abundant, while dissipative mergers produce compact remnants with UV size--luminosity relations matching observations at $z\simeq10-13$. At low redshift, galaxy sizes, cold-gas content, metallicities, radial profiles, and morphologies remain consistent with observations. Our results indicate that the tension between models and JWST observations at cosmic dawn can be substantially reduced by regulating star formation and feedback through the local gas surface density, without a modified initial mass function, non-standard cosmology, or new AGN physics.
\end{abstract}

\begin{keywords}
galaxies: formation -- galaxies: evolution -- galaxies: high-redshift --
methods: analytical -- methods: numerical
\end{keywords}



\section{Introduction}
Understanding how galaxies formed and evolved within the first few billion years of cosmic time is a major challenge in modern astrophysics \citep[e.g.,][]{Baugh2006, Silk2012, Naab2017, HerreraCamus2026}. Observations with the \textit{James Webb Space Telescope} (JWST) have transformed this field by revealing substantial populations of luminous, massive ($M_\star \gtrsim 10^{10}\,{\rm M}_\odot$), and in some cases morphologically evolved galaxies at redshifts $z \sim 4-15$ \citep{Straatman2014, Valentino2023, Carnall2023, deGraaff2025, Naidu2025, Pascalau2026}. These findings suggest that stellar mass assembly, structural transformation, and quenching can proceed rapidly in the early Universe. The high number density of these systems appears to be in tension with standard theoretical expectations \citep[e.g.,][]{BoylanKolchin2023, Vani2025, Lagos2025, Weller2025, Kim2025Agora}. 
Taken together, these observations motivate improved galaxy formation models, particularly in their treatment of star formation, feedback, and structural evolution during  cosmic dawn.

One of the early results from deep JWST surveys is the high number density of UV-bright galaxies at $z \gtrsim 10$ relative to theoretical predictions. Multiple JWST and Hubble fields have revealed galaxy populations within a few hundred million years of the Big Bang whose abundances and luminosities exceed expectations and indicate a slower decline at the bright end of the luminosity function at $z > 10$ \citep[e.g.,][]{Harikane2023, Harikane2024, Harikane2025, Finkelstein2023, Bouwens2023, Chemerynska2024, Donnan2024, Whitler2025} than predicted by many semi-analytic and hydrodynamical models of galaxy formation \citep[e.g.,][]{Kannan2023, Vijayan2024, Yung2024, Lagos2025, Basu2025}.

In parallel, recent observations have revealed a population of massive and structurally evolved galaxies in the early Universe. Several spectroscopic and photometric campaigns have identified galaxies with stellar masses approaching $M_\star \sim 10^{10}\,{\rm M}_\odot$ at $z \sim 7-14$ \citep[e.g.,][]{Harikane2023, Harikane2024, Harikane2025, Chemerynska2024, Donnan2024}. 
Notably, JWST/NIRSpec spectroscopy of candidate systems such as JADES-GS-z13-0, JADES-GS-z14-0, and other high-redshift sources confirms their high redshifts, while SED modelling indicates that galaxies of substantial stellar masses may already have been in place only a few hundred million years after the Big Bang \citep[e.g.,][]{CurtisLake2023, Robertson2023}. Many of these galaxies appear compact, show high specific star formation rates, and higher baryon conversion efficiencies than predicted by many current galaxy formation models \citep[e.g.,][]{Robertson2023, Tacchella2023, Jain2024, Shen2026Thesan}.

Perhaps most surprising in the early JWST data is the emergence of a sizeable population of quiescent, or ``red and dead'', galaxies at $z \sim 3-5$. Spectroscopic confirmation of quiescent systems with low specific star formation rates at these epochs indicates that stellar populations in some galaxies formed and quenched rapidly, with inferred formation redshifts of $z > 7-10$ in certain cases \citep[e.g.,][]{Weibel2024, Russell2025, AntwiDanso2025, Zhang2026, Ito2026}. 
These observations reveal galaxies with substantial stellar masses and little to no ongoing star formation within the first $\sim$2~Gyr of cosmic history, intensifying the challenge for galaxy formation models, which must allow for both rapid early mass assembly and an efficient quenching mechanism that will affect a significant subset of the population \citep[e.g.,][]{Carnall2023, Nanayakkara2025, Dekel2025FFB}.

Collectively, these empirical tensions have motivated a broad range of proposed theoretical and astrophysical explanations.
One class of models  which invokes bursty star formation histories, potentially driven by stochastic gas accretion or variability in baryon cycling, may enhance UV luminosities and star formation rates at high redshift \citep[e.g.,][]{FaucherGiguere2018, Iyer2020, Sun2023, Kravtsov2024arXiv, Bhagwat2024, McClymont2025, CarvajalBohorquez2025}. Alternative prescriptions for the stellar initial mass function (IMF), such as top-heavy IMFs or extended Population~III star formation, have also been proposed as mechanisms to boost the luminosity and ionizing output of early galaxies, thereby increasing their detectability and inferred abundances \citep[e.g.,][]{Chon2021, Chon2024, Menon2024, Trinca2024, Ma2025, Schaerer2025, Fontanot2026arXiv}. 
Additionally, scenarios invoking reduced stellar feedback at cosmic dawn suggest that high gas densities and low metallicities limit the efficiency of gas ejection, allowing galaxies to retain a larger fraction of their baryons and assemble stellar mass more rapidly \citep[e.g.,][]{Dekel2023, Menon2024, Li2024, Somerville2025, BenitezLlambay2026COLIBRE}. 

Other astrophysical pathways focus on the early quenching and structural transformation of galaxies, addressing the abundance of quiescent systems rather than the rapid assembly of massive galaxies. Powerful AGN feedback may expel or heat gas, thereby suppressing star formation in massive early galaxies \citep[e.g.,][]{Lagos2024, DeLucia2024, Kimmig2025, Weller2025}. Such feedback could be enabled by rapid black hole growth from heavy seeds \citep[e.g.,][]{Bonoli2025arXiv, Jeon2025, Ziparo2025} or by episodes of super-Eddington accretion \citep[e.g.,][]{Trinca2024arXiv_BHSuperEdd, Bonoli2025arXiv, Husko2025}. 

To explain the emergence of compact massive galaxies at high redshift, compaction events driven by violent disc instabilities, mergers, or nuclear inflows have been proposed to funnel gas into the central regions, triggering intense starbursts followed by inside-out quenching and the formation of dense ``red nuggets'' \citep[e.g.,][]{Dekel2014, Zolotov2015, Tacchella2016}. Such mechanisms have also been discussed in the context of the compact high-redshift galaxy population, including the recently identified ``Little Red Dots'' \citep[e.g.,][]{Pacucci2025, Akins2025, Kocevski2025}.

Theoretical interpretations of the high-$z$ tensions extend beyond baryonic physics to include modified cosmological scenarios. Variations of the standard $\Lambda$CDM framework such as early dark energy, altered growth histories, or deviations in the matter power spectrum have been explored as potential mechanisms to enhance early structure formation and mitigate the apparent abundance mismatch of bright galaxies \citep[e.g.,][]{Biagetti2023, Shen2024, Liu2024, Acharya2025, Blamart2026_Cosmicstrings}. 

In addition to modifications of star formation, feedback, or cosmology, internal dynamical processes provide further pathways for rapid bulge growth, black hole growth, and structural transformation at high redshift. Galaxy mergers and disc instabilities are expected to play an especially important role in the gas-rich, high-density environments of the early Universe. In particular, gas-rich mergers can trigger intense starbursts, redistribute angular momentum, and drive cold gas toward galactic centres, thereby fuelling both bulge growth and enhanced supermassive black hole (SMBH) accretion. Similarly, gravitational instabilities in stellar or gaseous discs can transport mass from rotationally supported components into central regions, promoting bulge formation and channelling gas onto SMBHs \citep[e.g., ][]{Somerville2001, Dekel2009, Zolotov2015, Tacchella2023, Omori2025, Mandelker2025, Cataldi2026}. Although these processes may not dominate in all systems, they likely contribute non-negligibly to the rapid buildup of central stellar mass and black hole growth at early cosmic times.

Within the broader context of rapidly expanding observational constraints and ongoing theoretical developments, this work investigates the role of high-efficiency star formation and feedback in dense gas regions, as well as disc instabilities and galaxy mergers, and their connection to quenching in shaping the earliest massive galaxies. Our goal is to understand how the physical conditions that drive rapid stellar mass assembly at high redshift influence the structural and morphological evolution of galaxies across cosmic time, while maintaining consistency with the observed galaxy population at low redshift ($0 \le z \lesssim 3$). 

To this end, we employ the \lgal{} cosmological semi-analytic model of galaxy formation and evolution, also known as the Munich semi-analytic model \citep{Kauffmann1998, Kauffmann1999clustering, Kauffmann1999highz, Springel2001populating}. Over the past three decades, this framework has evolved into a physically motivated and self-consistent tool capable of modelling key baryonic processes, including star formation, feedback, environmental quenching, and chemical enrichment across a wide range of cosmic environments \citep{Springel2005MilI, Henriques2015, Henriques2020, Ayromlou2021}. The most recent implementation by \citet{Barrera2023} integrates \lgal{} into the \gadgetfour{} framework \citep{Springel2021_G4} as a post-processing module, coupling it directly to the dark matter merger trees generated by this code, such as those from the MillenniumTNG simulations \citep{HernandezAguayo2023_G4DMO}. Overall, its modular structure makes it well-suited to exploring new physical prescriptions and calibrating them against a broad range of observations, while retaining the computational efficiency needed to investigate large cosmological volumes and extensive parameter spaces.

In this work, we extend the \lgal{} framework, built on top of \gadgetfour{} (referred to as \lgal{}, while earlier versions are denoted as \textit{legacy} models), to explore how enhanced star formation efficiencies in high gas-density environments, galaxy mergers with gas dissipation, gas and stellar disc instabilities, and regulating feedback processes affect the assembly, size evolution, and quenching of galaxies over the redshift range $0 \leq z \lesssim 15$. By incorporating physically motivated prescriptions for dense gas regulation, we test whether such processes can reproduce the abundance of massive, compact, and already quenched systems revealed by recent observations. Notably, we retain the standard \citet{Chabrier2003_IMF} initial mass function (IMF), assume a $\Lambda$CDM cosmology, and leave the AGN feedback model unchanged \citep[][]{Henriques2015, Henriques2020}, testing whether modifications to galaxy physics alone are sufficient to account for these recently observed populations.
In parallel, we update the \lgal{} codebase to align with the latest legacy release of \citet{Henriques2020}, incorporating several key developments, including the implementation of radial rings. 
Together, these developments extend \lgal{} beyond the low-redshift focus of the legacy versions to the early-Universe regime, introducing new high-redshift galaxy-physics prescriptions that together push the model's predictive range to $z\sim15$.

In this paper, we assume a $\Lambda$CDM cosmology with 
$\Omega_{\Lambda}=0.6911$, 
$\Omega_{\rm b}=0.0486$, 
$\Omega_{\rm m}=0.3089$, and 
$H_{0}=67.74\,\mathrm{km\,s^{-1}\,Mpc^{-1}}$ 
\citep{Planck2016}. All magnitudes are given in the AB photometric system \citep{OkeandGunn1983_ABphoto}, and a \citet{Chabrier2003_IMF} IMF is adopted throughout this analysis.

This paper is organised as follows. In \secref{sec:sim&subhalo}, we describe the dark matter simulation and subhalo identification. In \secref{Sec:Lgal_model}, we outline the new \lgal{} semi-analytic framework. In \secref{sec:modelproperties}, we summarise the key baryonic processes relevant to this study, including the updated density-dependent star formation model (\secref{sec:starformation}), stellar feedback and its regulation in dense gas environments (\secref{sec:SneFeedback} and \secref{sec:regulatingfeedback}), galaxy mergers with gas dissipation (\secref{sec:mergers}), and disc instabilities affecting both stellar and gaseous components, BH fuelling and central star formation (\secref{sec:DI}). 
Section~\ref{sec:MCMC_routines} describes the calibration procedure and the observational constraints used. In \secref{Sec:Results}, we compare model predictions with observed galaxy scaling relations of global properties (\secref{sec:global_prop}), quenched properties (\secref{sec:QG_prop}), morphological properties (\secref{sec:struct_prop}), and galaxy--halo properties (\secref{sec:halo_prop}). In \secref{sec:local_diagnostics}, we test the recalibrated model against additional local diagnostics, such as galaxy sizes and cold-gas content, mass–metallicity relations, and the resolved profiles of Milky Way-like galaxies, as independent consistency checks.
Finally, Section~\ref{sec:summary_discussion} summarises and discusses our findings. 
\appref{app:MCMCcalib} presents the details of the calibration framework.

\section{Simulations and galaxy formation model}
\label{Sec:Model}

\subsection{Simulations and subhalo identification}
\label{sec:sim&subhalo}
In this work, we employ the dark-matter-only (DMO) simulations from the MillenniumTNG project \citep[MTNG; for a detailed overview see][]{HernandezAguayo2023_G4DMO}, a next-generation suite of cosmological DMO and hydrodynamical simulations. The DMO runs used here were performed with the \gadgetfour{} code \citep{Springel2021_G4}\footnote{The hydrodynamical runs in the MTNG project use the AREPO code \citep{Springel2010pur}.}. In addition to dark matter halo catalogues and merger trees, the MTNG simulations provide continuous lightcone outputs and extended subhalo properties, including local background environment (LBE) estimates following \citet{Ayromlou2019new}\footnote{The LBE measurements are only available in later MTNG runs and are not used in the present work.}. For further details on the simulation suite and data products, we refer the reader to \citet{Barrera2023}. In this study, we use the primary DMO simulation, MTNG740-DM, which evolves $4320^3$ particles in a $(740\,\mathrm{Mpc})^3$ comoving box and has a dark matter particle mass of $m_\mathrm{DM}\simeq1.3\times10^8\,h^{-1}\,\mathrm{M}_\odot$. 

Dark matter haloes are identified using the Friends-of-Friends \citep[FoF;][]{Davis1985} algorithm, while gravitationally bound substructures (subhaloes) are extracted with the \textsc{Subfind-Hbt} algorithm \citep{Han2018, Springel2021_G4}, which improves the tracking of subhaloes across snapshots relative to earlier versions of \textsc{Subfind} \citep{Springel2001populating}. These structures are identified on the fly across 265 snapshots spaced logarithmically in scale factor, with $\Delta\log(a)=0.0081$ for $0\le z<3$, $\Delta\log(a)=0.0162$ for $3\le z<10$, and $\Delta\log(a)=0.0325$ for $10\le z<30$, including 32 snapshots in the highest-redshift interval. For our purposes, this spacing corresponds to approximately fixed fractions of the halo dynamical time, yielding temporal resolutions from $\sim116$ Myr at $z=0$ to $\sim4.8$ Myr at $z=30$. This is sufficient for robust merger-tree construction and for sampling early structure formation at high redshift. The resulting subhalo catalogues form the basis of the merger trees, which are constructed following the methodology described in \citet{Barrera2023}.

\subsection{\lgal{} model}
\label{Sec:Lgal_model}

\begin{figure*}
    \centering
    \includegraphics[width=0.9\linewidth]{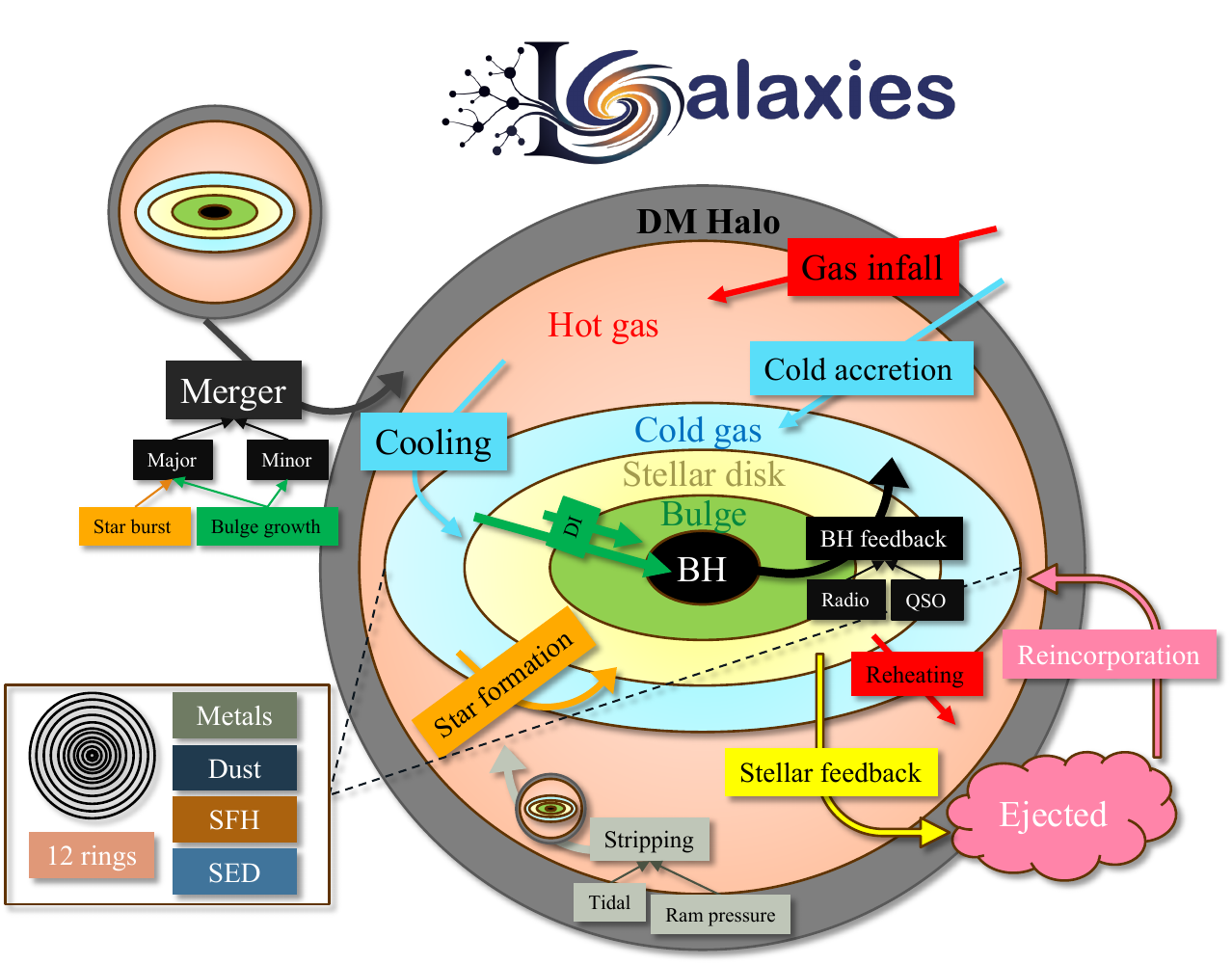}
    \caption{Schematic overview of some of the key physical processes implemented in \lgal{}. Gas accretes onto the dark matter halo and is shock-heated into a hot atmosphere, from which it cools onto the galactic disc. In low-mass haloes, a fraction of the accreted gas may also reach the cold-gas reservoir directly via cold accretion. Cold gas fuels star formation in the stellar disc, while stellar feedback reheats gas and can drive outflows into an external reservoir. Ejected gas may later be reincorporated into the halo. Black hole growth occurs through gas accretion and galaxy mergers, with AGN feedback regulating gas cooling. Galaxy mergers and environmental processes such as stripping further influence the redistribution of mass between components, while disc instabilities (DI) drive bulge growth and morphological transformation. Each galaxy is divided into 12 concentric rings, within which we track the evolution of stellar mass, cold gas, metals, dust, and the star formation history (SFH). Photometry or SED is then computed self-consistently from the SFH in each ring.}
    \label{fig:LGSchematic}
\end{figure*}

The semi-analytic framework, \lgal{}\footnote{\url{https://lgalaxiespublicrelease.github.io/}}, models baryonic processes on top of dark matter halo merger trees using a set of physically motivated equations \citep{White1978, White1991,Kauffmann1993formation,Kauffmann1999clustering,Springel2001populating}. In this framework, the baryonic content associated with each subhalo is partitioned into several components, including hot gas, cold gas (with atomic and molecular phases), stellar discs, bulges, halo stars, SMBHs, and an ejected gas reservoir. Gas cools from the halo onto galactic discs, forms stars, is redistributed through mergers and internal instabilities, and is regulated by stellar and AGN feedback (see \figref{fig:LGSchematic} for a schematic overview). In this way, \lgal{} follows the hierarchical growth of structure while modelling other processes such as galaxy mergers, black hole growth, and environmental processes like tidal and ram-pressure stripping \citep[e.g.,][]{Croton2006, Guo2011, Henriques2015,Henriques2020,Ayromlou2021}.

In this work, we use the \lgal{} model developed by \citet{Barrera2023}, implemented as a post-processing framework on top of \gadgetfour{}, enabling efficient processing of the large merger trees produced by the MTNG simulations. Within this framework, \lgal{} interfaces directly with halo catalogues through a fully integrated implementation in \gadgetfour{}\footnote{\url{https://wwwmpa.mpa-garching.mpg.de/gadget4/}}, sharing common \textsc{C++} classes for input/output and memory management while remaining logically distinct. This restructuring resulted in a more modular, efficient, and extensible codebase, with all input and output operations migrated to the \textsc{HDF5} format.

We refer to this updated implementation as the new \lgal{} model, while earlier versions are denoted as \textit{legacy} models. Although the code infrastructure has been significantly modernised, the underlying physical prescriptions in \citet{Barrera2023} largely follow the 2015 legacy model \citep{Henriques2015}. In this work, we extend this framework by incorporating the radial-ring structure introduced in later \lgal{} developments \citep{Fu2013_rings, Henriques2020}, together with the new physical prescriptions described below. The broader codebase also contains more recent \lgal{} modules for environmental processes \citep{Ayromlou2019new}, time-dependent chemical enrichment \citep{Yates2013}, and $\mathrm{H}_2$-based star formation \citep{Fu2013_rings, Krumholz2009HI_H2}; however, these modules are not enabled or tested in the present analysis and will be explored in future work.

In the ring-based framework, the stellar and gaseous discs are divided into concentric annuli (``rings'')\footnote{The outer edges of the rings are defined as $r_i = 0.01 \times 2.0^{i}\, h^{-1}\,\mathrm{kpc}$, for $i = 1, 2, \dots, 12$.}, enabling a radially resolved treatment of galaxy evolution \citep[][]{Fu2013_rings}. In this work, we further extend the star formation history (SFH) framework to operate on a ring-by-ring basis, allowing independent tracking of SFHs within each ring. The prescriptions for gas cooling and gas inflow are then revised to account consistently for the redistribution of material across the disc.

In this study, we introduce several additional physical prescriptions, including a gas-based disc-instability model that fuels SMBH growth, a merger treatment that incorporates gas dissipation, and a gas-density-dependent star formation prescription operating within the radial rings. We also implement a regulation scheme for stellar feedback in high gas-density environments (see \secref{sec:modelproperties}). Together, these processes affect black hole growth, star formation, and the structural and morphological evolution of galaxies.
Finally, building on the legacy Markov chain Monte Carlo (MCMC) methodology \citep{Henriques2009MCMC}, we incorporate an upgraded MCMC calibration to improve the flexibility and performance of our parameter adjustment (see \secref{sec:MCMC_routines}).
We note that the present model augments the \citet{Barrera2023} version with selected \citet{Henriques2020} features and the new prescriptions, while retaining simpler treatments of chemical enrichment and environmental processes. More advanced treatments of these latter aspects exist in other recent \lgal{} versions; we defer their integration to future work in order to focus here on the high-redshift regime without introducing confounding changes.

\subsection{Galaxy formation modelling}
\label{sec:modelproperties}

In this subsection, we describe the implementations of the physical processes most relevant to this study. We briefly explain the previous prescriptions and describe in more detail our specific model improvements. The standard \lgal{} models include comprehensive descriptions in the supplementary material of \citet{Henriques2020, Ayromlou2021}, while a detailed comparison between the model variants is given in \citet{Vani2025}. 

Sections~\ref{sec:starformation}--\ref{sec:DI} describe the new model developments introduced in this work. In contrast, Sections~\ref{sec:BHprocesses}--\ref{sec:colors} summarise physical processes inherited from earlier model versions, which are included here for completeness due to their relevance to the present study.

\subsubsection{Star formation}
\label{sec:starformation}
In legacy variants of \lgal{} \citep[e.g.,][]{Henriques2015}, star formation was assumed to proceed from the total cold gas reservoir in galactic discs, with the star formation rate (SFR) scaling with the global interstellar medium (ISM) content. Later updates \citep{Henriques2020, Ayromlou2021} introduced an $\mathrm{H}_2$-based star formation law in which the star formation surface density is proportional to the molecular gas surface density \citep{Fu2013_rings, Krumholz2009HI_H2}, with an inverse dependence on dynamical time. While this prescription naturally enhances star formation at early times when dynamical times are short, it does not explicitly allow the star formation efficiency (SFE) of dense gas to vary with surface density. Both simulations and observations suggest that cloud-scale efficiencies can increase in high-pressure and high-density environments \citep[e.g.,][]{Grudic2018, Tacconi2020, DessaugesZavadsky_2023, Menon2024, Wang2025, Leroy2025}.

Motivated by these findings, we adopt a density-modulated star formation efficiency model following \citet{Somerville2025}. In this framework, star formation within a molecular cloud proceeds until stellar feedback injects sufficient momentum to overcome self-gravity. This yields a critical surface density,
$\Sigma_{\rm crit} \approx 2.2 \times 10^3\, {\rm M}_\odot\,{\rm pc^{-2}}$ (used in \equref{eq:SFE}), above which cloud-scale star formation becomes highly efficient.

For each radial ring $j$, the mean cold gas surface density is computed as
\begin{equation}
\label{eq:ColdGasSurfacedensity}
\Sigma_{\rm ColdGas}^j =
\frac{M_{\rm ColdGas}^j}{A^j},
\end{equation}
where $M_{\rm ColdGas}^j$ and $A^j$ denote the cold gas mass and area of the $j^{\rm th}$ ring. The cloud surface density is assumed to scale with the local ISM surface density via a clumping factor $c$,
\begin{equation}
\Sigma_{\rm cl}^j = c\, \Sigma_{\rm ColdGas}^j,
\end{equation}
and we here adopt $c=1$. This choice assumes that the GMC-scale surface density traces the local mean ISM surface density. While this neglects unresolved clumping, potentially important in turbulent, gas-rich high-redshift environments, the clumping factor itself is uncertain and depends on local conditions such as metallicity and ISM structure \citep[e.g.,][]{Krumholz2009HI_H2, Grudic2018}. We therefore adopt $c=1$ as a reasonable first-order approximation for our purposes, given the poorly understood uncertainties.

The integrated cloud (cl)-scale star formation efficiency is
\begin{equation}
\label{eq:SFE}
\epsilon_{\star,\rm cl}^j =
\frac{\Sigma_{\rm cl}^j}
{\Sigma_{\rm cl}^j + \Sigma_{\rm crit}},
\end{equation}
which yields low efficiencies at GMC-like surface densities and approaches unity at high surface densities.

We assume a characteristic cloud mass $M_{\rm cl}=10^5\,{\rm M}_\odot$, consistent with GMC masses in the local Universe and adopted in \citet{Somerville2025}. This choice sets the cloud size and density scale for the free-fall time (Eq.~\ref{eq:time_ff}) at high surface densities. The cloud free-fall time is computed as
\begin{equation}
\label{eq:time_ff}
t_{\rm ff}^j =
\left( \frac{3\pi}{32 G \bar{\rho}^j} \right)^{1/2},
\end{equation}
where $\bar{\rho}^j$ is the mean cloud density, assuming a spherical cloud, given by $\bar{\rho}^j = 3 M_{\rm cl} / (4 \pi R_{\rm cl}^3)$, with the cloud radius expressed as $R_{\rm cl} = \left[M_{\rm cl}/(\pi \Sigma_{\rm cl}^j)\right]^{1/2}$.

Following simulation-based fits, the cloud lifetime depends on surface density as
\begin{equation}
t_{\rm cl}^j =
\tau_{\rm cl,0}\, t_{\rm ff}^j
\left[
\left( \frac{\Sigma_{\rm cl}^j}{\Sigma_{\rm cl,0}} \right)^{-\gamma_1}
+
\left( \frac{\Sigma_{\rm cl}^j}{\Sigma_{\rm cl,0}} \right)^{\gamma_2}
\right]^{-1},
\end{equation}
where we adopt $\tau_{\rm cl,0}=8.64$, $\Sigma_{\rm cl,0}=10^5\,{\rm M}_\odot\,{\rm pc^{-2}}$, $\gamma_1=0.4$, and $\gamma_2=0.14$, following \citet{Somerville2025}. The choice of $\Sigma_{\rm cl,0}$ sets the characteristic surface density at which the transition between long- and short-lived clouds occurs and is calibrated to match cloud-scale simulation results \citep[see][and references therein]{Somerville2025}. This formulation yields relatively long-lived clouds at low and intermediate surface densities, while producing short-lived, rapidly evolving clouds at high surface densities.

The star formation rate per cloud in a ring $j$ is
\begin{equation}
{\rm SFR}_{\rm cl}^j =
\frac{\epsilon_{\star,\rm cl}^j\, M_{\rm cl}}{t_{\rm cl}^j}.
\end{equation}
Using $f_{\rm dense}$ as an effective parameter that sets the amount of cold gas participating in cloud-scale star formation, the total SFR in ring $j$ is
\begin{equation}
{\rm SFR}_{\rm galaxy}^j =
{\rm SFR}_{\rm cl}^j \times
f_{\rm dense}\,
\frac{M_{\rm ColdGas}^j}{M_{\rm cl}},
\end{equation}
where the term
$f_{\rm dense}\,
M_{\rm ColdGas}^j/M_{\rm cl}$
represents the number of dense star-forming clouds in ring $j$.
The global galaxy SFR is obtained by summing over all radial rings. In this work, we treat $f_{\rm dense}$ as a free parameter and calibrate it within the MCMC framework. 
Here, $f_{\rm dense}$ should be interpreted as an effective cloud fraction that sets the normalisation of the star-formation law, rather than strictly as the physical fraction of cold gas residing in dense clouds \citep{Somerville2025}. Its calibrated value may therefore also absorb unresolved effects such as gas clumping and its metallicity or cosmic time dependence, which are not explicitly modelled here.

This implementation explicitly links local gas surface density to cloud-scale efficiency, cloud lifetime, and feedback regulation, naturally enhancing star formation in the high-density environments characteristic of high-redshift galaxies while maintaining low efficiencies in GMC-like conditions at low redshift. 

However, when calibrated primarily against low- and intermediate-redshift constraints ($z \lesssim 3$), this prescription alone does not substantially modify the predicted properties of galaxies at very high redshift in our framework. This motivates the inclusion of a regulated feedback model to produce the full range of high-redshift galaxy properties (\secref{sec:regulatingfeedback}).

\subsubsection{Stellar feedback}
\label{sec:SneFeedback}
Massive, short-lived stars release substantial amounts of mass, metals, and energy through stellar winds and core-collapse supernova (SN) explosions. In this work, following \citet{Henriques2015} and \citet{Barrera2023}, we adopt the instantaneous recycling approximation, in which the total mass, metals, and energy produced during a star formation episode are assumed to be returned to the ISM instantaneously. While the return is instantaneous, the total feedback energy budget remains tied to the amount of newly formed stellar mass. 

The total energy injected by supernovae and stellar winds is
\begin{equation}
\label{eq:Esn}
\Delta E_{\rm SN} =
\epsilon_{\rm halo}\,
\Delta M^{\rm eff}_{\star}\,
\eta_{\rm SN}\,
E_{\rm SN},
\end{equation}
where $\Delta M_{\star}^{\rm eff}$ is the effective stellar mass contributing to feedback \citep[as opposed to the total stellar mass formed used in][see \equref{eq:MstarEff}]{Henriques2015},  $\eta_{\rm SN}$ is the number of supernovae per unit stellar mass ($0.0149\,{\rm M}_\odot^{-1}$, assuming a universal \citealt{Chabrier2003_IMF} IMF), and $E_{\rm SN}$ is the energy released per supernova \citep[$10^{51}\,{\rm erg}$, based on][]{Leitherer1999}. The efficiency parameter is given by $\epsilon_{\rm halo} = \eta_{\rm eject}\left[0.5+\left(V_{\rm max}/V_{\rm eject}\right)^{-\beta_{\rm eject}}\right]$, where $\eta_{\rm eject}$ and $\beta_{\rm eject}$ are free parameters controlling the normalization and slope, $V_{\rm max}$ gives the maximum circular velocity of the halo, and $V_{\rm eject}$ is a free parameter which sets the characteristic velocity scale.

A fraction of this energy is used to reheat cold gas and transfer it to the hot halo, with the reheated mass given by
\begin{equation}
\label{eq:Mreheat}
\Delta M_{\rm reheat} =
\epsilon_{\rm disc}\,
\Delta M^{\rm eff}_{\star},
\end{equation}
where the mass-loading efficiency is parameterised as $\epsilon_{\rm disc} = \epsilon_{\rm reheat}\left[0.5+\left(V_{\rm max}/V_{\rm reheat}\right)^{-\beta_{\rm reheat}}\right]$. Here, $\epsilon_{\rm reheat}$ and $\beta_{\rm reheat}$ are free parameters controlling the normalization and slope, and $V_{\rm reheat}$ is a free parameter which sets the characteristic velocity scale.

The corresponding energy required for reheating is
$\Delta E_{\rm reheat} =
\frac{1}{2}\,\Delta M_{\rm reheat}\, V_{\rm 200}^2$.
If excess energy remains ($\Delta E_{\rm SN} > \Delta E_{\rm reheat}$), it is used to eject hot gas from the halo:
\begin{equation}
\frac{1}{2}\,\Delta M_{\rm eject}\, V_{\rm 200}^2 =
\Delta E_{\rm SN} - \Delta E_{\rm reheat},
\end{equation}
where $V_{\rm 200}$ is the virial velocity. The ejected gas ($\Delta M_{\rm eject}$) is subsequently reincorporated on a halo mass dependent timescale, $t_{\rm reincorporation} \propto M_{\rm 200}^{-1}$ \citep{Henriques2013winds}. We refer to \citet{Guo2011}, \citet{Henriques2015}, and \citet{Henriques2020} for further details on stellar feedback.

In the ring-based implementation adopted in this work, the reheating and ejection processes are computed locally for each radial ring. The total ejected and reheated mass is then obtained by summing the contributions from all rings.

In the regulated feedback model introduced in \secref{sec:regulatingfeedback}, only a fraction, $f_{\star\rm, fb}$ (see Eqs.~\ref{eq:RFB_function}-\ref{eq:fRFB}), of newly formed stars in each radial ring effectively contributes to driving feedback in high gas-density environments. This is implemented by defining an effective stellar mass participating in feedback,
\begin{equation}
\label{eq:MstarEff}
\Delta M_{\star}^{\rm eff} = f_{\star\rm, fb}\, \Delta M_{\star},
\end{equation}
where $\Delta M_{\star}$ is the total stars formed.

This formulation reduces the effective coupling of stellar feedback in high gas surface-density regions, leading to lower reheating and ejection efficiencies, while preserving underlying prescriptions for stellar winds, SN reheating, ejection, and reincorporation. Consequently, a larger fraction of gas remains in the cold phase, increasing the available fuel for star formation, enhancing central gas densities, and indirectly promoting subsequent processes such as galaxy mergers and black hole growth.

\subsubsection{Regulating stellar feedback}
\label{sec:regulatingfeedback}
Recent theoretical work has highlighted a regime of star formation in which newly formed stellar populations can give rise to subsequent generations of stars before stellar feedback from earlier generations has had sufficient time to disperse the parent cloud \citep[e.g.,][]{Dekel2023, Li2024}. This regime is expected to arise in extremely dense, gas-rich environments, particularly at high redshift in massive haloes, where cloud free-fall times can be shorter than the $\sim 3\,\mathrm{Myr}$ delay preceding the onset of the first core-collapse supernovae. Under such conditions, strong gravitational binding energy and high column densities can trap ionizing radiation and stellar winds, thereby limiting the ability of feedback to immediately disrupt the cloud \citep{Chevance2022, Dekel2023, Li2024, Andalman2025, BenitezLlambay2026COLIBRE}.

Motivated by this physical picture, we implement a regulated stellar feedback model (here focusing solely on stellar winds and supernovae) that reduces the effective coupling of stellar feedback to the surrounding ISM in regions of sufficiently high gas surface density. Rather than eliminating feedback, this prescription modulates the fraction of feedback energy that escapes the star-forming cloud and couples to the ISM as a continuous function of gas surface density. We hereafter refer to this implementation as regulated feedback.

Following \citet{Dekel2023} and \citet{Li2024}, we define a characteristic surface density threshold by requiring that the cloud free-fall time (Eq. \ref{eq:time_ff}) be shorter than a critical timescale $t_{\rm ff,crit}$ associated with the onset of stellar feedback. Assuming a uniform-density spherical cloud of radius $R_{\rm fb}$ and mean density $\bar{\rho}$, this condition defines a critical volume density, which can be converted into a mean surface density threshold
\begin{equation}
\Sigma_{\rm crit, fb} =
\frac{\pi R_{\rm fb}}
{8 G t_{\rm ff,crit}^2},
\end{equation}
where $G$ is the gravitational constant. In this work, we adopt $R_{\rm fb}\simeq 10\,\mathrm{pc}$ \citep[e.g.,][]{Lada2020, Li2024} as a fiducial characteristic size for the compact, dense clouds associated with the onset of the feedback-regulated regime. We treat $t_{\rm ff,crit}$ as a free parameter up to $\sim 3\,\mathrm{Myr}$, calibrated within the MCMC framework.
The best-fitting value, $t_{\rm ff,crit}=2.34\,\mathrm{Myr}$ (Table~\ref{tab:MCMC_free_params}), corresponds to $\Sigma_{\rm crit,fb}\simeq160\,{\rm M_\odot\,pc^{-2}}$. Notably, this is well below the characteristic surface density $\Sigma_{\rm crit}\simeq2200\,{\rm M_\odot\,pc^{-2}}$ at which the cloud-scale star-formation efficiency rises strongly (\equref{eq:SFE}). At $\Sigma_{\rm crit,fb}$, $\epsilon_{\star,\rm cl}\simeq0.07$, implying that feedback suppression begins while star formation remains moderately efficient and subsequently overlaps with the rise in cloud-scale efficiency at higher surface densities. The combination of these two effects, rather than enhanced star-formation efficiency alone, therefore enables rapid stellar mass assembly in dense systems.
For each radial ring $j$, we compute the local cold-gas surface density following \equref{eq:ColdGasSurfacedensity}.

The fraction of feedback that couples efficiently to the ISM is then modelled using a surface-density-dependent saturation function:
\begin{equation}
\label{eq:RFB_function}
f_{\rm eff}^j =
\frac{\left(\Sigma_{\rm ColdGas}^j / \Sigma_{\rm crit, fb}\right)^{n}}
{1 + \left(\Sigma_{\rm ColdGas}^j / \Sigma_{\rm crit, fb}\right)^{n}},
\end{equation}
where $n=4$ controls the sharpness of the transition. The effective stellar feedback coupling factor ($f_{\star\rm, fb}$) applied to stellar feedback is then
\begin{equation}
\label{eq:fRFB}
f_{\star\rm, fb}^j = 1 - f_{\rm eff}^j.
\end{equation}
At low surface densities ($\Sigma_{\rm ColdGas}^j \ll \Sigma_{\rm crit, fb}$), feedback suppression is negligible ($f_{\rm eff}^j \approx 0$), such that stellar feedback couples efficiently to the ISM ($f_{\star\rm, fb}^j \approx 1$). Conversely, at high surface densities ($\Sigma_{\rm ColdGas}^j \gg \Sigma_{\rm crit, fb}$), most of the feedback is trapped within dense, gravitationally bound clouds ($f_{\rm eff}^j \rightarrow 1$), reducing the effective coupling to the ISM ($f_{\star\rm, fb}^j \rightarrow 0$).

This prescription reflects the idea that in extremely dense environments, strong gravitational binding can counteract radiative and mechanical stellar feedback, while ionising photons and stellar winds may become trapped within optically thick, high-column-density clouds \citep{Grudic2018, Fukushima2020, Dekel2023, Li2024}. In such conditions, massive and dense star-forming clouds can remain shielded from the feedback of previous stellar generations, allowing continued gas consumption and star formation \citep{Li2024, Grudic2018, Lancaster2021}. Nevertheless, real molecular clouds are unlikely to be perfectly homogeneous or fully sealed; internal porosity, turbulence, and substructure may enable partial leakage of energy and radiation \citep[e.g.,][]{GarrattSmithson2018, Lancaster2021, Smith2021, Chevance2022, AlmadaMonter2024, Zimmermann2025}. The adopted \equref{eq:RFB_function} therefore provides a continuous transition between efficiently coupled and weakly coupled feedback regimes rather than imposing a sharp threshold, thereby capturing the expected gradual change in feedback effectiveness with increasing gas surface density\footnote{To prevent artificial synchronisation of galaxies evolving in phase, the onset of regulated feedback is randomly staggered across the integration sub-time-steps.}. This combined treatment allows dense, high-redshift systems to experience episodes of rapid stellar mass assembly, while maintaining consistency with lower-redshift constraints as gas surface densities decline.

\subsubsection{Galaxy mergers and bulge size evolution}
\label{sec:mergers}
In the standard \lgal{} framework, bulges form through three distinct channels:
major mergers, minor mergers, and disc instabilities. Mergers produce classical bulges, while disc instabilities give rise to pseudo-bulges.

Earlier implementations treated mergers as dissipationless, effectively assuming that the orbital and internal energies of the progenitors were redistributed without significant radiative losses. In gas-rich systems, however, colliding gas clouds can efficiently radiate away kinetic energy, allowing material to sink deeper into the potential well and form more compact, denser remnants. Neglecting this process, therefore, leads to systematically overestimated galaxy sizes, particularly for low-mass systems \citep[e.g.,][]{Irodotou2019, IzquierdoVillalba2019, Parente2023, Vani2025}. In this work, we compute bulge structural properties using energy conservation arguments that explicitly include radiative dissipation, following \citet{Covington2008, Covington2011} and \citet{Tonini2016}.

In major mergers, progenitor discs are destroyed. All pre-existing stars and those formed in the merger-induced burst (following the collisional starburst model of \citealt{Somerville2001}) are transferred to the bulge. The final bulge half-mass radius $R_{\rm fin}$ is obtained from solving the energy conservation equation given as
\begin{equation}
E_{\rm final} = E_{\rm initial} + E_{\rm orbital} + E_{\rm radiative},
\end{equation}
with
\begin{equation}
E_{\rm final} =
- G
\frac{(M_{\star,1}+M_{\star,2}+M_{\star,\rm burst})^2}
{R_{\rm fin}},
\end{equation}
\begin{equation}
E_{\rm initial} =
- G
\left(
\frac{M_{1}^2}{R_{1}}
+
\frac{M_{2}^2}{R_{2}}
\right),
\end{equation}
\begin{equation}
E_{\rm orbital} =
- G
\frac{M_{1} M_{2}}
{R_{1}+R_{2}},
\end{equation}
\begin{equation}
\label{eq:Erad}
E_{\rm radiative} =
C_{\rm rad}
f_{\rm ColdGas}
E_{\rm initial},
\end{equation}
Here, $M_{\star,i}$ denotes the stellar mass of each progenitor, $M_{\star,\rm burst}$ is the stellar mass formed during the merger-induced burst, and \mbox{$M_i = M_{\star,i} + M_{{\rm ColdGas},i}$} represents the total baryonic mass of each galaxy. The quantities $R_i$ correspond to the stellar half-mass radii of the progenitors. The gas fraction entering the dissipation term is defined as $f_{\rm ColdGas} =
{(M_{{\rm ColdGas},1}+M_{{\rm ColdGas},2})}/{(M_{1}+M_{2})}$. Cold gas contributes to the initial binding energy and to radiative dissipation during the merger, while the final binding term describes the stellar spheroid/bulge.

The value of $C_{\rm rad}$ can vary from $\sim 0$ to $\sim 2.5$ depending on galaxy morphology \citep[see][and references therein]{Irodotou2019}. In this work, motivated by the calibration of \citet{Covington2008}, we adopt $C_{\rm rad}\sim 1$ for quantifying the efficiency of radiative losses. Gas-rich (``wet'') mergers therefore yield more compact remnants due to dissipation.

In minor mergers, the central stellar disc is assumed to survive the encounter, while the stellar mass of the satellite is added to the bulge of the central galaxy. Thus, only the pre-existing central bulge and the stellar component of the satellite enter the energy calculation. The final binding energy of the remnant bulge is
\begin{equation}
E_{\rm final} =
- G
\frac{\left(M_{\rm bulge,1}+M_{\star,2}\right)^2}
{R_{\rm fin}},
\end{equation}
where the corresponding initial binding energy is
\begin{equation}
E_{\rm initial} =
- G
\left(
\frac{M_{\rm bulge,1}^2}{R_{\rm bulge,1}}
+
\frac{M_{\star,2}^2}{R_{2}}
\right),
\end{equation}
and the orbital energy contribution is
\begin{equation}
E_{\rm orbital} =
- G
\frac{M_{\rm bulge,1} M_{\star,2}}
{R_{\rm bulge,1}+R_{2}}.
\end{equation}
Here, $M_{\rm bulge,1}$ and $R_{\rm bulge,1}$ are the stellar mass and half-mass radius of the pre-existing central bulge, while $M_{\star,2}$ and $R_2$ are the stellar mass and half-mass radius of the satellite. The cold gas contributes through dissipation, which is accounted for via the radiative correction term (\equref{eq:Erad}). Mergers also trigger quasar-mode SMBH accretion, the corresponding prescriptions are described in \secref{sec:BHprocesses}.

In the gas-poor regime, we classify minor mergers as effectively dry when $f_{\rm ColdGas} < 0.1$. This threshold is implemented as a discrete transition in the current model. For such events, we apply the phenomenological size-growth model motivated by the dissipationless minor-merger scaling of \citet{Naab2009}. This is introduced to capture the strong size growth expected from collisionless accretion at late times and to better reproduce the observed sizes of massive galaxies. Dry minor mergers have long been argued to drive substantial radial growth with only modest stellar mass increase \citep[e.g.,][]{Fan2010}.

In this limit, assuming virial equilibrium and negligible relative orbital energy of the accreted material, the remnant size scales as
\begin{equation}
R_{\rm fin} =
R_{\rm cen}
\left(1 + \frac{M_{\rm sat}}{M_{\rm cen}}\right)^2,
\end{equation}
where $R_{\rm cen}$ is the pre-merger half-mass radius of the central galaxy and $M_{\rm sat}/M_{\rm cen}$ is the stellar mass ratio of the satellite to the central galaxy.

While the effects of mergers on galaxy sizes were previously explored at low redshift within the legacy \lgal{} framework \citep{Irodotou2019, IzquierdoVillalba2019, Parente2023}, here we revisit this mechanism across all cosmic epochs and assess its connection with the updated physical processes implemented in this work.

\subsubsection{Disc instabilities}
\label{sec:DI}
Disc instabilities arise when the self-gravity of a galactic disc overcomes rotational support, leading to inward mass transport onto the bulge and structural transformation \citep{Efstathiou1982, Toomre1964}. Previous \lgal{} implementations considered instabilities of the stellar disc only \citep{Henriques2015, Henriques2020}. In this work, we follow and extend the implementations of \citet{Irodotou2019} and \citet{Parente2023} by accounting for the stability of both the stellar and gaseous disc components.

The global stability parameter is defined as
\begin{equation}
\epsilon_{\rm total} =
\frac{M_{\rm disc,stars}\,\epsilon_{\rm stars}
+ M_{\rm disc,ColdGas}\,\epsilon_{\rm ColdGas}}
{M_{\rm disc,stars}+M_{\rm disc,ColdGas}},
\end{equation}
where
\begin{equation}
\epsilon_i =
c_i
\left(
\frac{G M_{\rm disc,i}}
{V_{\rm circ}^2 R_{\rm disc,i}}
\right)^{1/2},
\end{equation}
with $i=\{\mathrm{stars},\mathrm{ColdGas}\}$, $V_{\rm circ}$ the halo circular velocity, $M_{\rm disc,i}$ and $R_{\rm disc,i}$ the mass and scale length of each component. The dimensionless constants $c_{\rm stars}=1$ and $c_{\rm ColdGas}=1.1$ account for structural and dynamical differences between stellar and gaseous discs, following the stability criteria for isolated discs \citep{Efstathiou1982, Christodoulou1995}.

The disc is considered unstable when $\epsilon_{\rm total} > 1$, corresponding to the regime where self-gravity becomes comparable to or exceeds rotational support. In this case, the disc is susceptible to the growth of non-axisymmetric perturbations (e.g. bars and inflows), which drive angular momentum redistribution and mass transport toward the central regions.

The unstable cold gas mass in each radial ring is computed using a free-fall timescale derived from the local cold gas density. In this case, the dense cold gas fraction $f_{\rm dense}$ is treated as a proxy for the star-forming gas fraction in the disc.

For each ring $j$, the dense cold gas surface density is
\begin{equation}
\Sigma_{\rm DenseColdGas}^j =
f_{\rm dense}\,
\frac{M_{\rm ColdGas}^j}{A^j},
\end{equation}
where $A^j$ is the area of the ring. Assuming vertical hydrostatic equilibrium for a thin, isothermal dense-gas layer, we estimate the gas scale height in ring $j$ using the self-gravitating gas-sheet approximation,
\begin{equation}
H^j =
\frac{k_B T_{\rm DenseColdGas}}
{2\pi G \mu m_p \Sigma_{\rm DenseColdGas}^j},
\end{equation}
where $k_B$ is the Boltzmann constant, $m_p$ the proton mass, $\mu \approx 2.3$ the mean molecular weight appropriate for molecular gas, and we adopt a characteristic molecular gas temperature of $T_{\rm DenseColdGas}\approx 10\,\mathrm{K}$ \citep{Mo2010_Book}. This prescription depends only on the local dense-gas surface density and the adopted effective temperature of the gas, and should therefore be regarded as an approximate scale-height estimate.
The characteristic cold-gas volume density is then approximated as
\begin{equation}
\rho^j = 
\frac{M_{\rm ColdGas}^j}{A^j H^j},
\end{equation}
from which the free-fall time follows as $t^j_{\rm ff} = \sqrt{3\pi/(32 G \rho^j)}$. Given the approximate treatment of the vertical gas structure, $t_{\rm ff}^j$ should be regarded as an effective instability timescale rather than the exact free-fall time of the dense gas.

The unstable gas mass during a timestep $\Delta t$ is
\begin{equation}
M_{\rm gas,unstable}^j =
\frac{\Delta t}{t_{\rm ff}^j}
\, M_{\rm ColdGas}^j.
\end{equation}
where the unstable gas mass is capped at the available cold gas mass in each ring.
A fraction of this unstable cold gas accretes onto the central SMBH following the velocity-dependent prescription adopted for merger-driven growth \citep{Kauffmann2000, Henriques2020, Parente2023}:
\begin{equation}
f_{\rm BH,unst} =
\frac{f_{\rm BH}}
{1 + \left(V_{\rm BH,DI}/V_{\rm 200}\right)^2},
\end{equation}
where $f_{\rm BH}$ controls the accretion efficiency and is a free parameter, and the velocity-dependent term introduces a dependence on the depth of the potential well. The saturation velocity is defined as
\begin{equation}
V_{\rm BH,DI} = \alpha_{\rm BH,DI}\, V_{\rm BH},
\end{equation}
where $\alpha_{\rm BH,DI}$ is treated as a free parameter in the MCMC. We adopt $\alpha_{\rm BH,DI} > 1$, such that $V_{\rm BH,DI}$ is larger than the value used for merger-driven growth ($V_{\rm BH}$, also a free parameter; see \secref{sec:BHprocesses}). This choice suppresses excessive SMBH growth and prevents the over-quenching of intermediate-mass galaxies \citep{Parente2023}.
The remaining unstable gas forms new disc stars, after which stellar feedback is applied to the newly formed stellar mass. The stability criterion is then re-evaluated.

If the disc remains unstable after cold gas consumption, stellar mass is transferred from the disc to the bulge in a ring-based iterative manner until $\epsilon_{\rm total} \lesssim 1$. This process preferentially redistributes low-angular-momentum material inward while preserving the outer disc structure \citep{Guo2011}.

If a galaxy already possesses a bulge of mass $M_{\rm initial}$ and half-mass radius $R_{\rm initial}$, and an unstable stellar mass $M_{\rm unstable}$ with half-mass radius $R_{\rm unstable}$ is transferred, the final bulge size is computed using energy conservation:
\begin{equation}
\frac{M_{\rm final}^2}{R_{\rm final}} =
\frac{M_{\rm initial}^2}{R_{\rm initial}}
+
\frac{M_{\rm unstable}^2}{R_{\rm unstable}}
+
\frac{\alpha}{C}
\frac{M_{\rm initial} M_{\rm unstable}}
{R_{\rm initial}+R_{\rm unstable}},
\end{equation}
where $\alpha/C=4.0$ is a structural parameter regulating the interaction energy term. If no bulge existed prior to the instability ($M_{\rm initial}=0$), we set $R_{\rm final}=R_{\rm unstable}$ \citep[see ][]{Guo2011}.

After formation via either mergers or disc instabilities, bulges are assumed to follow a \citet{Jaffe1983} density profile. The half-mass radius obtained from the energy formalism above defines the structural scale of the spherical bulge component.

\subsubsection{Black hole related processes}
\label{sec:BHprocesses}
In this work, just like in legacy \lgal{} variants, stellar feedback efficiently regulates star formation in low-mass galaxies but is insufficient to suppress cooling in massive haloes. To quench star formation in these systems, the model follows \citet{Croton2006} and \citet{Henriques2013winds}, in which central SMBHs regulate galaxy growth through two accretion modes: ``quasar mode'' and ``radio mode''.

In the quasar mode, SMBHs grow via accretion of cold gas funnelled to the halo centre during galaxy mergers. The accreted mass is given by
\begin{equation}
\Delta M_{\rm BH,Quasar} =
\frac{f_{\rm BH} \left(M_{\rm sat}/M_{\rm cen}\right) M_{\rm ColdGas}}
{1 + \left(V_{\rm BH}/V_{200}\right)^2},
\end{equation}
where $M_{\rm sat}$ and $M_{\rm cen}$ are the baryonic masses of the merging galaxies, $M_{\rm ColdGas}$ is their total cold gas mass, $V_{200}$ is the halo virial velocity, and $f_{\rm BH}$ and $V_{\rm BH}$ are free parameters controlling the growth efficiency and velocity scaling. This mode primarily contributes to BH mass growth and is associated with merger-driven starbursts.

In the radio mode, SMBHs undergo continuous low-level accretion from the surrounding hot halo gas. The accretion rate is parameterised as
\begin{equation}
\dot{M}_{\rm BH} =
\kappa_{\rm AGN}
\left(\frac{M_{\rm HotGas}}{10^{11}\,{\rm M}_\odot}\right)
\left(\frac{M_{\rm BH}}{10^8\,{\rm M}_\odot}\right),
\end{equation}
where $\kappa_{\rm AGN}$ is a free parameter. The corresponding energy injection rate is
\begin{equation}
\dot{E}_{\rm radio} = \eta \, \dot{M}_{\rm BH} c^2,
\end{equation}
with efficiency $\eta=0.1$. This energy heats the hot halo gas and suppresses further cooling onto the disc, thereby regulating star formation in massive systems.

In the new \lgal{} model, we adopt a simple phenomenological prescription for early black hole seeding. Once a halo reaches $M_{\rm 200}\geq10^{10}\,h^{-1}{\rm M}_\odot$, it is assumed to host a central black hole with a minimum mass of $10^5\,h^{-1}{\rm M}_\odot$. This choice is motivated by heavy-seed scenarios, in particular direct-collapse black hole models, which predict seed masses of $\sim10^4-10^6\,{\rm M}_\odot$ in atomic-cooling haloes and facilitate the rapid assembly of early SMBHs. 
This approach is broadly consistent with halo-threshold seeding prescriptions adopted in large-volume simulations such as IllustrisTNG \citep{Weinberger2017} and the BRAHMA simulations \citep{Bhowmick2025}. We further assume that efficient black hole growth is suppressed in lower-mass haloes, as stellar feedback can efficiently heat and expel gas in shallow potential wells, limiting sustained accretion. For a more detailed treatment of black hole seeding and growth within \lgal{}, we refer to \citet{Bonoli2025arXiv}.

\subsubsection{Environmental processes}
\label{sec:env_process}
Galaxies are subject to environmental processes such as tidal interactions, ram-pressure stripping, and encounters within dense environments, all of which can significantly alter their structure and star formation activity. In this work, as in legacy versions of \lgal{} \citep{Henriques2015, Henriques2020}, we implement gradual stripping of the hot gas reservoir in satellite galaxies, allowing them to retain a fraction of their hot halo after infall.

Tidal stripping operates on satellites within the virial radius of their host halo. The current implementation does not yet include the LBE prescriptions introduced by \citet{Ayromlou2019new, Ayromlou2021}. Consequently, environmental quenching is limited to the legacy tidal stripping and disruption prescriptions. The treatment of tidal disruption of the stellar and cold-gas components follows \citet{Guo2011} and is applied exclusively to orphan galaxies, i.e. systems that have lost their dark matter subhalo and associated hot gas reservoir.

\subsubsection {Processing of photometry}
\label{sec:colors}
The model stores detailed star-formation and metal-enrichment histories, following the updated implementation described in \citet{Barrera2023}. In the present framework, SFHs can be further tracked on a ring-by-ring basis, enabling spatially resolved post-processing of galaxy properties such as colours, luminosities, metallicities, dust attenuation, and surface-brightness profiles. This allows luminosities, colours, and spectral energy distributions to be computed using any stellar population synthesis model. We adopt the stellar population synthesis models of \citet{MillanIrigoyen2021, MillanIrigoyen2025}. These models provide broad-band photometry broadly consistent with other works \citep[e.g.,][]{Maraston2005_stellar_pop_syn} while improving the treatment of massive stars, low-metallicity populations, and longer wavelength emission. Model snapshots include emission in selected photometric bands, computed on-the-fly based on the SFH. Dust attenuation from the diffuse ISM and molecular clouds is incorporated using the prescriptions of \citet{Devriendt1999_extinction} and \citet{Charlot2000extinction}, respectively, identical to those adopted in previous \lgal{} implementations \citep[e.g.,][]{Henriques2020, Barrera2023}.

\section{Model calibration}
\label{sec:MCMC_routines}

\subsection{Formalism for parameter determination}

Like all semi-analytic and hydrodynamical simulation models of galaxy formation, the new \lgal{} framework contains a number of free parameters that must be set to match observational data. We therefore calibrate the model used in this work using a Markov chain Monte Carlo (MCMC) approach, building a new methodology inspired by \citet{Henriques2009MCMC}.

The MCMC chains need to be evolved for several tens of thousands of steps, requiring repeated executions of the semi-analytic model across parameter space. Running the semi-analytic model on the full $(740\,\mathrm{Mpc})^3$ MTNG box with $4320^3$ particles at each MCMC iteration would therefore be computationally prohibitive. To render the calibration procedure tractable, the MCMC algorithm is instead executed on a randomly selected representative subset of halo merger trees. These trees are selected in halo mass bins such that the halo mass function is reproduced within $5\%$ of that of the full simulation, thereby ensuring that derived galaxy statistics are likewise preserved (see Eq.~\ref{eq:Treeselection} and \appref{app:MCMCcalib} for details). This selection corresponds to an effective volume of approximately $(60\,\mathrm{Mpc})^3$, significantly reducing the computational cost while maintaining the relevant statistical properties of the full simulation. 
The updated MCMC framework also takes advantage of the parallel task-scheduling architecture of the \gadgetfour{} framework, enabling the simultaneous execution of multiple independent MCMC chains and further improving computational efficiency.

All free parameters listed as variable in Table~\ref{tab:MCMC_free_params} are varied simultaneously during the calibration process. The model is constrained using twelve independent observational data sets: the stellar mass function and the fraction of quenched galaxies at $z \simeq 0, 1, 2, 3,$ and $4$, as well as the UV luminosity function at $z \simeq 11$ and $12$. We note that structural observables, radial profiles, metallicities, as well as halo-scale properties, are not included in the calibration and therefore serve as independent tests of the model.

Selecting appropriate observational constraints and assigning suitable weights is non-trivial. Equal weighting of all datasets results in a poor reproduction of the $z=0$ stellar mass function, particularly in the mass range $10 \lesssim \log_{10}(M_{\star}/{\rm M_{\odot}}) \lesssim 12$, where the population is dominated by central galaxies and is strongly regulated by AGN feedback \citep{Vani2025}. Similar to the previous works \citep[e.g.][]{Ayromlou2021}, we therefore adopt a weighting scheme that prioritises the $z=0$ constraints. Within $z=0$, the stellar mass function is given higher weight than the quenched fraction. At higher redshifts ($z>0$), all observational constraints are assigned equal weights, with two exceptions. The quenched fraction at $z=4$ is given increased weight to better constrain the onset of early quenching. Similarly, the UV luminosity functions at $z=11$ and $z=12$ are assigned comparable weights, as they provide key constraints on the density-dependent star formation and regulated stellar feedback prescriptions introduced in this work. These observables are particularly important for calibrating the efficiency of early star formation during the first few hundred million years of cosmic history. 

Our model introduces three additional free parameters associated with the new density-based star formation prescription (see \secref{sec:starformation}), regulated stellar feedback in dense gas (see \secref{sec:regulatingfeedback}), and the updated disc instability treatment (see \secref{sec:DI}). Table~\ref{tab:MCMC_free_params} lists the best-fitting parameters in comparison with previous model variants, along with their corresponding governing equations.
The updated model provides a significant improvement over previous-generation models, particularly at high redshift. The results are discussed in detail in Section~\ref{Sec:Results}. 

\subsection{Observational data for calibration}
\label{sec:obs_constraints}
We calibrate our model using a comprehensive compilation of stellar mass functions (SMFs), red/quenched fractions, and high-redshift ultraviolet luminosity functions (UVLFs). These data are drawn from wide-area low- and high-redshift surveys, multiwavelength deep fields, and recent JWST programmes. Rather than relying on a single observational dataset, we combine multiple independent measurements, following the approach adopted in legacy \lgal{} studies \citep[e.g.,][]{Henriques2009MCMC, Henriques2013winds, Henriques2015, Henriques2020}. Given the systematic uncertainties in stellar mass estimates, survey selection effects, and sample variance, individual datasets can exhibit non-negligible biases. By combining results from different surveys, we aim to construct a more representative and less biased set of constraints for the calibration, thereby reducing the impact of dataset-specific systematics. The observational compilation broadly follows that used for model comparisons in \citet{Vani2025}, but is expanded here to construct the calibration dataset.

At $z\simeq0$, the SMF constraints are anchored to measurements from the Galaxy And Mass Assembly (GAMA) survey, as reported by \citet{Baldry2012, Kelvin2014, Wright2017, Driver2011}, and \citet{Driver2022}. We also incorporate recent refinements by \citet{Sbaffoni2026}. For the local red/quenched fraction, we adopt SDSS-based determinations from \citet{Bell2003} and \citet{Baldry2004}. We additionally incorporate updated low-redshift SMF and quenched fraction constraints from \citet{Xu2025DESI}, who derive revised stellar mass functions and quenched fractions based on the DESI Legacy Imaging Surveys \citep{DESICollaboration2016arXiv, Dey2019}.

At $1 \lesssim z \lesssim 4$, the SMFs and quenched fractions are taken from multi-field, multiwavelength catalogues, including COSMOS/UltraVISTA-based measurements from \citet{Ilbert2013} and ZFOURGE, a FourStar near-infrared medium-band survey covering well-studied extragalactic fields (CDFS, COSMOS, UDS), as analysed by \citet{Tomczak2014}. We further include CANDELS-based compilations that utilise HST multi-band photometry over GOODS-S, GOODS-N, EGS, COSMOS, and UDS, as presented by \citet{Santini2022}. At the highest redshifts used in the calibration ($z\sim3-4$), we incorporate recent JWST-enabled measurements in the COSMOS field from COSMOS-Web NIRCam imaging, based on the works of \citet{Weaver2023, Weibel2024}, and \citet{Shuntov2025}.

For the $z\simeq11-12$ UVLF constraints, we use recent JWST/NIRCam-based determinations combining multiple deep programmes and survey fields to improve dynamic range and mitigate field-to-field variance. These include UVLF measurements derived from early JWST ERO/ERS data and Cycle-1 imaging (e.g. CEERS, GLASS, SMACS0723, and JADES), as reported by  \citet{Donnan2023, Donnan2024} and \citet{Adams2024}, as well as independent determinations from JWST lensing and blank-field campaign studies from \citet{Harikane2023} and \citet{Chemerynska2026}. Collectively, these data provide direct constraints on both the bright end and, in several cases, the faint end of the UVLF at $z\gtrsim10$, which is particularly sensitive to the early efficiency of star formation and feedback in the model.

For each observational constraint used here, we combine the published parametric fits by sampling the reported functional forms at 0.25~dex intervals and propagating the quoted uncertainties to construct a representative constraint used in the MCMC likelihood evaluation described in \appref{app:MCMCcalib}.

\renewcommand{\arraystretch}{1.3}
\setlength{\tabcolsep}{2.5pt}
\begin{table*}
\centering
\caption{Free parameters used in the MCMC calibration in this work and in legacy \lgal{} versions \citep[][here referred to as \citetalias{Henriques2015}, \citetalias{Henriques2020}, \citetalias{Ayromlou2021}]{Henriques2015, Henriques2020, Ayromlou2021}. The parameter set of \citet{Henriques2015} was also adopted in \citet[][]{Barrera2023} (hereafter \citetalias{Barrera2023}).}
\label{tab:MCMC_free_params}
\begin{tabular}{|*{7}{c|}}
\hline 
Model Parameter & Equation in \textsc{L-Galaxies} & \citetalias{Henriques2015}/\citetalias{Barrera2023} & \citetalias{Henriques2020} & \citetalias{Ayromlou2021} & This Work & Units \\
\hline 

\hline

\multicolumn{7}{|c|}{{Star formation}} \\
\hline

$f_{\rm dense}$ (Dense cold gas fraction)&
${\rm SFR}_{\rm galaxy} =
{\rm SFR}_{\rm cl} \times
f_{\rm dense}\,
{M_{\rm ColdGas}}/{M_{\rm cl}}$ 
& -- & -- & -- & $0.63$$^\ddagger$ & \\

$\alpha_{\rm SF}$ (Star formation efficiency)& $\Sigma_{\rm SFR} = \alpha_{\rm SF}\, \Sigma_{\rm ColdGas}/t_{\rm dyn}$  & 0.025$^{**}$ & 0.06$^{**}$ & 0.073$^{**}$ & --$^{**}$ & \\

$\alpha_{\rm SF,burst}$ (Star formation burst efficiency)&
$M_{\star,\rm burst} = \alpha_{\rm SF,burst}\left(\frac{M_1}{M_2}\right)^{\beta_{\rm SF,burst}}M_{\rm ColdGas}$ 
& 0.60 & 0.50 & 0.116 & $0.10$ & \\

$\beta_{\rm SF,burst}$ (Star formation burst slope)&
same as above 
& 1.9 & 0.38 & 0.674 & $0.14$ & \\

\hline
\multicolumn{7}{|c|}{{Stellar feedback$^*$}} \\
\hline

$\epsilon_{\rm reheat}$ (Mass-loading efficiency)&
$\epsilon_{\rm disc} = \epsilon_{\rm reheat}\left[0.5+\left(\frac{V_{\rm max}}{V_{\rm reheat}}\right)^{-\beta_{\rm reheat}}\right]$ 
& 2.6 & 5.6 & 9.7 & $1.67$ & \\

$V_{\rm reheat}$ (Mass-loading scale)&
same as above 
& 480 & 110 & 119 & $112.11$ & ${\rm km\,s^{-1}}$ \\

$\beta_{\rm reheat}$ (Mass-loading slope)&
same as above 
& 0.72 & 2.9 & 2.9 & $3.39$ & \\

$\eta_{\rm eject}$ (Supernova ejection efficiency)&
$\epsilon_{\rm halo} = \eta_{\rm eject}\left[0.5+\left(\frac{V_{\rm max}}{V_{\rm eject}}\right)^{-\beta_{\rm eject}}\right]$ 
& 0.62 & 5.5 & 9.56 & $0.68$ & \\

$V_{\rm eject}$ (Supernova ejection scale)&
same as above 
& 100 & 220 & 172 & $114.70$ & ${\rm km\,s^{-1}}$ \\

$\beta_{\rm eject}$ (Supernova ejection slope)&
same as above 
& 0.80 & 2.0 & 1.88 & $2.94$ & \\

$\gamma_{\rm reinc}$ (Ejecta reincorporation) &
$t_{\rm reinc} = \gamma_{\rm reinc}\frac{10^{10}\,\rm M_{\odot}}{M_{\rm 200}}$ 
& $3.0\times 10^{10}$ & $1.2\times 10^{10}$ & $6.6\times 10^{9}$ & $5.45\times10^{10}$ & ${\rm yr}$ \\

\hline
\multicolumn{7}{|c|}{{Regulated stellar feedback}} \\
\hline

$t_{\rm ff,crit}$ (Critical free-fall time)&
$\Sigma_{\rm crit,fb} =
{\pi R_{\rm fb}}/
{8 G t_{\rm ff,crit}^2}$ 
& -- & -- & -- & $2.34$$^\ddagger$ & Myr\\

\hline
\multicolumn{7}{|c|}{{Black hole growth and AGN feedback}} \\
\hline

$\kappa_{\rm AGN}$ (Radio feedback efficiency)&
$\dot{M}_{\rm BH} = \kappa_{\rm AGN}\left(\frac{M_{\rm HotGas}}{10^{11}\,\rm M_{\odot}}\right)\left(\frac{M_{\rm BH}}{10^8\, \rm M_{\odot}} \right)$ 
& $5.3\times 10^{-3}$ & $2.5\times 10^{-3}$ & $8.5\times 10^{-3}$ & $6.25\times10^{-3}$ & ${\rm M_{\odot}}\,{\rm yr}^{-1}$ \\

$f_{\rm BH}$ (Black hole growth efficiency)&
$\Delta M_{\rm BH,Q} = \frac{f_{\rm BH}(M_{\rm sat}/M_{\rm cen})\, M_{\rm ColdGas}}{1+(V_{\rm BH}/V_{\rm 200})^2}$ 
& 0.041 & 0.066 & 0.011 & $0.08$ & \\

$V_{\rm BH}$ (Quasar growth scale) &
same as above 
& 750 & 700 & 1068 & $720.53$ & ${\rm km\,s^{-1}}$ \\

$\alpha_{\rm BH, DI}$ (Disc instability cutoff factor) &
$f_{\rm BH,unst} =
\frac{f_{\rm BH}}
{1 + \left(\alpha_{\rm BH,DI}\, V_{\rm BH}/V_{\rm 200}\right)^2},$
& -- & -- & -- & $5.84$$^\ddagger$ &  \\

\hline
\multicolumn{7}{|c|}{{Galaxy mergers}} \\
\hline

${R}_{\rm merger}$ (Major-merger threshold) & -- 
& 0.1$^\dagger$ & 0.1$^\dagger$ & 0.1$^\dagger$ & 0.1$^\dagger$ & \\

$\alpha_{\rm friction}$ (Dynamical friction) &
$\alpha_{\rm friction}\frac{V_{\rm 200}r_{\rm sat}^2}{GM_{\rm sat}\ln(1+M_{\rm 200}/M_{\rm sat})}$ 
& 2.5 & 1.8 & 0.312 & $0.42$ & \\

\hline
\multicolumn{7}{|c|}{{Gas inflow}} \\
\hline

$v_{\rm inflow}$ (Gas inflow velocity) &
$v_{\rm inflow}=r/t_v$ 
& --$^{***}$ & 1000$^\dagger$ & 1000$^\dagger$ & 1000$^\dagger$ & ${\rm km\,s^{-1}Mpc^{-1}}$ \\

\hline
\end{tabular}
\vspace{0.3cm}
\begin{minipage}{\textwidth}
\footnotesize
\noindent
$\dagger$ Parameters fixed during the MCMC calibration and not allowed to vary.\\
$^{*}$ For stellar feedback parameters, the effective coupling to the interstellar medium is additionally regulated in this work by the factor $f_{\star\rm, fb}$ (see Eq.~\ref{eq:fRFB}), which modulates the fraction of stellar wind and supernova energy that couples to the gas as a function of cold gas surface density.\\
$^{**}$ In \citetalias{Henriques2015}, star formation is based on the total cold gas surface density. In contrast, \citetalias{Henriques2020} and \citetalias{Ayromlou2021} adopt an $\mathrm{H}_2$-based star formation law. In those models, $\Sigma_{\rm ColdGas}$ effectively corresponds to $\Sigma_{\mathrm{H}_2}$. This parameter is not required in this work.\\
$^{***}$ This parameter was introduced in \citetalias{Henriques2020} and \citetalias{Ayromlou2021} within the radial ring framework, and is therefore not applicable to the non-ring model variant.\\
$^{\ddagger}$ New free parameter introduced in this work.
\end{minipage}
\end{table*}
\renewcommand{\arraystretch}{1}

\section{Results}
\label{Sec:Results}
In \citet{Vani2025}, we showed that legacy \lgal{} variants systematically underpredict the abundance of massive quenched galaxies at high redshift. Building on that work, we first investigated whether the discrepancy could be resolved through an extended recalibration of the \citet{Ayromlou2021} model, including additional high-redshift constraints, in particular the SMF and quenched fraction at $z\simeq 3-4$. However, even with this extended calibration, the model failed to produce a significant population of massive quenched systems. This suggests that the discrepancy is not primarily driven by calibration choices, but instead points to the need for updated, physically motivated, and self-consistent prescriptions governing high-redshift galaxy evolution.

In this work, we therefore revise several key model components, as described in \secref{sec:modelproperties}. After calibrating the updated model and identifying the best-fitting parameter set (see \secref{sec:MCMC_routines}), we run the new \lgal{} model on the full simulation volume, generating a galaxy catalogue containing about 122 million galaxies. Here, we compare the updated model to the \citet{Barrera2023} implementation, which serves as the baseline \lgal{} model within the \gadgetfour{} framework and is based on the legacy \citet{Henriques2015} model. The results are presented in this section.

\subsection{Evolution of global galaxy properties}
\label{sec:global_prop}
\subsubsection{The stellar mass function}

The stellar mass function (SMF) is computed by counting galaxies in logarithmic stellar mass bins, dividing by the comoving simulation volume, and normalising by the bin width. A sliding-bin approach is adopted to reduce binning noise, following \citet{Vani2025}.

\begin{figure*}
    \centering
    \includegraphics[width=0.24\linewidth]{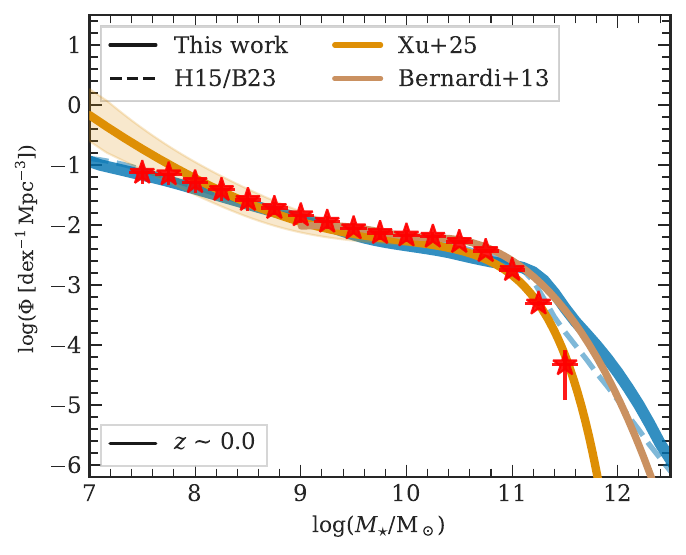}
    \includegraphics[width=0.24\linewidth]{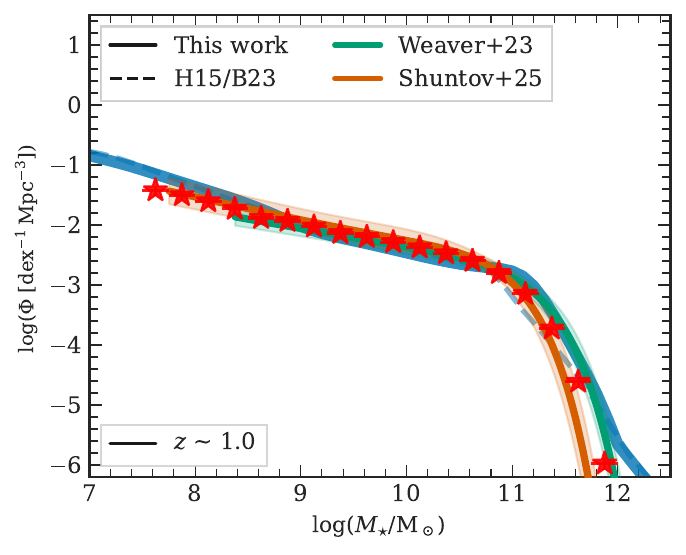}
    \includegraphics[width=0.24\linewidth]{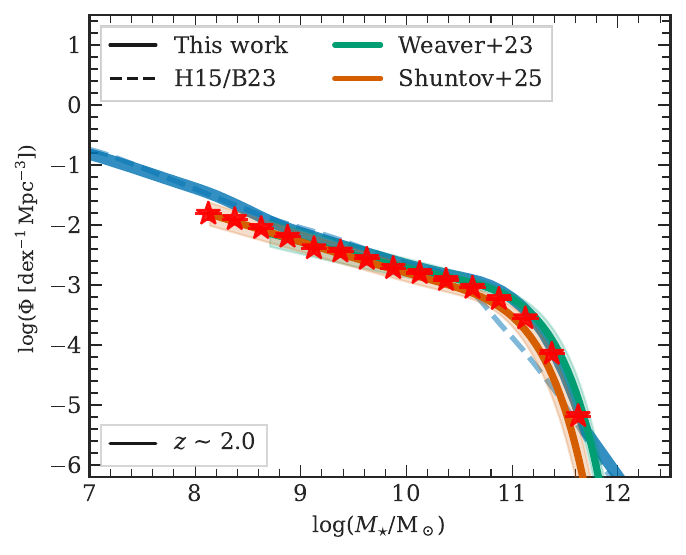}
    \includegraphics[width=0.24\linewidth]{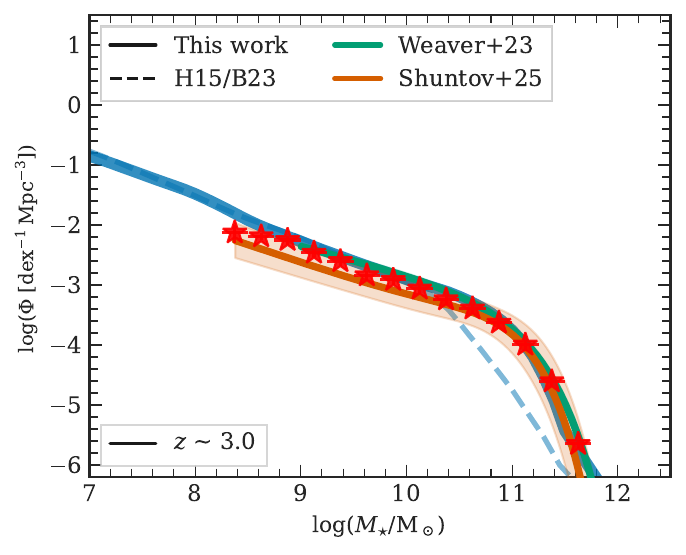}
    \includegraphics[width=0.24\linewidth]{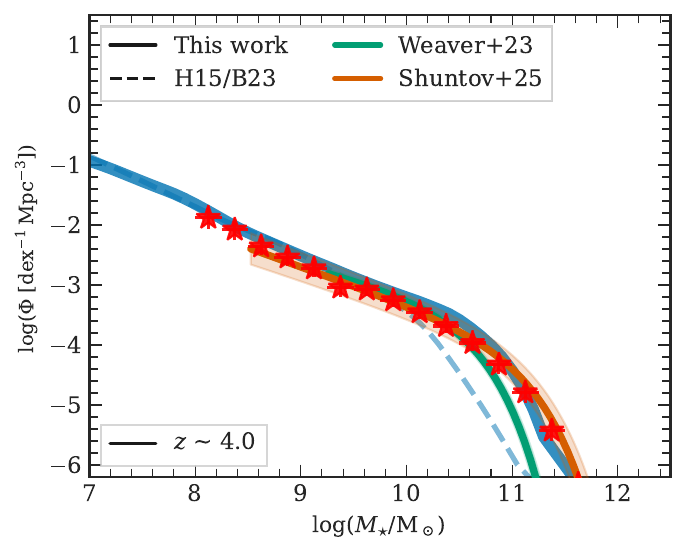}
    \includegraphics[width=0.24\linewidth]{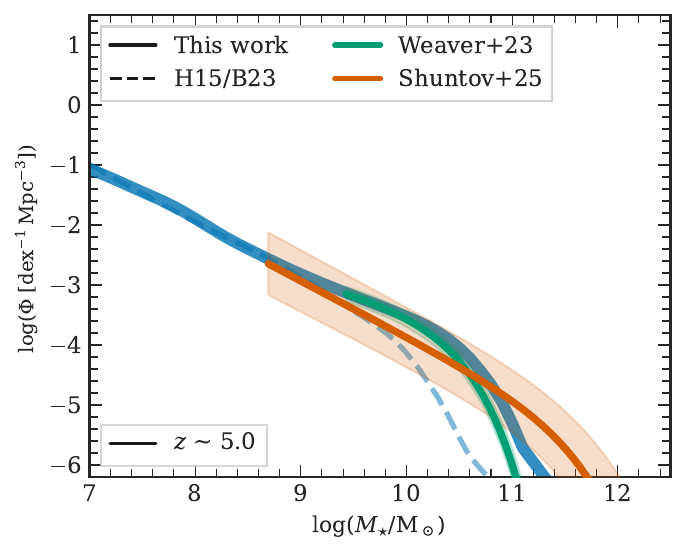}
    \includegraphics[width=0.24\linewidth]{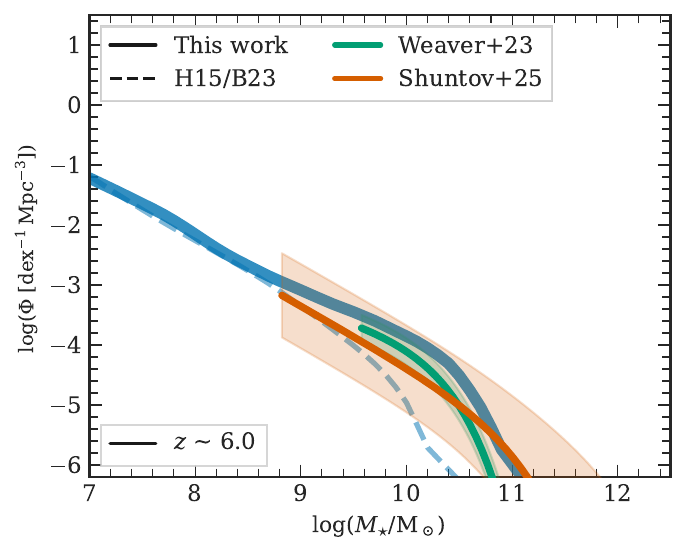}
    \includegraphics[width=0.24\linewidth]{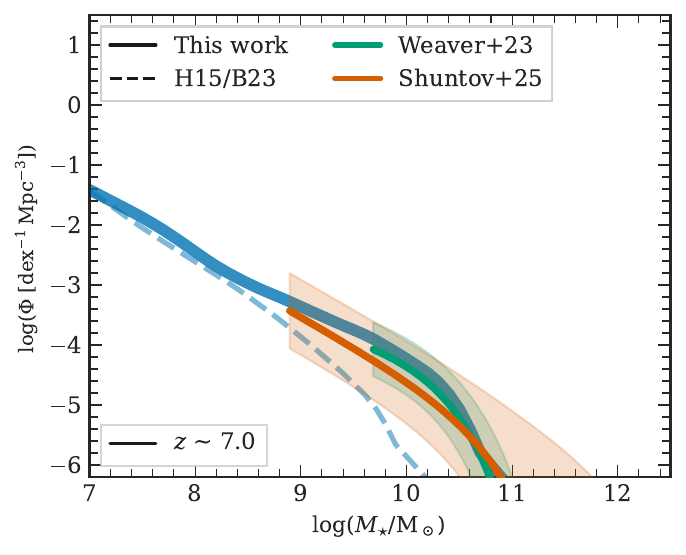}
    \includegraphics[width=0.24\linewidth]{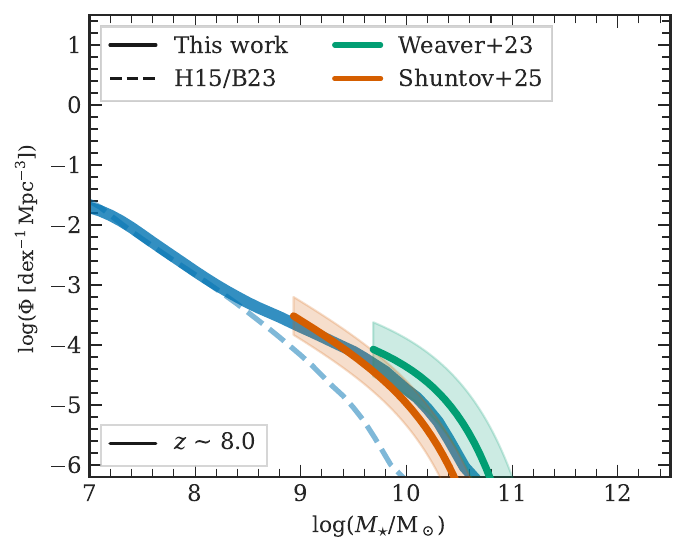}
    \includegraphics[width=0.24\linewidth]{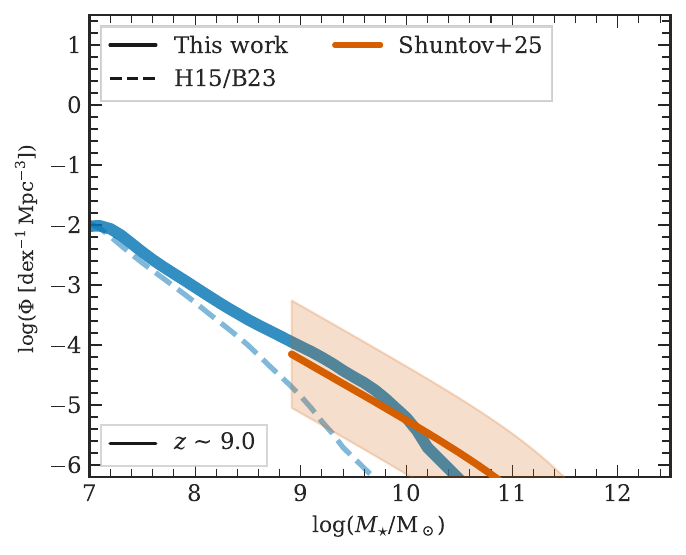}
    \includegraphics[width=0.24\linewidth]{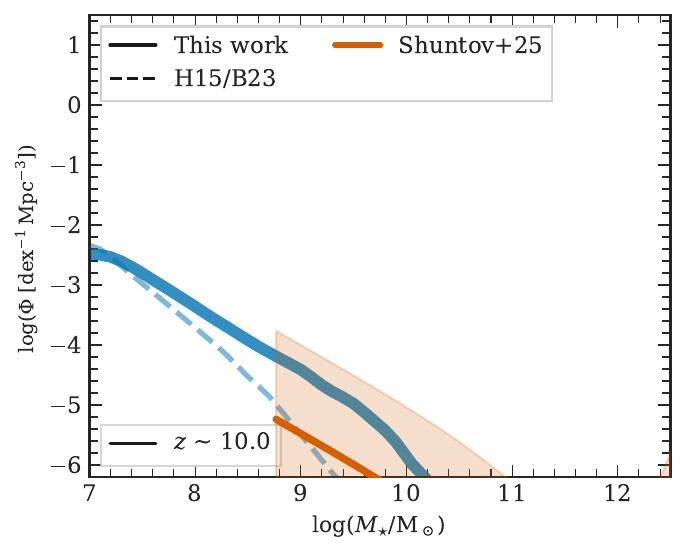}
    \includegraphics[width=0.24\linewidth]{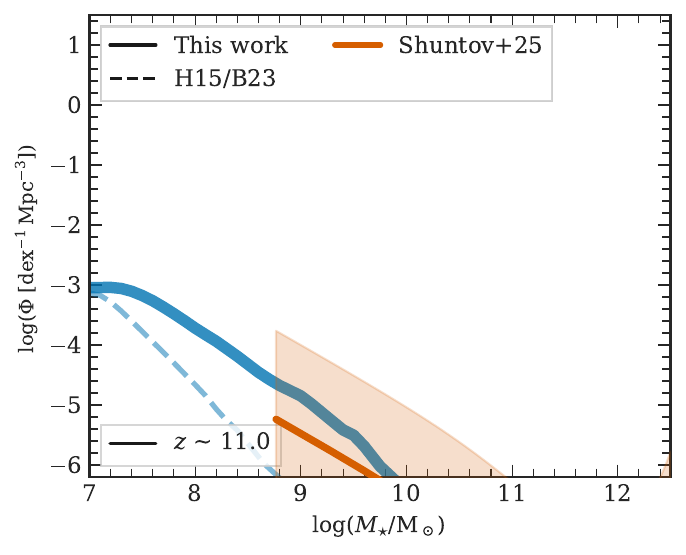}
    \caption{Total stellar mass function from $z=0$ to $z=11$. Solid blue and dashed grey curves show the predictions of this work and the \citetalias{Henriques2015}/\citetalias{Barrera2023} model, respectively. Observational constraints are taken from \citet{Xu2025DESI} and \citet{Bernardi2013} at $z=0$, and from \citet{Weaver2023} and \citet{Shuntov2025} at higher redshifts, where available with quoted uncertainties. Red star symbols indicate the observational datasets included in the MCMC calibration, with the corresponding error bars.}
    \label{fig:SMF}
\end{figure*}

The evolution of the total stellar mass function is shown in \figref{fig:SMF}. The solid blue line represents this work, the dashed line indicates legacy work, and other colours correspond to observations. Overall, the updated model reproduces the observed evolution of the SMF over more than five orders of magnitude in stellar mass and from $z=0$ to $z\approx 11$. Compared to the legacy \citetalias{Henriques2015}/\citetalias{Barrera2023} implementation, the new model predicts systematically larger stellar masses at early times while maintaining good agreement with low-redshift observations.

At $z=0$, the predicted SMF is in good agreement with the recent DESI measurements of \citet{Xu2025DESI}, reproducing the low-mass slope and the overall shape of the mass function. Around the characteristic knee ($M_\star\sim10^{11}\,{\rm M_\odot}$), the updated model closely follows the legacy implementation and remains in reasonable agreement with the observations.
The small remaining offset in the knee reflects the balance struck by the joint calibration, which prioritises simultaneous agreement across a wide redshift range ($z\simeq0-4$ and $z\simeq11-12$) and multiple observables over an exact match to any single one. The adopted parameter set, therefore, represents a deliberate compromise; a marginally better knee could likely be obtained by refined weighting of the calibration datasets. At the massive end ($M_\star\gtrsim10^{11}\,{\rm M_\odot}$), the updated model exhibits a decline similar to the legacy implementation and remains in reasonable agreement with the local SMF of \citet{Bernardi2013}, derived using Sérsic--exponential profile fitting and characterised by a relatively shallow high-mass end due to the recovery of extended stellar envelopes.

At $1\lesssim z\lesssim4$, where the model is also calibrated against the observed evolution of the SMF (red stars representing the calibration data), the agreement with the measurements of \citet{Weaver2023} and \citet{Shuntov2025} remains good over the full stellar mass range. The \citet{Weaver2023} measurements, based on the COSMOS-Web survey, provide robust constraints over a wide stellar mass range, while \citet{Shuntov2025} extend these measurements to higher redshifts using deep JWST observations, offering improved constraints on the abundance of massive galaxies during the first few billion years of cosmic time. The updated model reproduces both the evolution of the characteristic knee and the abundance of massive galaxies, generally providing a better match than the legacy implementation, particularly around the exponential cutoff. 

The most significant differences between the two model variants emerge at $z\gtrsim 5$, where the stellar mass function provides a largely independent test of the model, modulo the inclusion of the $z=11-12$ UVLF in the MCMC calibration. Compared with the legacy implementation, the updated prescriptions produce a systematically larger population of massive galaxies, increasing the number densities by $\sim0.6$ dex at intermediate redshift and intermediate stellar masses, rising to as much as $\sim2$ dex for the most massive galaxies ($M_\star\gtrsim10^{10}\,{\rm M_\odot}$) at the highest redshifts ($z\gtrsim9$). Consequently, the updated model is able to reproduce the observed abundance of galaxies with stellar masses approaching $10^{10}$--$10^{11}\,{\rm M_\odot}$ out to $z\sim9-11$. The improvement is particularly evident at the exponential cutoff of the SMF, where the updated model follows the observations of \citet{Weaver2023} and \citet{Shuntov2025} considerably more closely. This enhanced agreement arises naturally from the combination of density-dependent star formation and regulated stellar feedback, which together promote more rapid stellar mass assembly in the dense, gas-rich environments characteristic of the early Universe.

The predicted low-mass slope remains broadly consistent with the available observations within the current uncertainties. It should also be noted that observational constraints become increasingly incomplete below $M_\star\sim10^{9.5}\,{\rm M_\odot}$ at $z\gtrsim 5$, limiting the strength of conclusions that can presently be drawn in this regime.

Overall, the updated model simultaneously maintains reasonable agreement with the well-constrained low-redshift stellar mass function while significantly improving the predicted abundance of massive galaxies during the first billion years of cosmic time.

\subsubsection{The UV luminosity function}
\begin{figure*}
    \centering
    \includegraphics[width=0.33\linewidth]{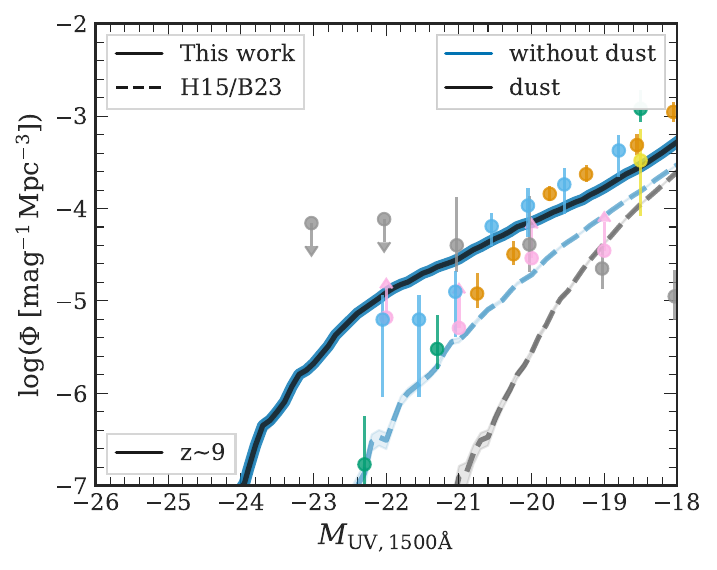}
    \includegraphics[width=0.33\linewidth]{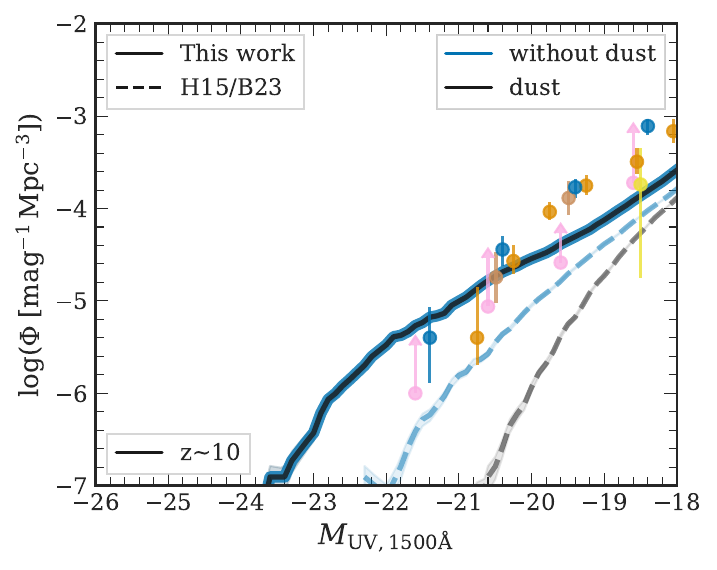}
    \includegraphics[width=0.33\linewidth]{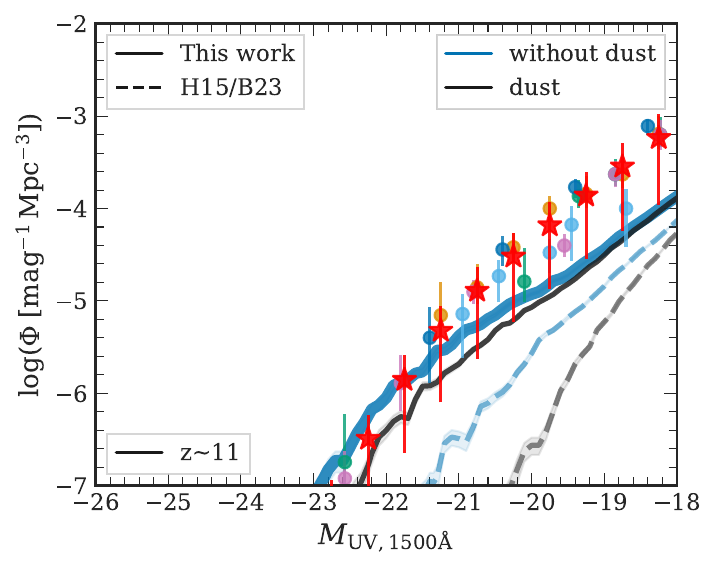}
    \includegraphics[width=0.33\linewidth]{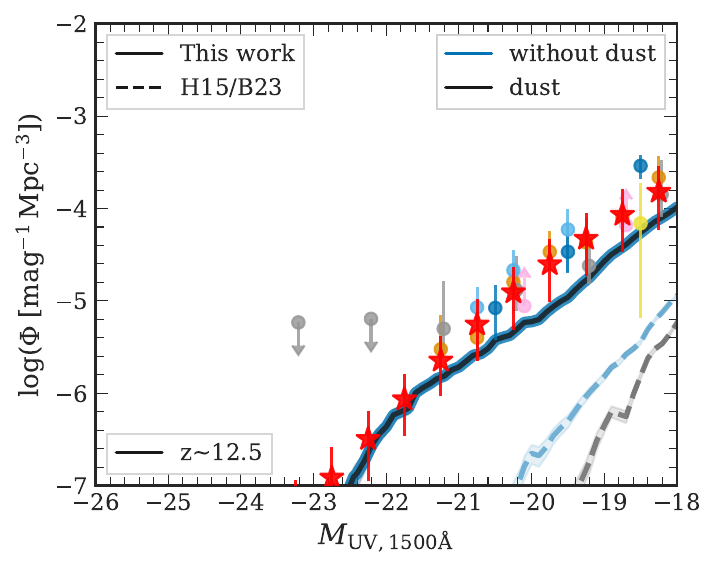}
    \includegraphics[width=0.33\linewidth]{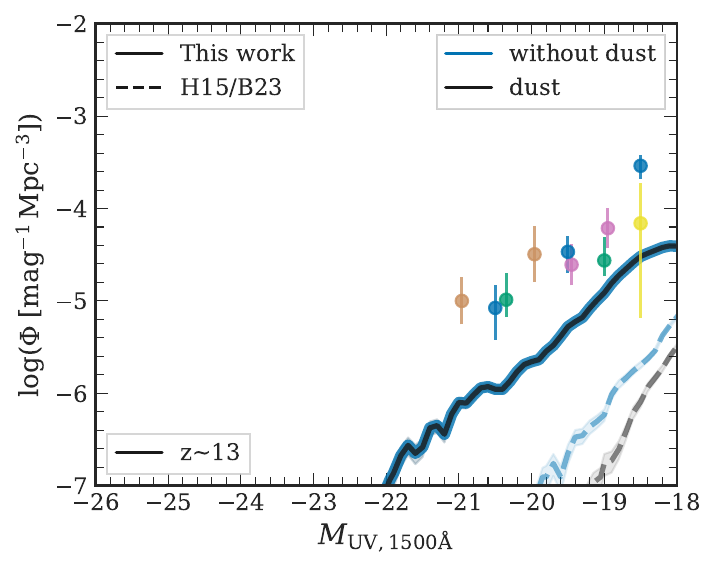}
    \includegraphics[width=0.33\linewidth]{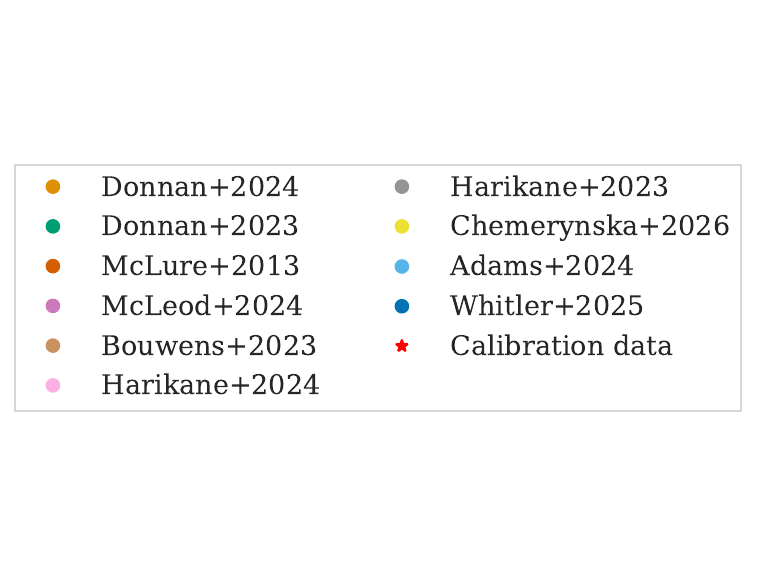}
\caption{Rest-frame UV luminosity function at $z\simeq9-13$. Solid and dashed curves show the predictions of this work and the \citetalias{Henriques2015}/\citetalias{Barrera2023} model, respectively. Black and blue curves denote the UVLF, including and excluding dust attenuation. Coloured symbols with error bars show observational measurements from \citet{Adams2024, Bouwens2023, Chemerynska2024, Donnan2023, Donnan2024, Harikane2023, Harikane2024, McLure2013}, and \citet{Whitler2025}, where available. Red star symbols indicate the observational datasets included in the MCMC calibration, with the corresponding error bars representing the adopted observational uncertainties.}
    \label{fig:UVLF}
\end{figure*}

The UV luminosity function (UVLF) is constructed in a manner analogous to the SMF. We compute the rest-frame UV luminosity at 1500\,\AA{} from the star formation histories of individual galaxies using the adopted stellar population synthesis (SPS) model (see \secref{sec:colors}). We present the UVLF both with and without dust attenuation, shown as black and blue lines, respectively, in \figref{fig:UVLF}. The solid lines correspond to the updated model presented in this work, while the dashed lines show the predictions of the legacy implementation. Observational measurements from various studies are overplotted as coloured points.

We caution that the dust correction is applied in post-processing using attenuation prescriptions calibrated primarily at lower redshifts (see \secref{sec:colors}). Such prescriptions may not fully capture the dust properties of galaxies at $z\gtrsim10$, where the dust content, grain-size distribution, dust-star geometry, and covering fraction are still highly uncertain. Recent JWST studies indicate that attenuation curves can evolve significantly with redshift and may differ from commonly adopted local templates \citep[e.g.,][]{Markov2025, Shivaei2025arXiv}. In addition, very high-redshift galaxies may have low dust attenuation, rapidly forming dust dominated by supernova production, or highly anisotropic and porous dust distributions \citep[e.g.,][]{Langeroodi2024arXiv, Burgarella2025}. Therefore, the dust-attenuated UVLF should be interpreted as an approximate estimate, while the dust-free prediction provides a useful reference for the intrinsic UV output of the model.

Figure~\ref{fig:UVLF} shows that the updated model predicts a substantially larger abundance of UV-bright galaxies than the legacy \citetalias{Henriques2015}/\citetalias{Barrera2023} implementation across the entire redshift range $z\simeq9-13$. The improvement is most pronounced at the bright end ($M_{\rm UV}\lesssim-20$), where the updated prescriptions increase the predicted number densities by up to $\sim2$ dex, corresponding to a factor of $\sim100$x more galaxies than the legacy model. Consequently, the updated model reproduces the observed abundance of luminous galaxies detected with JWST far more successfully, particularly at $z\gtrsim10$ where the legacy model predicts very few galaxies brighter than $M_{\rm UV}\sim-20$.

The enhanced bright-end UVLF follows naturally from the increased efficiency of early star formation in dense gas. The density-dependent star formation law allows gas-rich galaxies to assemble stellar mass more rapidly, while regulated stellar feedback delays the disruption of dense star-forming clouds, enabling sustained star formation during the first few hundred million years of cosmic time. Together, these effects produce both a larger population of massive galaxies and higher intrinsic UV luminosities, yielding much better agreement with recent JWST observations.

Despite this significant improvement, some tension remains at the brightest magnitudes, particularly at $z\gtrsim12.5$, where the model still predicts fewer galaxies than inferred from the most recent observational measurements. Reducing this discrepancy will likely require additional physical ingredients beyond those considered here, including improved modelling of early dust attenuation, or variations in the stellar initial mass function \citep[e.g., ][]{Chon2021, Gelli2024, Markov2025}. Nevertheless, the updated model reduces the discrepancy at the bright end of the UVLF by nearly two orders of magnitude relative to the legacy implementation while remaining consistent with intermediate-redshift observations.

\subsubsection{Cosmic star formation rate density}

\begin{figure}
    \centering
    \includegraphics[width=1\linewidth]{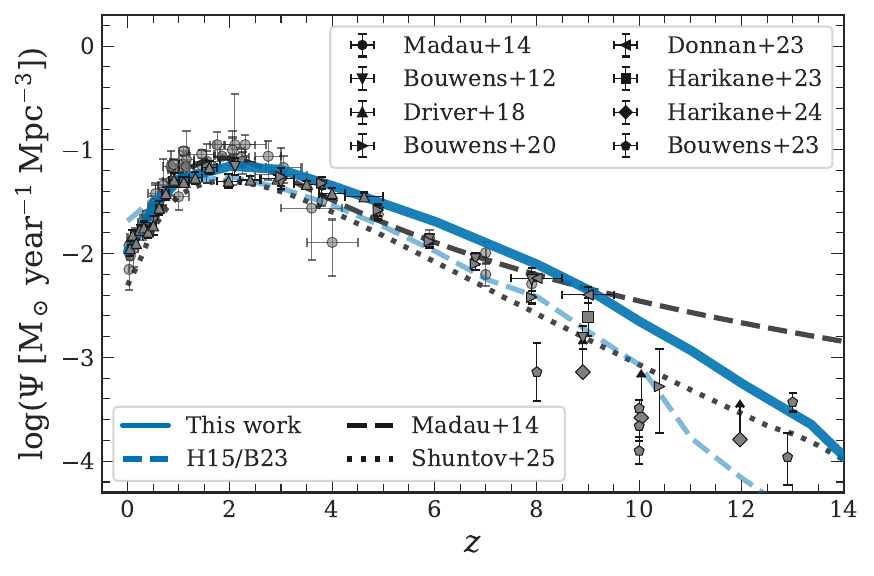}
\caption{Evolution of the cosmic star formation rate density. The solid blue curve shows the predictions of the updated \lgal{} model, while the dashed blue curve corresponds to the legacy \citetalias{Henriques2015}/\citetalias{Barrera2023} implementation. The dashed black and dotted black curves show the empirical compilations of \citet{Madau2014} and \citet{Shuntov2025}, respectively. Observational measurements from the literature are shown as symbols with their quoted uncertainties.}
    \label{fig:SFRD}
\end{figure}

The evolution of the cosmic star formation rate density (CSFRD) is shown in \figref{fig:SFRD}. It is computed by summing the instantaneous star formation rates of all galaxies in the simulation volume, normalised by the comoving volume. We compare the updated model with the legacy \citetalias{Henriques2015}/\citetalias{Barrera2023} implementation, the compilation of \citet{Madau2014}, the empirical reconstruction of \citet{Shuntov2025}, and observational measurements from the literature \citep{Madau2014, Bouwens2012_SFH, Bouwens2023, Driver2018_SFH, Donnan2023, Harikane2023, Harikane2024}.

At low redshift, the updated model reproduces the observed rise of the CSFRD from $z=0$ to a broad peak at $z\sim2-3$, consistent with the established cosmic star formation history \citep{Madau2014}. However, at $z\gtrsim5$, the model predicts a systematically higher CSFRD than most observational estimates and the legacy model. This excess becomes increasingly apparent towards $z\sim6-12$.

The enhanced CSFRD is a direct consequence of the updated dense-gas prescriptions. In high-redshift galaxies, high gas surface densities increase cloud-scale star formation efficiency and reduce the effective coupling of stellar winds and supernova feedback to the ISM in dense regions. Together, these effects allow cold gas to be converted into stars more rapidly before it is reheated or ejected, boosting the global star formation rate density.

This behaviour reflects the production of more massive and UV-bright galaxies at early times (see \figref{fig:UVLF}), but at face value it may also indicate that the model now forms stars too efficiently at $z\gtrsim5$. Moderating an excess CSFRD at high redshift could require  additional regulatory mechanisms  such as  more efficient stellar or AGN feedback, delayed reincorporation of ejected gas, stronger AGN-driven gas removal, or a more detailed treatment of gas accretion and outflows in low-mass haloes \citep[e.g.,][]{Bonoli2025arXiv, Shen2026Thesan}. However, it is also important to note that observational estimates of the CSFRD at $z\gtrsim6$ remain subject to significant uncertainties arising from sample incompleteness, cosmic variance, dust corrections, and the conversion of UV luminosities into star formation rates \citep[e.g.,][]{Madau2014, Bouwens2023, Harikane2024, Shuntov2025}.

\subsection{Properties of quenched galaxies}
\label{sec:QG_prop}
We next examine the population of quenched galaxies predicted by the new \lgal{} model. We classify quenched systems using two complementary approaches. First, we apply a rest-frame NUVrJ colour-colour selection \citep{Ilbert2013}, following observational studies that use broad-band colours to separate quiescent galaxies from dusty star-forming systems \citep[e.g.,][]{Weaver2023, Shuntov2025, Vani2025}. Second, we adopt a specific star formation rate criterion \citep[e.g.,][]{Franx2008},
${\rm sSFR} < {0.2}/{t_{\rm age}(z)}$,
where $t_{\rm age}(z)$ is the age of the Universe at that redshift. This definition provides a physically motivated, redshift-dependent threshold for identifying galaxies whose current star formation is negligible relative to their past average star formation.

Both selections have advantages and limitations. Colour--colour criteria such as NUVrJ are closely tied to observational classifications and help reduce contamination from dusty star-forming galaxies. However, at high redshift the separation between quiescent, dusty, and recently quenched systems becomes more uncertain, and the inferred colours can depend on the adopted stellar population synthesis model \citep[e.g.,][]{MillanIrigoyen2021, Gould2023, Valentino2023, Liu2024_SEDSSP}. 
The sSFR-based criterion is more directly linked to the intrinsic star formation activity of model galaxies and is therefore useful for comparing different model variants. In particular, it uses the instantaneous SFR predicted by the model, which is closely related to observational tracers of recent star formation, such as H$\alpha$ emission.

\begin{figure*}
    \centering
    \includegraphics[width=0.33\linewidth]{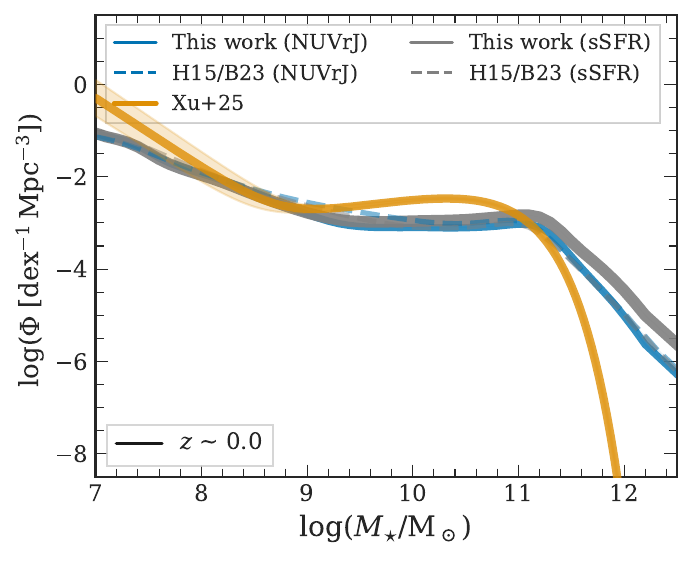}
    \includegraphics[width=0.33\linewidth]{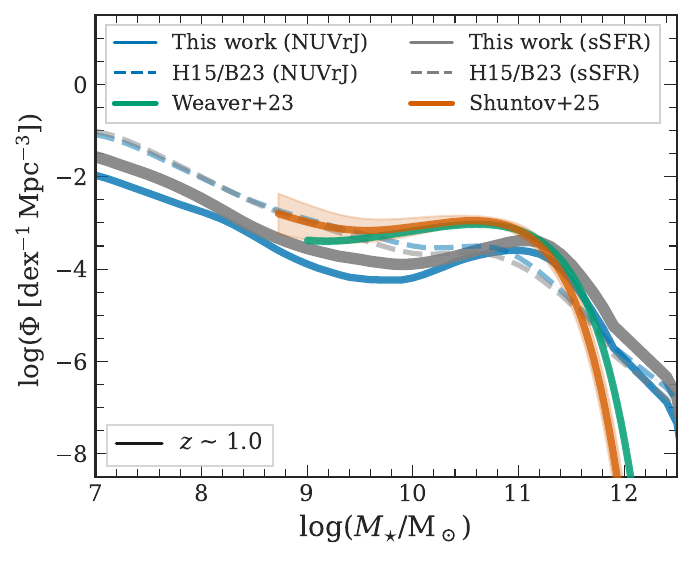}
    \includegraphics[width=0.33\linewidth]{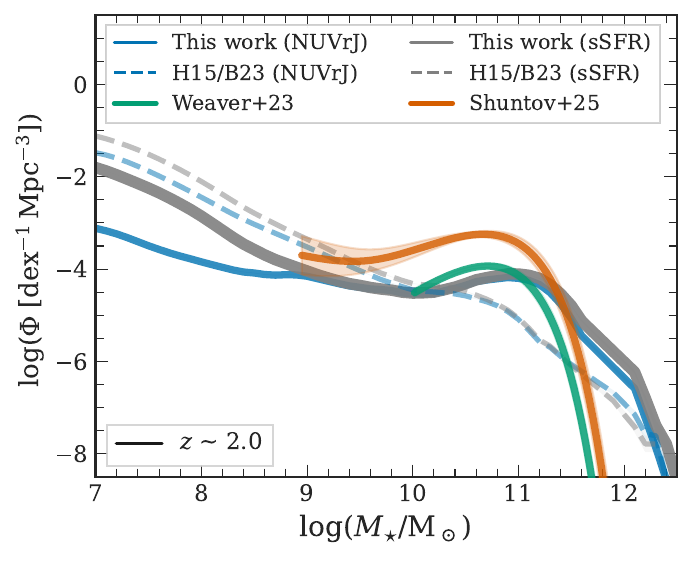}
    \includegraphics[width=0.33\linewidth]{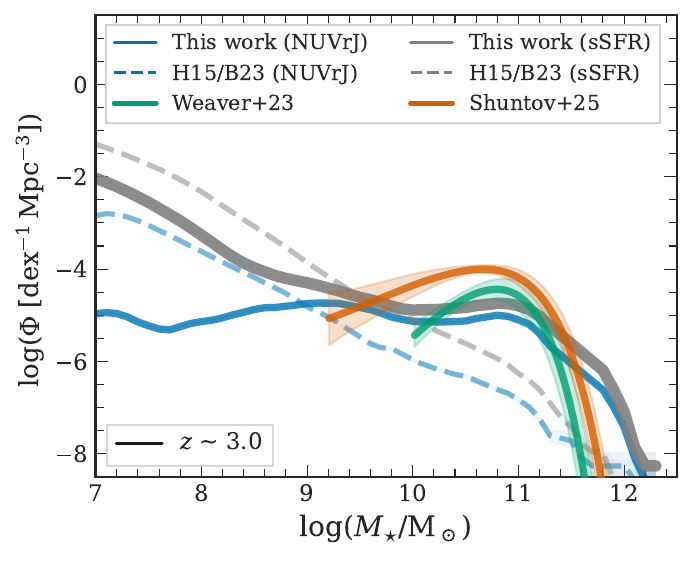}
    \includegraphics[width=0.33\linewidth]{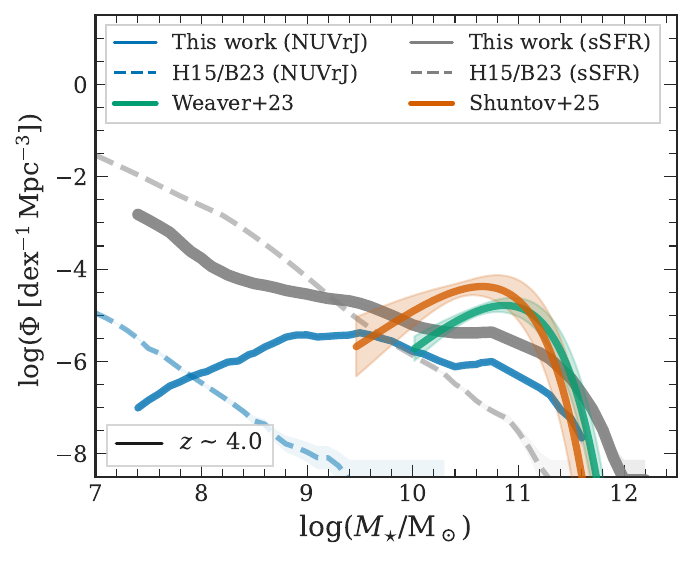}
    \includegraphics[width=0.33\linewidth]{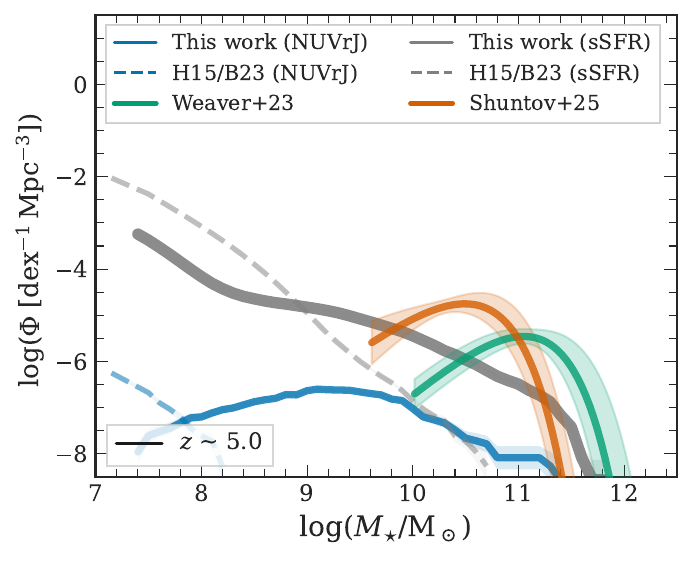}
    \caption{Quenched galaxy stellar mass function at $z=0-5$. Blue curves show the QGSMF obtained using the NUVrJ colour-colour selection, while grey curves show the QGSMF selected using the specific star formation rate criterion, $\mathrm{sSFR}<0.2/t_{\rm age}$. Solid lines correspond to the predictions of this work and dashed lines to the \citetalias{Henriques2015}/\citetalias{Barrera2023} model. Observational measurements are taken from \citet{Xu2025DESI} at $z=0$, and from \citet{Weaver2023} and \citet{Shuntov2025} at $z>0$, where available.}
    \label{fig:QGSMF}
\end{figure*}

Figure~\ref{fig:QGSMF} presents the evolution of the quenched galaxy stellar mass function (QGSMF) from $z=0$ to $z\approx5$. The panels show quenched galaxies selected using the observational NUVrJ colour-colour criterion (blue lines) and the intrinsic instantaneous star formation rate criterion, ${\rm sSFR}<0.2/t_{\rm age}$ (grey lines). Overall, the updated model predicts a substantially larger population of quenched galaxies than the legacy \citetalias{Henriques2015}/\citetalias{Barrera2023} implementation, particularly at intermediate and high redshift for the most massive galaxies.

At $z=0$, both selection methods yield QGSMFs that are in good agreement with the recent DESI measurements of \citet{Xu2025DESI}. The differences between the updated and legacy models are relatively modest, reflecting the fact that both models are calibrated to reproduce local galaxy statistics. The updated model produces a slightly larger abundance of massive quenched galaxies, yielding a steeper decline at the massive end and a sharper exponential cutoff.

The largest differences emerge at $z\gtrsim2$. While the legacy model does not form many massive quenched galaxies with increasing redshift, the updated implementation maintains a substantial population of quenched systems with $M_\star\gtrsim10^{10}\,{\rm M}_\odot$, yielding much better agreement with the measurements of \citet{Weaver2023} and \citet{Shuntov2025}. The improvement is particularly pronounced at $z\simeq3-5$, where the updated model increases the abundance of massive quenched galaxies by about two orders of magnitude relative to the legacy implementation. A pronounced feature also develops around $M_\star\sim10^{10}\,{\rm M}_\odot$, producing a shoulder in the QGSMF that persists to high redshift \citep[e.g.,][]{Santini2022}. This behaviour follows naturally from the revised physical prescriptions introduced in this work. Density-dependent star formation enables rapid early stellar mass assembly, while regulated stellar feedback allows dense gas clouds to sustain star formation for longer. As galaxies assemble their stellar mass more rapidly, they also exhaust or stabilise their gas reservoirs earlier, allowing the existing AGN feedback prescription to quench a substantially larger fraction of massive systems. The enhanced bulge growth driven by gas-rich mergers and disc instabilities further contributes to black hole growth and the maintenance of quiescence.

The two selection criteria produce qualitatively similar trends but differ systematically at high redshift. The sSFR selection generally predicts a larger abundance of quenched galaxies, particularly at lower stellar masses, because it is based directly on the instantaneous star formation activity of the model galaxies. In contrast, the NUVrJ selection depends on synthetic broad-band colours, which are influenced by the adopted stellar population synthesis (SPS) model, dust attenuation, metallicity, and the recent star formation history. At high redshift, these effects become increasingly uncertain, and the inferred colours may vary between different SPS models and dust prescriptions \citep[e.g.,][]{Conroy2013, Gould2023, Valentino2023}. Consequently, while the NUVrJ selection facilitates direct comparison with observations, the sSFR criterion provides a cleaner theoretical definition of quenching for comparing different model variants.

At the low-mass end ($M_\star\lesssim10^{8}\,{\rm M}_\odot$), the agreement with the observed quenched population remains less satisfactory. This is partly due to the simplified environmental model inherited from the \citetalias{Henriques2015}/\citetalias{Barrera2023} implementation, as the existing MTNG merger trees do not provide the halo properties required by the more advanced ram-pressure and tidal stripping prescriptions of \citet{Ayromlou2019new, Ayromlou2021}. These environmental processes are expected to increase the abundance of quenched low-mass satellites and, through a more self-consistent evolution of satellite populations and merger histories, also improve the agreement with the observed population of massive quenched galaxies \citep{Vani2025}. Furthermore, the downturn of the QGSMF at low stellar masses reflects that most galaxies with $M_\star\lesssim10^{9}-10^{9.5}\,{\rm M}_\odot$ remain actively star forming at high redshift, while observational constraints in this regime are also affected by incompleteness and uncertainties in quiescent-galaxy selection \citep[see][]{Baker2026,Doven2026arXiv}. Incorporating the improved environmental model within the MTNG framework will be explored in future work.

It is important to note that the MCMC calibration was performed using the evolution of the quenched fraction, with the best-fitting model compared directly with the calibration data in \figref{Fig:f_qg}. Consequently, the agreement shown here is not directly imposed by the calibration but emerges naturally from reproducing the observed quenched fractions. Compared with the legacy implementation, the updated model produces significantly more quenched galaxies across all redshifts and stellar masses, yielding a quenched fraction that is more consistent with the observational constraints. 

\begin{figure*}
    \centering
    \includegraphics[width=0.45\linewidth]{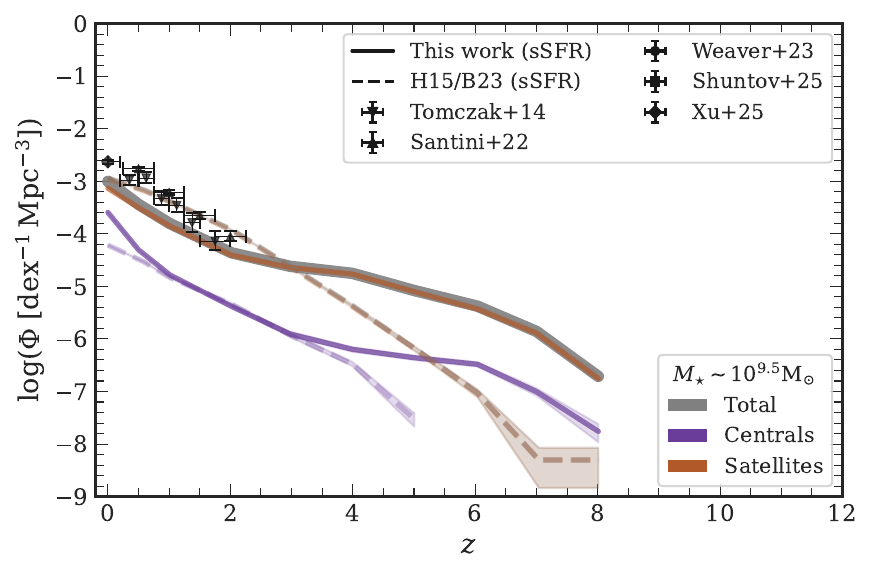}
    \includegraphics[width=0.45\linewidth]{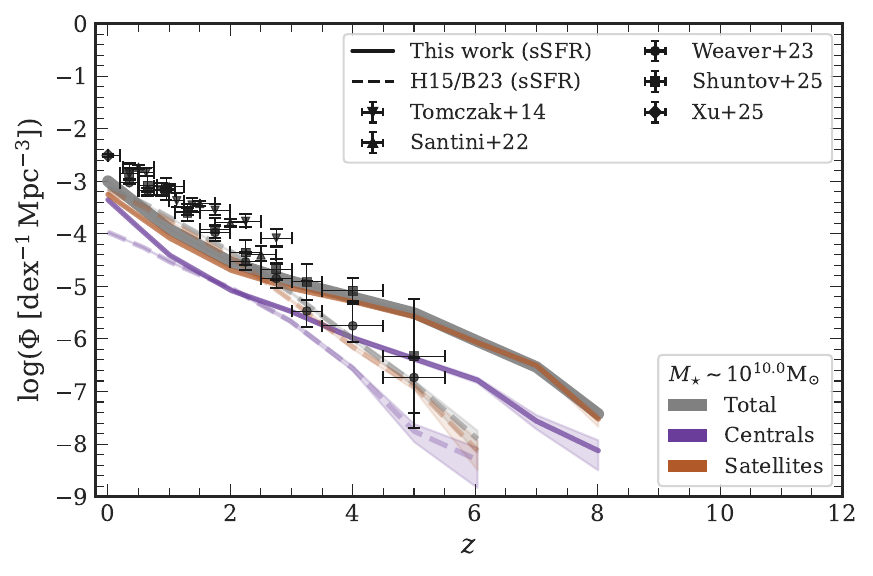}
    \includegraphics[width=0.45\linewidth]{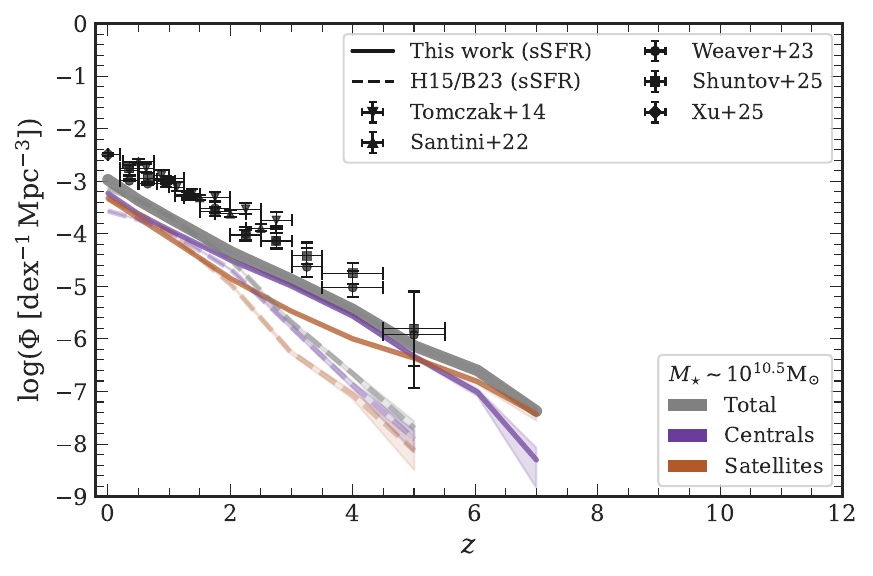}
    \includegraphics[width=0.45\linewidth]{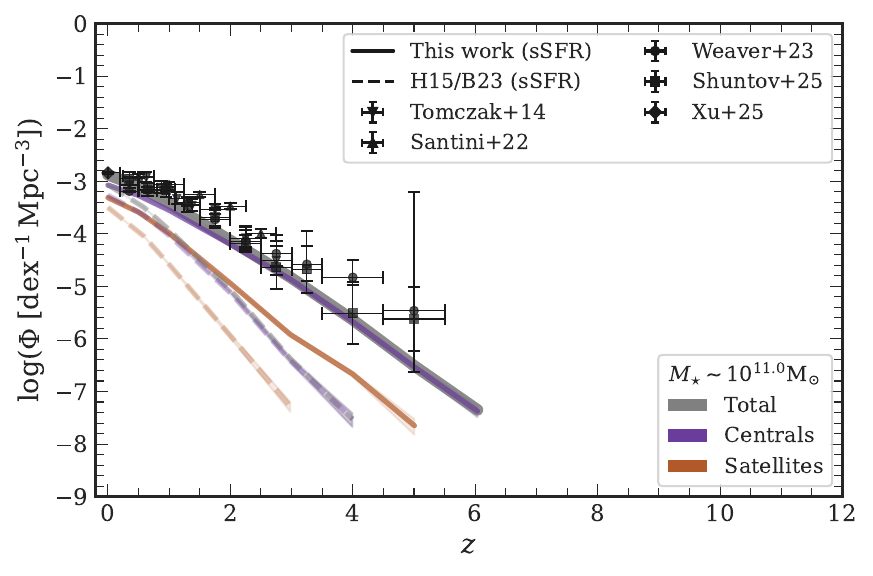}
    \caption{Evolution of the number density of quenched galaxies of stellar mass of $M_\star\sim 10^{9.5}$, $10^{10}$, $10^{10.5}$, and $10^{11}\,{\rm M_\odot}$. The panels show quiescent galaxies selected using the sSFR criterion (${\rm sSFR}<0.2/t_{\rm age}$). Solid and dashed curves correspond to the updated model and the \citetalias{Henriques2015}/\citetalias{Barrera2023} model, respectively. Thick curves show the total quenched population, while the purple and brown curves indicate the contributions from central and satellite galaxies. Observational measurements from \citet{Tomczak2014, Santini2022, Weaver2023, Xu2025DESI}, and \citet{Shuntov2025} are shown where available, with quoted uncertainties.}
    \label{fig:N_qg}
\end{figure*}

Figure~\ref{fig:N_qg} shows the evolution of the number density of quenched galaxies above stellar mass thresholds of $M_\star\sim 10^{9.5}$, $10^{10}$, $10^{10.5}$, and $10^{11}\,{\rm M}_\odot$ as a function of redshift. The panels adopt the intrinsic sSFR criterion (solid grey line) to separate the quenched galaxies. In all cases, we separate the contributions from central and satellite galaxies.

The updated model predicts a substantially higher abundance of quenched galaxies than the legacy \citetalias{Henriques2015}/\citetalias{Barrera2023} implementation (dashed lines), with the largest improvements at high stellar masses and high redshift. For galaxies with $M_\star\gtrsim10^{10.5}\,{\rm M}_\odot$, the updated model maintains a significant quenched population out to $z\sim5-7$, whereas the legacy model predicts a much more rapid decline in number density beyond $z\sim2$. The improvement is particularly evident for the most massive galaxies ($M_\star\gtrsim10^{11}\,{\rm M}_\odot$), where the updated model is broadly consistent with the observational constraints from \citet{Tomczak2014, Santini2022, Weaver2023, Xu2025DESI}, and \citet{Shuntov2025}, while the legacy implementation underpredicts their abundance by more than an order of magnitude.

The decomposition into central and satellite galaxies shows that the high-redshift quenched population is dominated by centrals, with satellites contributing only a minor fraction. This is expected, as environmental quenching becomes increasingly important only at later cosmic times \citep{Doven2026arXiv}, while the massive quenched galaxies observed at high redshift are predominantly central systems whose evolution is governed by rapid stellar mass assembly, bulge growth, and AGN feedback. The updated model predicts a much more gradual decline in the number density of massive quenched galaxies with increasing redshift than the legacy implementation, resulting in a steady build-up of the quenched central population from the first billion years of cosmic history onwards. For galaxies with $M_\star\gtrsim10^{11}\,{\rm M}_\odot$, both the evolution of the number density and the slope of the mass assembly are broadly consistent with current observational constraints, although the abundance of the most massive quenched systems remains slightly underrepresented at the highest redshifts.

The enhanced population of massive quenched galaxies is accompanied by a modest increase in black hole growth at early times. Although black hole masses remain somewhat lower than those predicted by models that incorporate more sophisticated black hole seeding and accretion physics, such as \citet{Bonoli2025arXiv}, they are systematically larger than in the legacy \lgal{} implementation. The increased availability of cold gas in the updated model fuels more efficient quasar-mode accretion, thereby enhancing black hole growth during the early stages of galaxy assembly.

It is important to note that the primary aim of this work is not to modify the AGN feedback model itself, but rather to investigate whether physically motivated changes to star formation and stellar feedback can indirectly enhance the quenching of massive galaxies. 
Although the AGN parameters are jointly recalibrated with the other model parameters, their best-fitting values remain close to those adopted in the legacy models, including \citetalias{Henriques2015, Henriques2020} and \citetalias{Ayromlou2021} (see Table~\ref{tab:MCMC_free_params}). For example, we obtain $\kappa_{\rm AGN}=6.25\times10^{-3}$ and $V_{\rm BH}=721\,{\rm km\,s^{-1}}$, compared with $5.3\times10^{-3}$ and $750\,{\rm km\,s^{-1}}$, respectively, in \citetalias{Henriques2015}.
Within this framework, density-dependent star formation and regulated stellar feedback promote rapid early stellar mass assembly, increasing both the stellar and black hole masses at earlier times. The resulting enhancement in AGN activity allows the existing feedback prescription to quench a substantially larger population of massive galaxies. Nevertheless, further improvements, particularly in reproducing the most massive and earliest quenched galaxies, will likely require more sophisticated modelling of black hole seeding, accretion, and AGN feedback, including the coupling between different feedback modes and the surrounding circumgalactic medium \citep[e.g.,][]{Habouzit2022, Ayromlou2023b, Lagos2024, DeLucia2024, Bonoli2025arXiv}.

It is also worth noting that the observational census of high-redshift quiescent galaxies remains uncertain. Recent JWST observations have revealed a population of compact ``Little Red Dots'' (LRDs) at $z\sim3-8$ whose nature is still under active debate. While some LRDs may represent genuinely quiescent or rapidly quenching galaxies, an increasing body of evidence suggests that many are powered by heavily obscured accreting black holes rather than evolved stellar populations. Consequently, photometrically selected quiescent galaxy samples at high redshift may be contaminated by LRDs, potentially biasing the inferred abundance of early quenched galaxies and the resulting observational constraints on galaxy formation \citep[e.g.,][]{Kocevski2025, Matthee2024, Akins2025}.

\subsection{Structural properties}
\label{sec:struct_prop}
We next examine the structural properties of galaxies through the stellar mass--size relation, the UV luminosity--size relation, and the bulge-to-total stellar mass ratio. Together, these observables probe the assembly of galaxy structure, the spatial distribution of recent star formation, and morphological evolution. They are directly influenced by the updated star formation and regulated stellar feedback mechanisms, merger and disc instability prescriptions, as gas-rich mergers produce compact remnants, dry mergers drive size growth, and disc instabilities transfer low-angular-momentum material from the disc to the bulge.

\subsubsection{Galaxy mass--size relation}

\begin{figure*}
    \centering
    \includegraphics[width=0.24\linewidth]{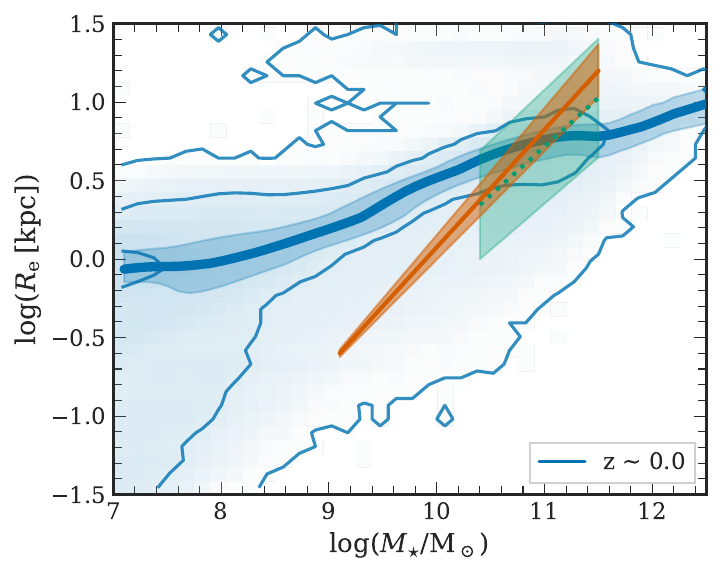}
    \includegraphics[width=0.24\linewidth]{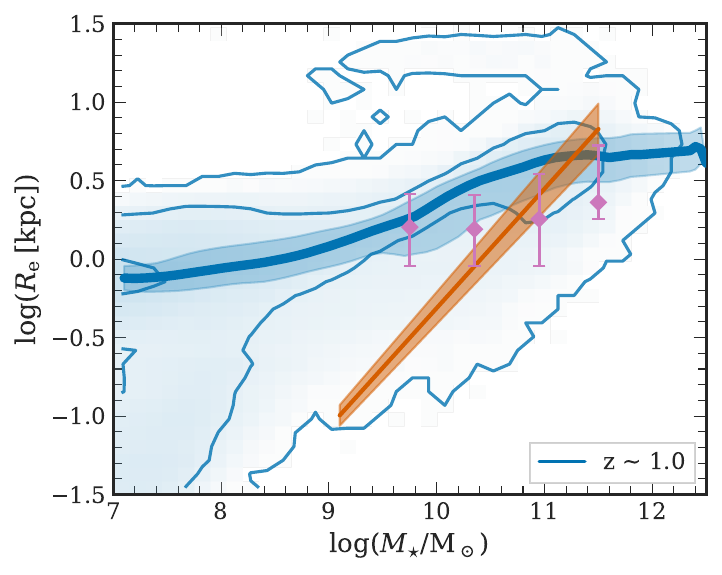}
    \includegraphics[width=0.24\linewidth]{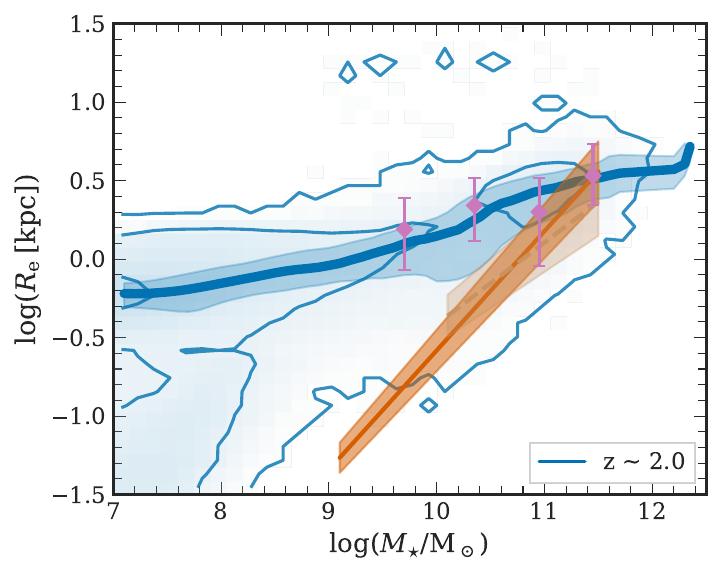}
    \includegraphics[width=0.24\linewidth]{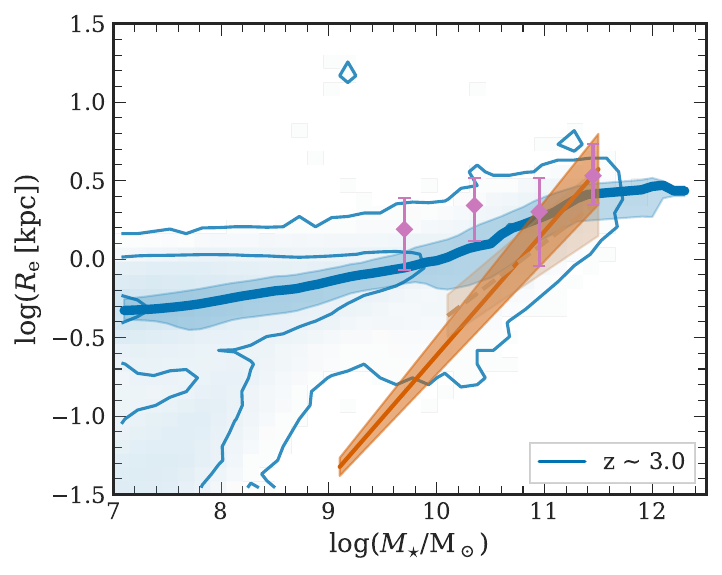}
    \includegraphics[width=0.24\linewidth]{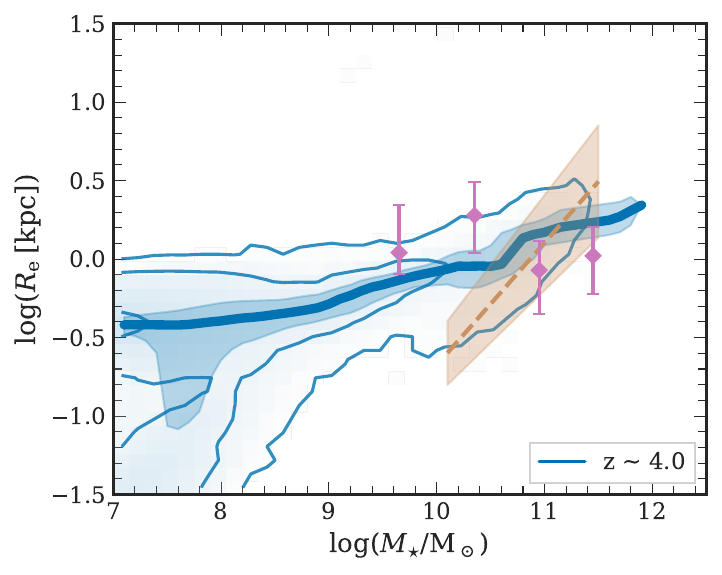}
    \includegraphics[width=0.24\linewidth]{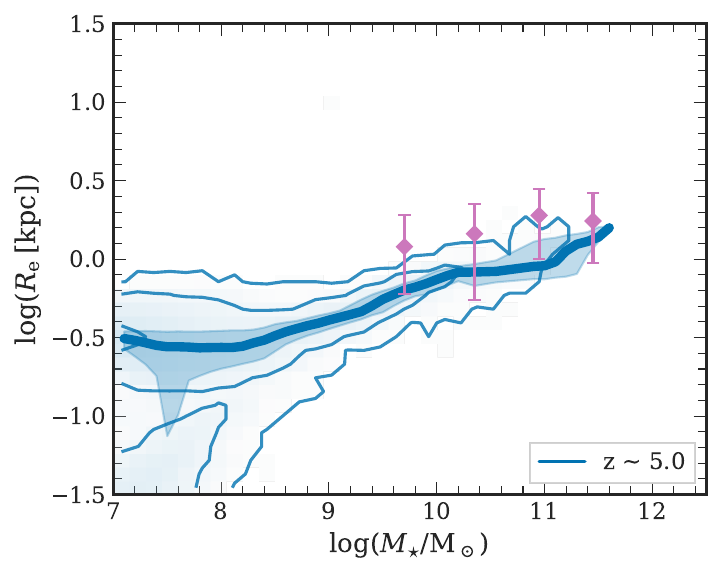}
    \includegraphics[width=0.24\linewidth]{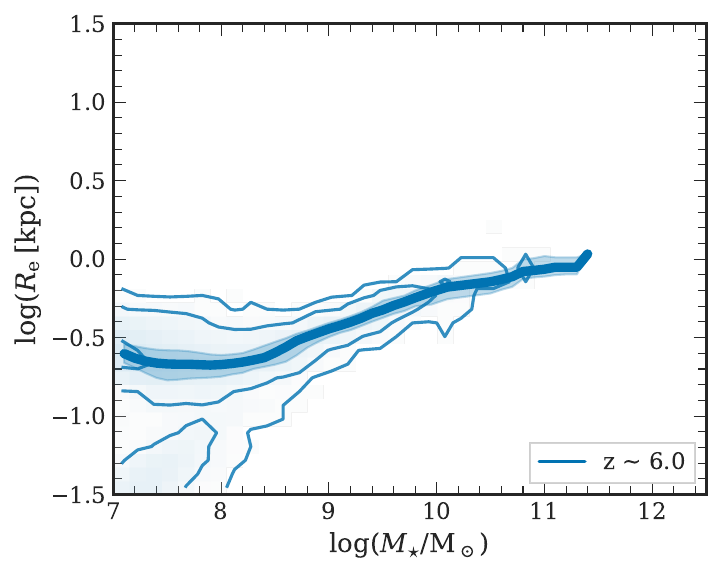}
    \includegraphics[width=0.24\linewidth]{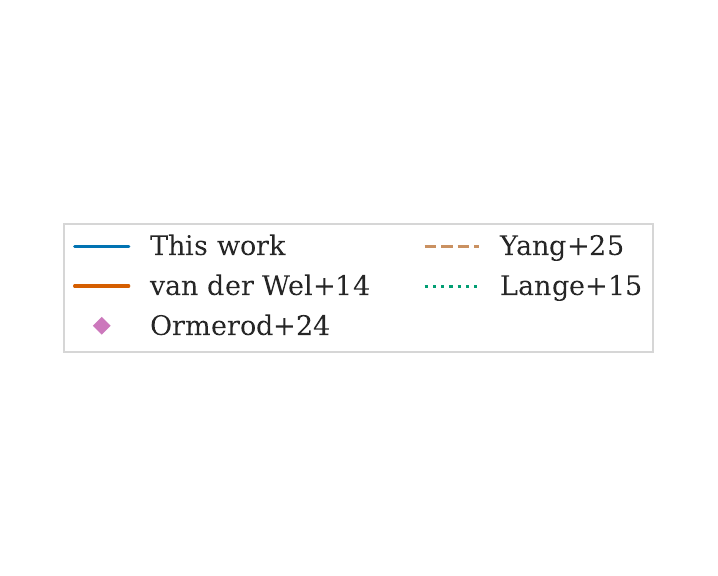}
    \caption{Galaxy stellar mass--size relation for quenched galaxies selected using the sSFR criterion from $z=0$ to $z\simeq6$. The blue density maps and the contours indicating increasing galaxy number density, corresponding to the $1$, $3$, and $6\sigma$ scatter of model population. The solid orange line and shaded region denote the observed median relation and intrinsic scatter from \citet{vanDerWel2014} at $z\lesssim3$. Purple symbols with error bars show recent JWST measurements of quenched galaxies from \citet{Ormerod2024} at higher redshifts, while the dashed orange and dotted green curves indicate the mass--size relations from \citet{Yang2025} and \citet{Lange2015}, respectively, where available.}
    \label{fig:MSR}
\end{figure*}

Figure~\ref{fig:MSR} presents the stellar mass--size relation of quenched galaxies selected using the instantaneous specific star formation rate criterion (${\rm sSFR}<0.2/t_{\rm age}$) from $z=0$ to $z\approx6$. The model predictions are compared with the observed mass--size relation of \citet{vanDerWel2014} at $z\lesssim3$, recent JWST measurements from \citet{Ormerod2024} at higher redshifts, and the empirical relations from \citet{Yang2025} and \citet{Lange2015}, where available.

Overall, the updated model reproduces the observed trend that more massive quenched galaxies are systematically larger than lower-mass systems. At $z\lesssim3$, the predicted distribution follows the observed relation of \citet{vanDerWel2014}. The model also captures the gradual evolution towards smaller galaxy sizes with increasing redshift, producing increasingly compact quenched systems that remain broadly consistent with the available measurements.

At $z\gtrsim4$, observational constraints become sparse and are currently limited to a small number of massive quenched galaxies. Nevertheless, the predicted sizes are generally consistent with the measurements of \citet{Ormerod2024}, while following the mass--size evolution by \citet{Yang2025}. The model predicts that the most massive quenched galaxies ($M_\star\gtrsim10^{10.5}\,{\rm M}_\odot$) remain compact, with typical effective radii of order $0.7-1.5\,{\rm kpc}$ even at $z\sim5$, reflecting their rapid assembly in dense environments.

Despite this overall agreement at the massive end, the model predicts relatively few compact quenched galaxies at low and intermediate stellar masses. While the observed relation is reproduced within its intrinsic scatter, the simulated distribution remains broader and contains an excess of comparatively extended systems below $M_\star\sim10^{10.5}\,{\rm M}_\odot$. This suggests that the current prescriptions do not yet produce sufficiently efficient central mass growth in a subset of lower-mass galaxies.

One possible explanation is that additional compaction channels are required beyond the merger- and disc-instability-driven mechanisms implemented here. Processes such as violent gas compaction, clump migration, and strong central gas inflows have been proposed to rapidly build dense stellar cores prior to quenching \citep{Dekel2014, Zolotov2015, Tacchella2016}. Incorporating such mechanisms may increase the abundance of compact quenched galaxies, although exploring these processes lies beyond the scope of the present work.

It is also important to note that the observational mass--size relation remains subject to significant systematic uncertainties, particularly at high redshift. Measured galaxy sizes depend on surface-brightness sensitivity, point spread function corrections, profile fitting, and the adopted rest-frame wavelength, while the small number of currently known quiescent galaxies at $z\gtrsim4$ leads to substantial statistical uncertainties \citep{vanDerWel2014, Suess2019, Ormerod2024, Yang2025}. Consequently, part of the apparent discrepancy at the compact end may reflect current observational limitations in addition to remaining uncertainties in the physical modelling.

Finally, we emphasise that galaxy sizes were not included among the observational constraints used during the MCMC calibration. The parameters governing structural evolution, including those controlling merger-driven size growth and disc instabilities, were constrained indirectly through other observables. A future calibration that includes the stellar mass--size relation, together with improved modelling of dissipative compaction and structural evolution, may further improve agreement with the observed distribution of quenched galaxies.

\subsubsection{UV size--luminosity relation}

\begin{figure*}
    \centering
    \includegraphics[width=0.24\linewidth]{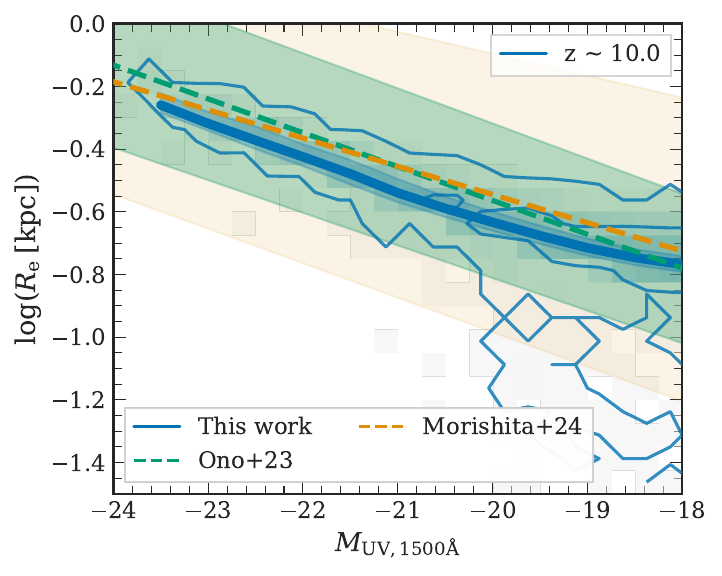}
    \includegraphics[width=0.24\linewidth]{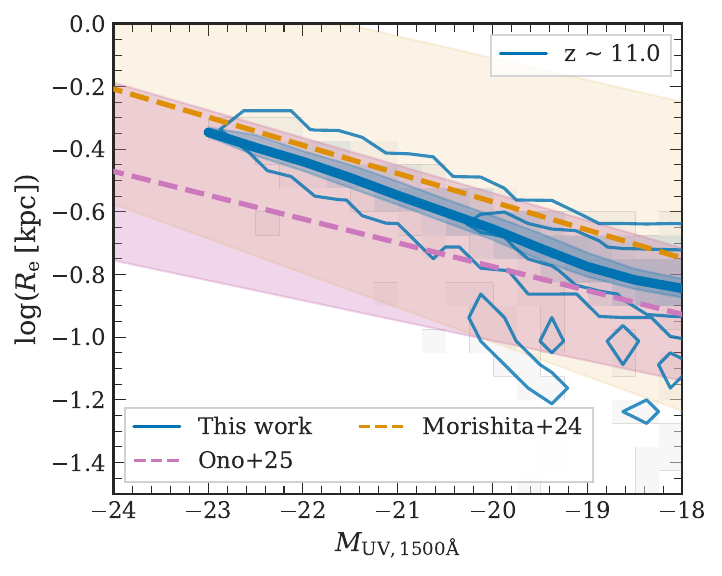}
    \includegraphics[width=0.24\linewidth]{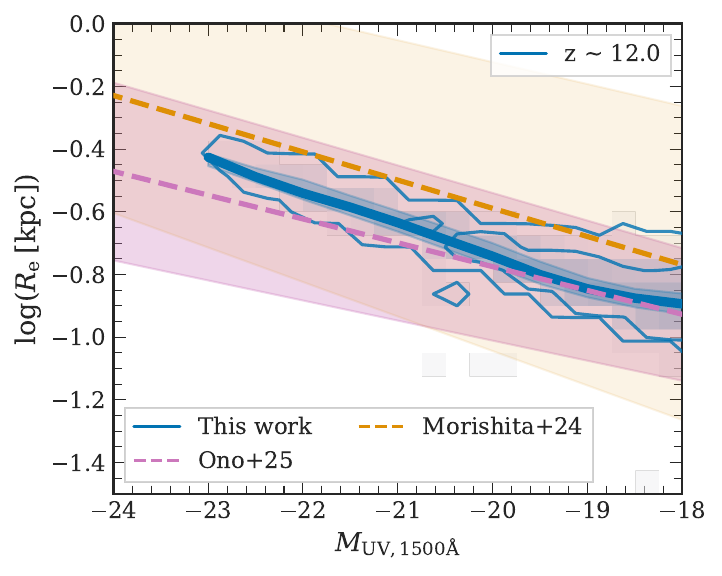}
    \includegraphics[width=0.24\linewidth]{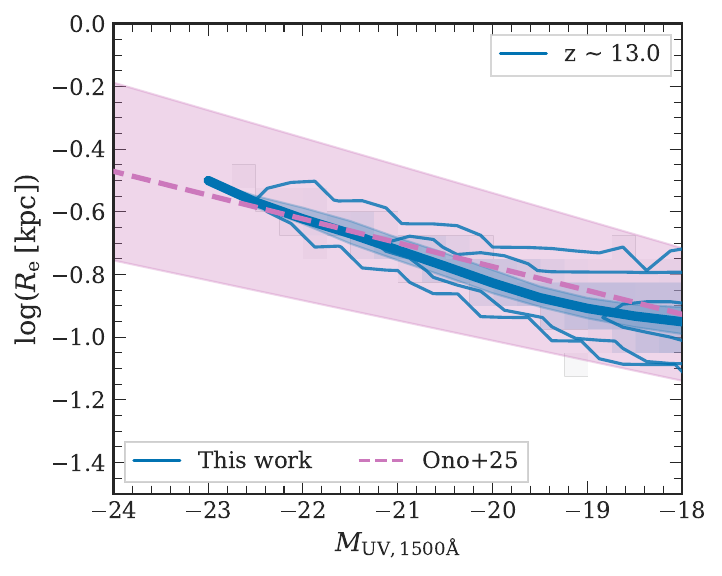}
    \caption{UV half-light radius as a function of absolute UV magnitude at 1500\,\AA{} from $z\simeq10$ to $z\simeq13$. The blue shaded distribution and contours show the predictions of this work, where contours indicate increasing galaxy number density, corresponding to the $1$, $3$, and $6\sigma$ scatter of the model populations. Dashed lines and shaded regions show the observational size--luminosity relations and associated uncertainties from \citet{Ono2023} at $z\sim9$, \citet{Morishita2024} at $z\sim10-12$ and \citet{Ono2025} at $z\gtrsim11$, where available.}
    \label{fig:MSR_UV}
\end{figure*}

The rest-frame UV size--luminosity relation provides an important probe of the spatial distribution of recent star formation in galaxies during the epoch of reionisation. Unlike the stellar mass--size relation, the UV half-light radius traces the young stellar populations responsible for the observed rest-frame UV emission, making it directly comparable to high-redshift JWST observations. In the model, the UV half-light radius at 1500\,\AA{} is computed from the radial UV luminosity profile.

Figure~\ref{fig:MSR_UV} shows the predicted relation between the rest-frame UV half-light radius and UV absolute magnitude at $z\simeq10-13$. The shaded blue density maps and contours indicate the distribution of model galaxies, while the dashed curves show recent observational constraints from \citet{Morishita2024} at $z\sim10-12$ and \citet{Ono2023, Ono2025} at  $z\sim9$ and $z\gtrsim11$, respectively, together with their quoted uncertainties. Overall, the model reproduces the observed trend that brighter galaxies exhibit systematically larger UV sizes, while galaxies become increasingly compact toward fainter UV magnitudes. The predicted relation also evolves only weakly over the relatively short time interval between $z\simeq10$ and $z\simeq13$, consistent with the emerging observational picture from JWST.

The model predicts typical UV half-light radii of $\sim0.2-0.6$\,kpc for the brightest galaxies ($M_{\rm UV}\lesssim-21$ mag) and $\lesssim0.2$\,kpc for galaxies fainter than $M_{\rm UV}\sim-19$ mag at $z\sim12$. The simulated galaxy population occupies a broad locus around the observed relations, with an intrinsic scatter comparable to that inferred from the current observations. At all redshifts, the majority of galaxies are predicted to lie close to the observational relation, indicating that the model forms compact star-forming systems in the dense environments of the early Universe. Such compact UV morphologies naturally arise from the high gas surface densities and centrally concentrated star formation expected in rapidly assembling galaxies at these redshifts.

Although the model broadly reproduces the observed UV size--luminosity relation, the current observational constraints remain limited by small sample sizes, uncertainties in profile fitting, and surface-brightness sensitivity at the faint end \citep{Shibuya2015, Morishita2024, Ono2025, Yang2025}. Future observations will therefore be crucial for determining whether the compact UV morphologies predicted by the model accurately reflect the distribution of star-forming regions in the earliest galaxies.

\subsubsection{Galaxy bulge-to-total stellar mass ratios}

\begin{figure*}
    \centering
    \includegraphics[width=0.33\linewidth]{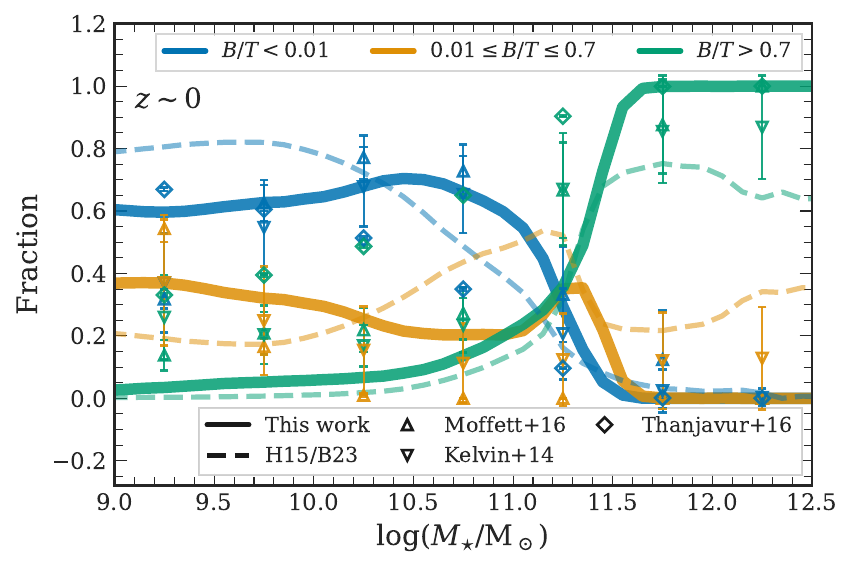}
    \includegraphics[width=0.33\linewidth]{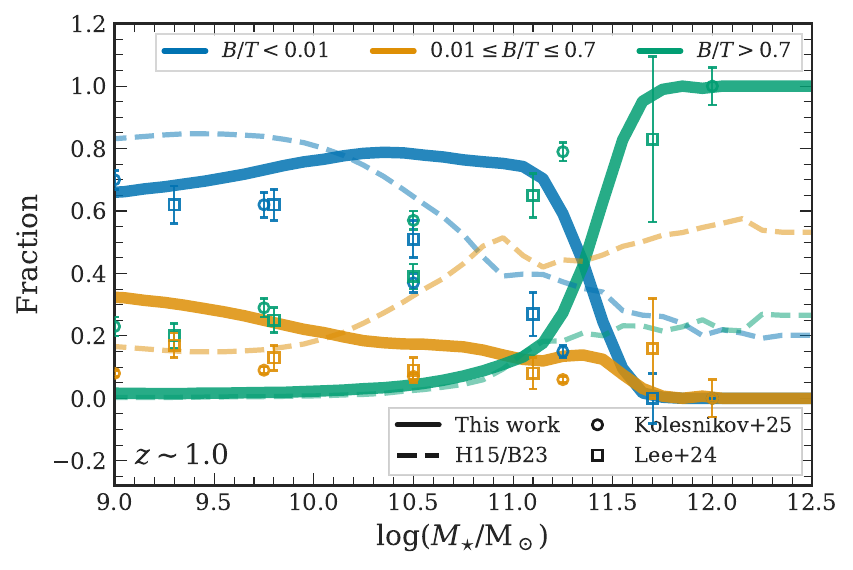}
    \includegraphics[width=0.33\linewidth]{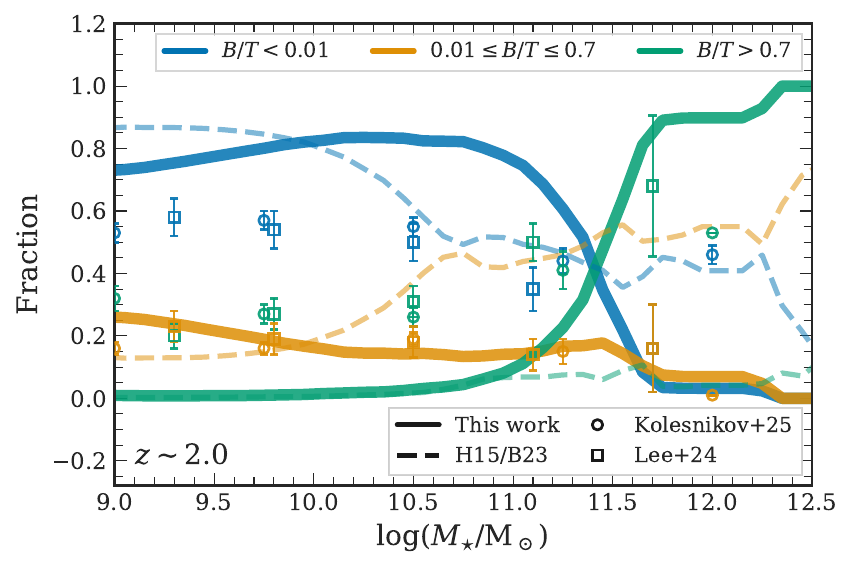}
    \includegraphics[width=0.33\linewidth]{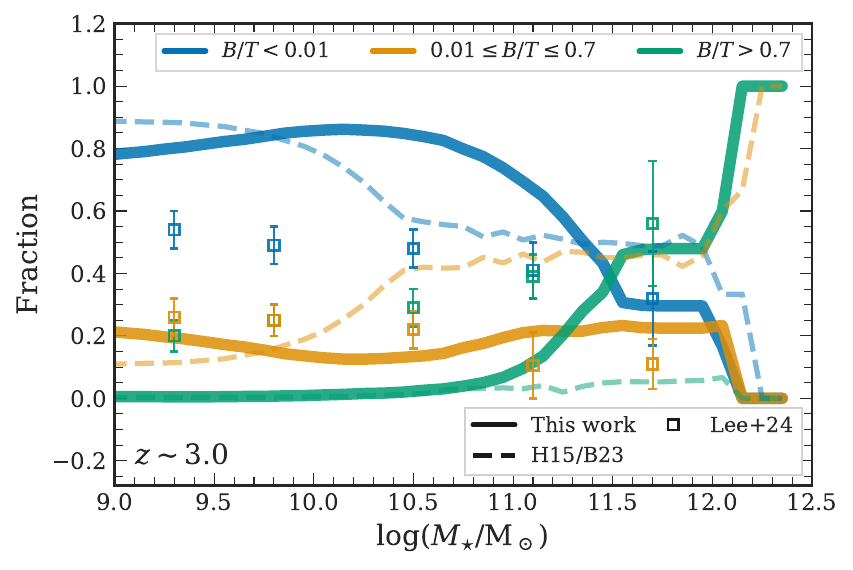}
    \includegraphics[width=0.33\linewidth]{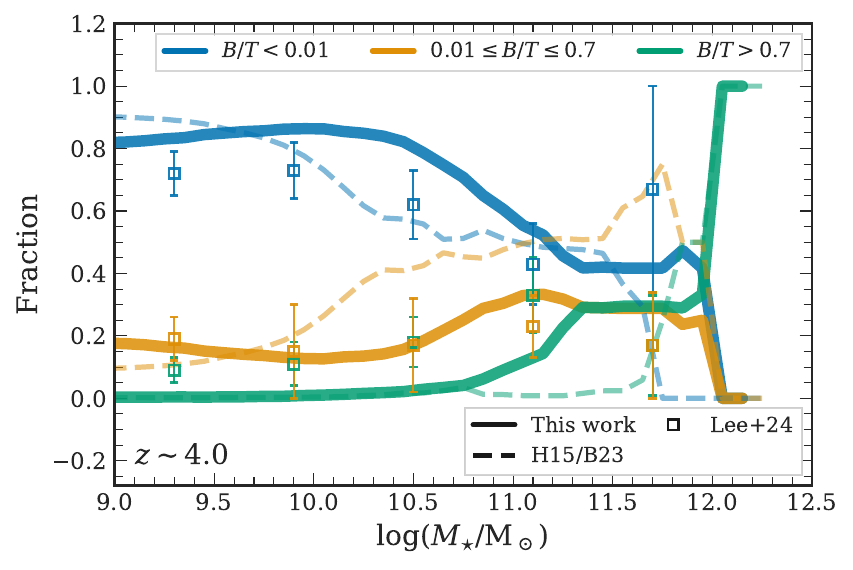}
    \includegraphics[width=0.33\linewidth]{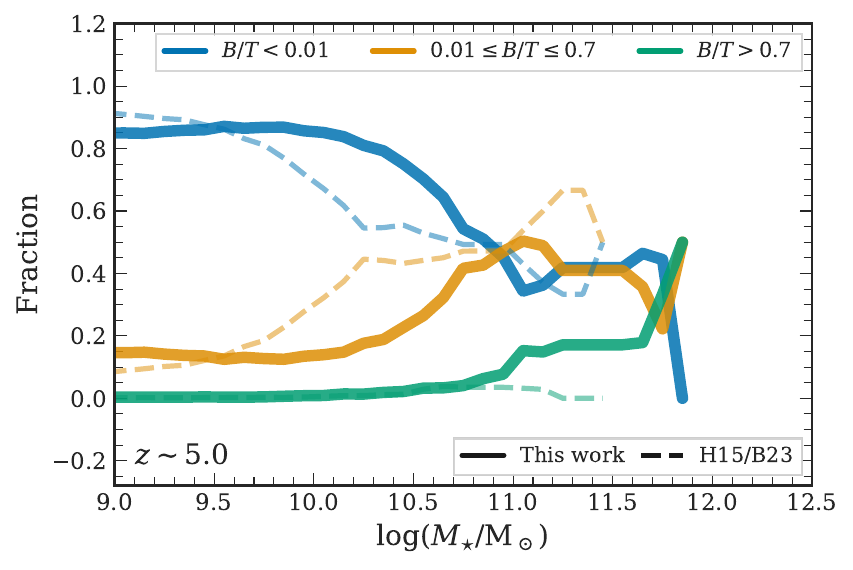}
    \caption{Fraction of galaxies in three bulge-to-total stellar mass ratio ($B/T$) bins as a function of stellar mass at $z=0-5$. Blue, orange, and green curves correspond to disc-dominated ($B/T<0.01$), intermediate ($0.01\leq B/T\leq0.7$), and bulge-dominated ($B/T>0.7$) galaxies, respectively. Solid lines show the predictions of this work, while dashed lines show the \citetalias{Henriques2015}/\citetalias{Barrera2023} model. Symbols with error bars denote observational measurements from \citet{Moffett2016, Thanjavur2016, Kelvin2014} at $z\simeq0$, and \citet{Kolesnikov2025} and \citet{Lee2024} at higher redshifts, where available.}
    \label{fig:fBT}
\end{figure*}

Figure~\ref{fig:fBT} shows the fraction of galaxies in three bulge-to-total stellar mass ratio ($B/T$) bins as a function of stellar mass from $z=0$ to $z\approx5$. The three populations correspond to disc-dominated ($B/T<0.01$), intermediate ($0.01\leq B/T\leq0.7$), and bulge-dominated ($B/T>0.7$) systems. We compare the predictions of the updated model with the \citetalias{Henriques2015}/\citetalias{Barrera2023} implementation and with available observational measurements.

At all redshifts, the model predicts the expected transition from predominantly disc-dominated galaxies at low stellar masses to increasingly bulge-dominated systems at higher masses. The fraction of intermediate-morphology galaxies peaks around the transition mass, while the bulge-dominated population becomes progressively more important toward the massive end. This behaviour reflects the increasing importance of mergers and disc instabilities in driving central mass growth and morphological transformation.

Compared to the legacy model, the updated implementation predicts a systematically larger fraction of bulge-dominated galaxies and a corresponding reduction in the fraction of disc-dominated systems, particularly at intermediate and high stellar masses. The differences become more pronounced at $z\gtrsim1$, where the combination of enhanced star formation, gas-rich mergers, and more efficient disc instabilities promotes rapid bulge growth. These changes substantially improve the predicted morphological mix while preserving agreement with the stellar mass function and quenched galaxy population. We caution, however, that observational morphological classifications are not always directly equivalent to the stellar-mass-based $B/T$ definition used here, particularly at high redshift where structural decomposition is challenging.

At $z\simeq0$, the predicted morphological fractions are in good agreement with the measurements of \citet{Kelvin2014, Moffett2016}, and \citet{Thanjavur2016} across the full stellar mass range, representing a clear improvement over the legacy model. This provides an important independent validation of the updated physical prescriptions, since the $B/T$ distribution or galaxy sizes were not included in the MCMC calibration.

At higher redshifts, the available constraints from \citet{Lee2024} and \citet{Kolesnikov2025} remain sparse and are subject to relatively large observational uncertainties. Nevertheless, the model reproduces the observed fraction of intermediate ($0.01\leq B/T\leq0.7$) systems reasonably well. The primary discrepancy is an overabundance of disc-dominated galaxies and a corresponding deficit of bulge-dominated systems at lower stellar masses. This suggests that, despite the enhanced bulge growth introduced through gas-rich mergers and disc instabilities, morphological transformation in low-mass galaxies may still proceed too slowly at early times.

The excess of disc-dominated galaxies is noteworthy because rotationally supported discs are the default outcome of gas cooling in semi-analytic models. In contrast, cosmological hydrodynamical simulations predict that many galaxies at high redshift first appear as turbulent, clumpy, and irregular systems before settling into ordered discs or undergoing rapid central compaction \citep{Dekel2009, Ceverino2010, Zolotov2015, Mandelker2025}. Importantly, the hydrodynamical simulations show a dearth of discs at high redshift and for low halo masses, in contrast to our model \citep[e.g., ][]{Benavides2025}. The remaining excess of low-mass discs in our model may nevertheless indicate that some early transformation channels, such as violent compaction events or additional clump-driven bulge growth, are still not fully captured. Nevertheless, the updated model represents a significant improvement over the legacy implementation and provides a considerably better match to the observed evolution of galaxy morphologies over a wide range of stellar masses.

\subsection{Halo baryon conversion efficiency}
\label{sec:halo_prop}
The halo baryon conversion efficiency, defined as $\epsilon = {M_\star}/{f_{\rm b}M_{200}}$, measures the fraction of the available baryonic mass within a halo ($M_{200}$) that has been converted into stars, where $f_{{\rm b}} = {M_{{\rm baryon}}(<R_{\rm 200})}/{M_{200}}$ is the baryon fraction calculated for each halo. Here, $M_{{\rm baryon}}(<R_{\rm 200})$ is obtained by summing the stellar, cold-gas, hot-gas, black-hole, and intra-cluster-light components of all model galaxies within the virial radius. It therefore provides a direct diagnostic of the competition between gas accretion, star formation, stellar feedback, AGN feedback, and environmental processes \citep[e.g.,][]{Wechsler2018, Moster2010}.

\begin{figure*}
    \centering
    \includegraphics[width=0.45\linewidth]{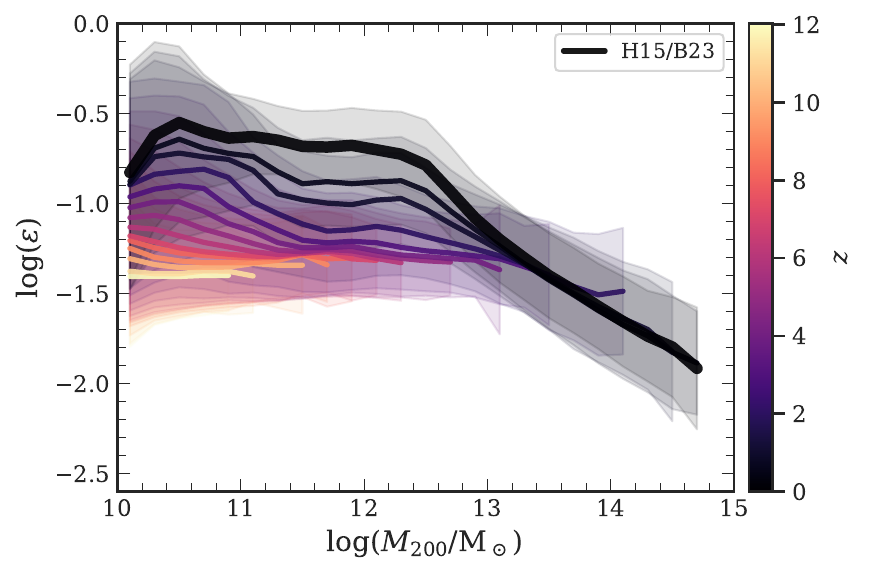}
    \includegraphics[width=0.45\linewidth]{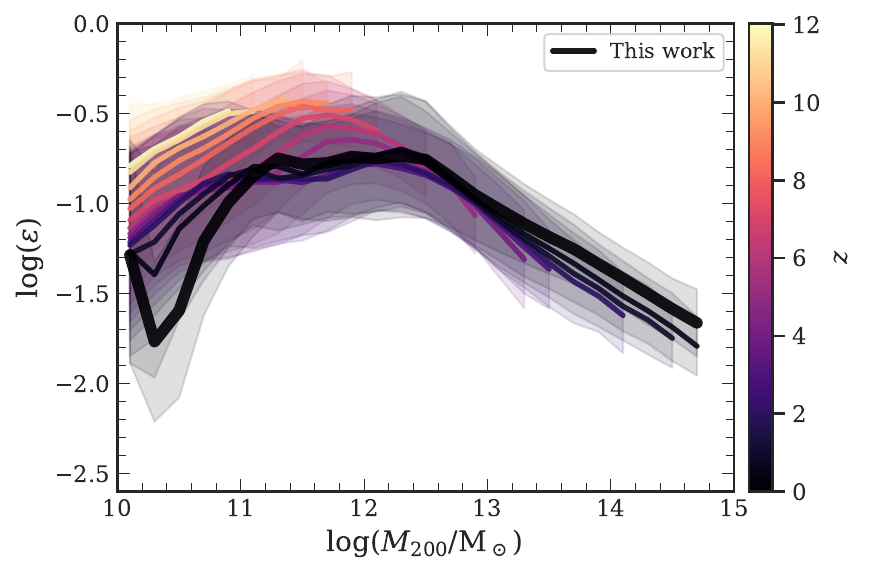}
    \caption{Evolution of the halo baryon conversion efficiency, $\epsilon=M_\star/(f_{\rm b}M_{200})$, as a function of halo mass, where $f_{{\rm b}} =
    {M_{{\rm baryon}}(<R_{\rm 200})}/{M_{200}}$ is the baryon fraction calculated for each halo. The left panel shows the predictions of the legacy \citetalias{Henriques2015}/\citetalias{Barrera2023} model, while the right panel shows the updated model presented in this work. Coloured curves correspond to different redshifts, as indicated by the colour bar, and the thick black curve denotes the $z=0$ relation.}
    \label{fig:eff_halo}
\end{figure*}

Figure~\ref{fig:eff_halo} compares the evolution of the halo baryon conversion efficiency in the legacy implementation and the updated model. At $z=0$, the legacy model reaches its maximum efficiency around $M_{200}\sim10^{12}\,{\rm M}_\odot$ and remains relatively flat towards lower halo masses, before declining at the massive end. The updated model shows a stronger mass dependence, with a broad maximum over a narrower intermediate-mass range and declining efficiencies towards both lower- and higher-mass haloes. The suppression at low masses is primarily associated with stellar feedback, while inefficient cooling and AGN feedback regulate the massive end.

The principal difference between the models is their redshift evolution. In the legacy implementation, the efficiency decreases towards higher redshift and retains a relatively flat dependence on halo mass below the peak. In contrast, the updated model develops a pronounced peaked distribution at high redshift, declining towards both lower and higher halo masses. Moreover, the peak efficiency increases towards earlier cosmic times, reflecting the more efficient conversion of dense gas into stars in the updated high-redshift model.

The model reaches a peak efficiency of approximately $35\%$ in haloes of $M_{200}\sim10^{11}\,{\rm M}_\odot$ at $z\sim12$, which decreases to $\sim15\%$ after the first billion years of cosmic evolution and further declines to only $\sim13\%$ by $z=0$. Thus, most of the evolution occurs at early times, while the peak efficiency evolves only weakly thereafter. At the same time, the characteristic halo mass associated with the highest efficiency shifts towards higher masses with cosmic time. This behaviour reflects the enhanced conversion of baryons into stars in intermediate-mass haloes at early times introduced by the updated star formation and feedback prescriptions.

The resulting halo baryon conversion efficiencies remain consistent with the model's broader galaxy population predictions. In particular, the updated model simultaneously reproduces the evolution of the stellar mass function (\figref{fig:SMF}) and quenched galaxy stellar mass function (\figref{fig:QGSMF}), while broadly recovering the cosmic star formation rate density (\figref{fig:SFRD}), apart from the high-redshift excess discussed above.

\section{Additional local galaxy diagnostics}
\label{sec:local_diagnostics}
The calibration spans $z\simeq0-4$ and $z\simeq11-12$, a substantially broader redshift range than previous \lgal{} calibrations. We therefore test additional, non-calibrated observables to verify that the improved high-redshift predictions preserve the model's consistency with local galaxy properties.

\subsection{Galaxy sizes and cold-gas content}

\begin{figure*}
    \centering
    \includegraphics[width=0.45\linewidth]{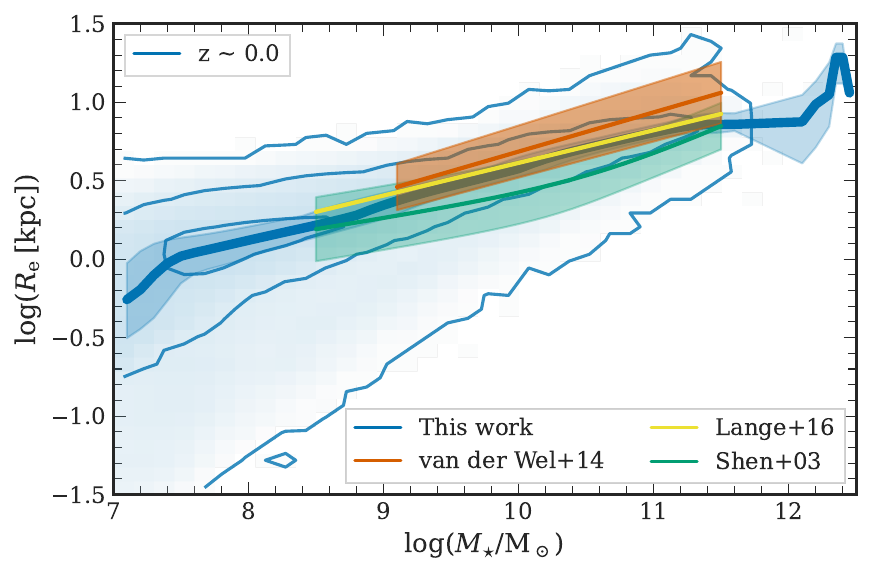}
    \includegraphics[width=0.45\linewidth]{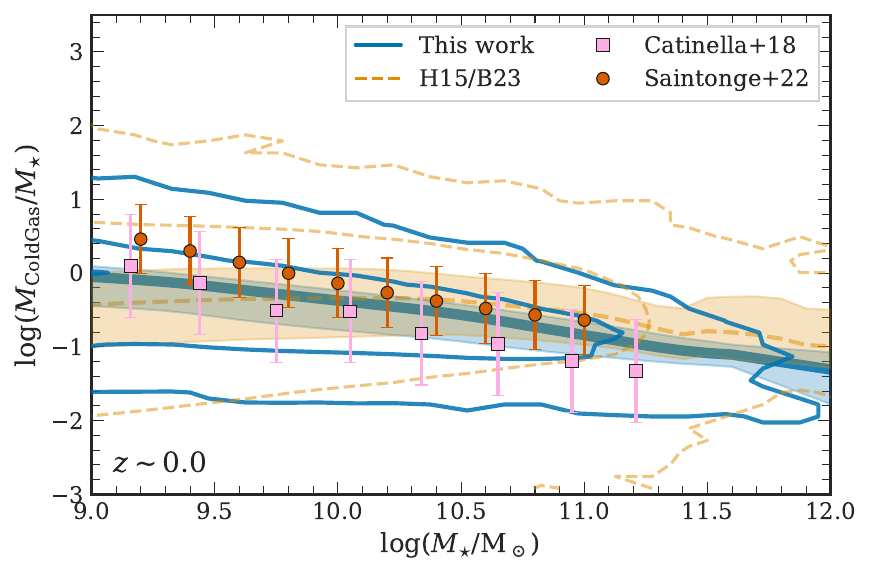}
    \caption{Structural and cold-gas properties of star-forming galaxies at $z=0$.  \textit{Left:} Stellar half-light radius as a function of stellar mass, compared with \citet{Shen2003}, \citet{vanDerWel2014}, and \citet{Lange2016}.  \textit{Right:} Cold-gas-to-stellar mass ratio, compared with \citet{Catinella2018} and \citet{Saintonge2022}.  The blue (solid) and orange (dashed) lines show the median predictions of this work and the \citetalias{Henriques2015}/\citetalias{Barrera2023} model, respectively, with shaded regions indicating the corresponding 16th--84th percentile ranges. Contours show the $1$, $3$, and $6\sigma$ scatter of the model populations.}
    \label{fig:local_size_gas}
\end{figure*}

Figure~\ref{fig:local_size_gas} shows two complementary properties of star-forming galaxies, the stellar mass--size relation and the cold-gas-to-stellar mass ratio.

The left panel shows the stellar half-light radius, $R_{\rm e}$, as a function of stellar mass. At $z\simeq0$, we compare with the local SDSS size--mass relation of \citet{Shen2003} and the GAMA-based measurements of \citet{Lange2016}, which derive structural properties from large samples of nearby galaxies. We additionally compare with \citet{vanDerWel2014} separated star-forming populations. The updated model follows the observed increase of size with stellar mass and provides a substantially improved description relative to the previous implementation \citep[see][]{Vani2025}. Importantly, we have verified that this agreement persists beyond the local Universe; the predicted size--mass relation of star-forming galaxies remains consistent with the \citet{vanDerWel2014} and \citet{Ormerod2024} measurements out to $z\sim4$.

The right panel shows the ratio of total cold-gas mass to stellar mass, $M_{\rm ColdGas}/M_\star$, for local star-forming galaxies. We compare with \citet{Catinella2018}, based on the xGASS survey spanning a broad range of stellar masses in the local Universe and including both atomic and molecular gas components. We additionally compare with the compilation of local cold-gas scaling relations presented by \citet{Saintonge2022}, based primarily on the xGASS and xCOLD GASS surveys. The atomic and molecular gas measurements are combined and corrected for helium to match the total cold-gas mass followed by the model. Both model variants reproduce the observed decline in cold-gas content with increasing stellar mass, with the updated model showing a slightly better agreement with the observed normalisation and mass dependence at $z\simeq0$.

\subsection{Local mass--metallicity relations}
Metallicity provides an additional consistency check because it reflects the integrated interplay between star formation, gas accretion, chemical enrichment, and the removal and recycling of material by feedback.

\begin{figure}
    \centering
    \includegraphics[width=1\linewidth]{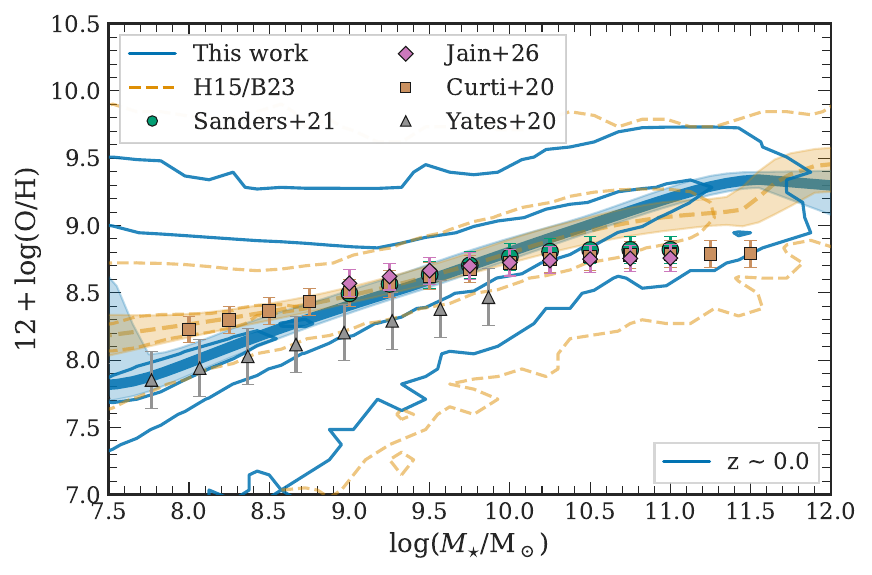}
    \caption{Gas-phase metallicity as a function of stellar mass of local star-forming galaxies, compared with the relations of \citet{Yates2020, Curti2020, Sanders2021} and \citet{Jain2026}. Solid and dashed lines show the median predictions of this work and \citetalias{Henriques2015}/\citetalias{Barrera2023} models, respectively, with shaded regions indicating the corresponding 16th--84th percentile ranges. Contours show the $1$, $3$, and $6\sigma$ scatter of the model populations.}
    \label{fig:local_mzr}
\end{figure}

Figure~\ref{fig:local_mzr} shows the gas-phase metallicity of local star-forming galaxies as functions of stellar mass. We compare with the oxygen abundance relations of \citet{Curti2020, Yates2020, Sanders2021} and \citet{Jain2026}, derived from rest-frame optical emission lines, with metallicity calibrations that account for evolving ISM conditions. Together, these studies trace the gas-phase mass-metallicity relation from the local Universe. To compare with these measurements, the model gas metallicity is converted to oxygen abundance assuming solar scaling, $12+\log({\rm O/H}) = 8.69+\log(Z/Z_\odot)$, with $Z_\odot=0.0134$ \citep[][]{Asplund2009}. Both models find the characteristic increase of metallicity with stellar mass and a flattening towards the massive end. The updated model reproduces this qualitative shape but predicts systematically higher gas-phase metallicities over part of the stellar-mass range. A comparable offset is present in the previous \citetalias{Henriques2015}/\citetalias{Barrera2023} implementation, indicating that it stems from the inherited instantaneous-recycling enrichment scheme rather than from the new prescriptions introduced here.  At low stellar masses, the updated model predicts lower metallicities than the legacy model, bringing it into better agreement with \citet{Yates2020}, although some tension remains with the other observational determinations. We have additionally verified that the updated model shows improved agreement with the observed mass--metallicity relations of \citet{Isobe2026arXiv} and \citet{Sanders2021} out to $z\sim3$ for the massive galaxies.  However, we defer a detailed interpretation of its high-redshift evolution to future work incorporating a more advanced chemical-enrichment treatment.

We caution, however, that these comparisons should not be interpreted as a detailed validation of the chemical-enrichment model. The present implementation adopts the instantaneous recycling approximation inherited from the earlier \lgal{} model \citep[][]{Henriques2015}. A more complete treatment, including the time-dependent enrichment channels of \citet{Yates2013, Yates2021}, will be tested and explored in future work.

\subsection{Radial profiles of Milky Way-like galaxies}

\begin{figure*}
    \centering
    \includegraphics[width=0.45\linewidth]{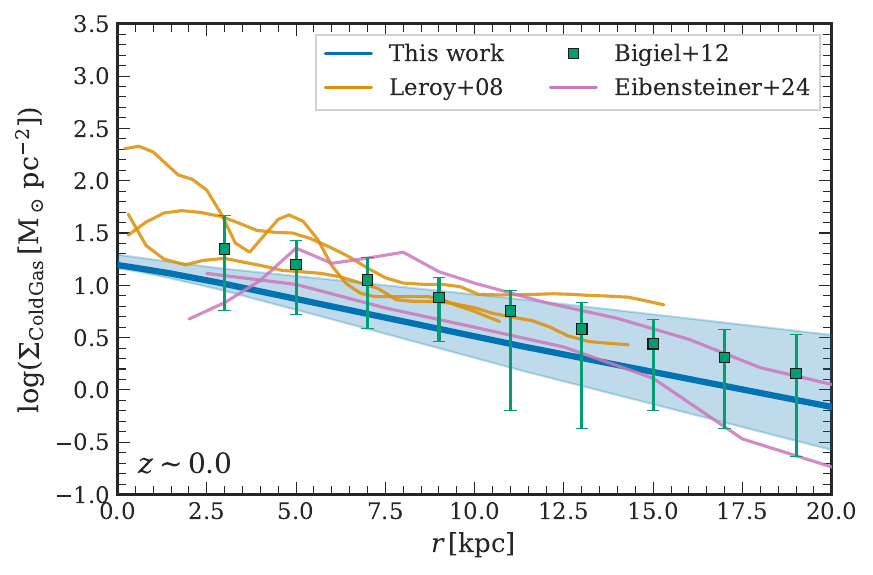}
    \includegraphics[width=0.45\linewidth]{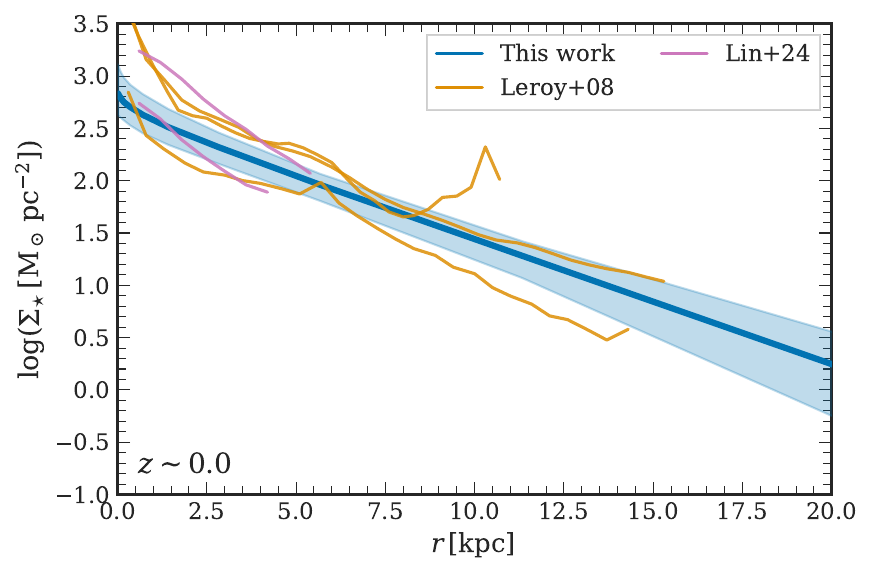}
    \caption{Radial surface-density profiles of Milky Way-like galaxies at $z\simeq0$. \textit{Left:} Total cold-gas surface density, compared with the resolved nearby-galaxy measurements of \citet{Leroy2008}, \citet{Bigiel2012}, and \citet{Eibensteiner2024}. \textit{Right:} Stellar surface density, compared with \citet{Leroy2008} and the compact and extended galaxy profiles of \citet{Lin2024}. The solid blue curve and shaded region show the median and 16th--84th percentile range of the updated model.}
    \label{fig:RadialProf}
\end{figure*}

The radial-ring implementation provides a further opportunity to test the model against spatially resolved observations, rather than only integrated galaxy properties. Figure~\ref{fig:RadialProf} shows the radial cold-gas and stellar surface-density profiles of Milky Way-like galaxies at $z=0$. Following the selection adopted by \citet{Henriques2020}, we select star-forming, disc-dominated galaxies with $M_{\rm bulge}/M_\star<0.15$, $200<V_{\rm vir}\,[{\rm km\,s^{-1}}]<235$, and $10.3<\log(M_\star/{\rm M_\odot})<10.7$. The profiles are measured directly from the stellar- and cold-gas masses stored in the radial rings, and the median and 16th--84th percentile range are computed at each radius. For a consistent comparison, we restrict the observational samples to galaxies that satisfy comparable selection criteria and for which the required properties are available.

For the cold-gas profiles in the left panel, we compare with spatially resolved measurements of nearby disc galaxies from \citet{Leroy2008}, \citet{Bigiel2012}, and \citet{Eibensteiner2024}. The \citet{Leroy2008} compilation combines multiwavelength observations of nearby galaxies to derive spatially resolved atomic and molecular gas, and stellar-mass surface densities, while \citet{Bigiel2012} derives mean neutral-gas profiles for nearby spirals and finds an approximately exponential decline in their outer discs. More recently, \citet{Eibensteiner2024} combined atomic hydrogen (H\,{\sc i}) and ALMA CO observations for nearby star-forming galaxies, tracing the transition from molecular-dominated inner regions to H\,{\sc i}-dominated outer discs. The atomic and molecular components are combined and corrected for helium to provide a quantity comparable to the total cold-gas mass followed by \lgal{}. The updated model reproduces both the normalisation and declining shape of the observed profiles reasonably well. Relative to \citet[][]{Henriques2020}, the radial distribution of cold gas is improved, with the updated model following the observed decline over a larger fraction of the disc.

The right panel shows the corresponding stellar surface-density profiles. The model predictions are compared with the stellar profiles from \citet{Leroy2008} and the MaNGA-based resolved stellar-mass profiles from \citet{Lin2024}, revealing compact and extended systems. For the latter, the reported $R/R_{\rm e}$ profiles are converted to physical radii assuming $R_{\rm e}=3\,{\rm kpc}$, based on \citet{Shen2003}. The updated model reproduces well a smooth decline in stellar surface density with radius and shows improved agreement with the observed profile shape compared with \citet{Henriques2020}, particularly across the main stellar disc.

These resolved comparisons provide an independent test of the radial-ring framework and, together with the local size, cold-gas, and metallicity relations, show that the modifications introduced to improve the high-redshift galaxy population preserve the broad consistency with the present-day galaxy population.

\section{Summary and discussion}
\label{sec:summary_discussion}

Recent high-redshift observations have revealed a population of massive, UV-bright, compact, and, in some cases, already-quenched galaxies within the first couple of billion years of cosmic history. These discoveries challenge standard galaxy formation models calibrated primarily at low redshift and motivate a re-examination of the physical processes governing early star formation, feedback regulation, structural transformation, and quenching.

In this work, we develop and extend the \lgal{} semi-analytic framework of \citet{Barrera2023} (implemented as a post-processing module on top of \gadgetfour{}), and introduce new prescriptions to enhance early stellar mass assembly while preserving agreement with well-constrained low-redshift observables. We expand upon the work of \citet{Henriques2015}, incorporating features from \citet{Henriques2020} within the \gadgetfour{} framework to introduce: (i) a density-modulated star formation prescription linking local gas surface density to cloud-scale efficiency (\secref{sec:starformation}); (ii) a regulated stellar feedback model in which the effective coupling of supernova and stellar wind energy to the ISM changes smoothly from high to low cold gas surface-density environments (\secref{sec:regulatingfeedback}); (iii) an updated ring-based disc instability treatment that self-consistently couples gas inflow, central star formation, bulge growth, and black-hole fuelling (\secref{sec:DI}); (iv) a dissipative merger-driven size model, allowing gas-rich mergers to produce compact remnants while dry mergers drive late-time size growth (\secref{sec:mergers}); and (v) a new fast, modular, and computationally efficient \lgal{} framework with an improved MCMC calibration routine (\secref{sec:MCMC_routines} and \appref{app:MCMCcalib}).

The model is calibrated against stellar mass functions and quenched fractions over $z\simeq0-4$, together with UV luminosity functions at $z\simeq11-12$.
Our results in \secref{Sec:Results} demonstrate that:
\begin{itemize}
    \item The updated model preserves good agreement with the observed low-redshift stellar mass function, while producing substantially more massive galaxies at $z\gtrsim4$ than the legacy \lgal{} implementation, in agreement with observations (see \figref{fig:SMF}).
    \item The abundance of UV-bright galaxies at $z\simeq9-13$ increases by up to nearly two orders of magnitude relative to the legacy model, significantly reducing the previous bright-end UVLF tension with the observations (see \figref{fig:UVLF}).
    \item Massive quenched galaxies are produced much earlier and in larger numbers, especially at $z\simeq3-8$, because rapid early stellar mass assembly indirectly enhances bulge growth, black-hole fuelling, and AGN-driven quenching (see Figs. \ref{fig:QGSMF}--\ref{fig:N_qg}).
    \item Gas dissipation in wet mergers produces compact quenched remnants, improving the qualitative agreement with observed mass--size relations, although other compaction mechanisms may still be necessary (see \figref{fig:MSR}). The model produces a population of compact, UV-bright galaxies at $z\gtrsim10$, with rest-frame UV half-light radii of $0.2\lesssim R_{\rm e}\lesssim 0.6\,{\rm kpc}$ in agreement with the observations (see \figref{fig:MSR_UV}).
    \item The updated model predicts a more realistic morphological mix, with enhanced bulge formation at intermediate and high stellar masses, while overproducing low-mass disc systems at high redshift (see \figref{fig:fBT}).
    \item The updated model predicts a peaked evolution of the halo baryon conversion efficiency, with haloes of fixed mass becoming increasingly efficient towards higher redshifts (see \figref{fig:eff_halo}).
    \item The new model reproduces the local size, cold-gas, and metallicity scaling relations of star-forming galaxies, as well as the resolved gas and stellar profiles of Milky Way-like galaxies, preserving consistency with $z=0$ observations (see \figref{fig:local_size_gas}--\ref{fig:RadialProf}).
\end{itemize}

These results support the idea that the early Universe may have hosted a phase of unusually efficient star formation in dense, gas-rich systems. In this respect, our model is closely related to the density-modulated star formation scenario of \citet{Somerville2025}, and to feedback-free burst models proposed by \citet{Dekel2023} and \citet{Li2024}. However, rather than imposing a fully feedback-free phase, our implementation introduces a continuous surface-density-dependent reduction in the effective stellar feedback coupling. This allows the model to boost early star formation while preserving the self-regulation required to reproduce the low-redshift galaxy population.

The updated model also complements recent cosmological hydrodynamical and semi-analytic simulations that have revisited the efficiency of early galaxy formation. Recent \textsc{Thesan} studies have shown that increasing the efficiency of star formation in dense gas substantially improves the abundance of massive galaxies and quenched systems at high redshift \citep{McClymont2025, Shen2026Thesan}. Similarly, the \textsc{FLARES} simulations demonstrate that high gas fractions and dense interstellar media naturally promote rapid stellar mass assembly in the most overdense regions of the early Universe \citep{Lovell2020_FLARES}. More recently, the \textsc{COLIBRE} simulations have shown that updated star formation and feedback prescriptions improve the predicted high-redshift stellar mass function, although they continue to underpredict the bright end of the dust-attenuated UV luminosity function \citep{Schaye2026COLIBRE, Lu2026arXivCOLIBRE}. Together with these studies, our results suggest that increasing the efficiency of star formation in dense gas is a promising route towards reconciling the observed abundance of massive galaxies at cosmic dawn while maintaining agreement with low-redshift galaxy populations.

Several limitations nevertheless remain. First, the model overpredicts the cosmic star formation rate density at $z\gtrsim 5$ (\figref{fig:SFRD}), suggesting that star formation in dense gas may now be too efficient in some halo mass ranges. Second, the dust attenuation model is applied in post-processing and needs to be updated to account for dust properties, geometry, and attenuation curves at $z\gtrsim10$. Third, although the revised star formation and stellar feedback prescriptions indirectly enhance black hole growth and substantially increase the abundance of quenched galaxies, the underlying AGN model is inherited from previous \lgal{} versions \citep{Croton2006, Henriques2015}. 
In addition, satellite quenching still relies on the gradual hot-gas stripping inherited from the legacy model, while the current chemical-enrichment treatment remains based on instantaneous recycling.
Finally, the excess of low-mass disc-dominated galaxies and deficit of compact quenched systems suggest that the model may still lack efficient mechanisms for redistributing low-angular-momentum gas towards galaxy centres.
The resulting residuals appear predominantly at the low-mass end and do not affect the high-redshift, high-mass conclusions that are the focus of this paper.

The observational comparison should also be interpreted with appropriate caution. Although JWST has transformed our view of galaxy formation during the first billion years, the inferred abundances and masses of massive and quiescent galaxies continue to evolve as larger samples become available. In particular, contamination from dusty star-forming galaxies, obscured AGN, and compact ``Little Red Dots'' remains an important source of systematic uncertainty in photometrically selected quiescent galaxy samples. Future spectroscopic observations together with JWST Mid-Infrared Instrument (MIRI)  and Atacama Large Millimeter/submillimeter Array (ALMA) constraints on dust and gas will be essential for establishing the true abundance of massive quiescent galaxies at high redshift \citep[e.g., ][]{Kocevski2025, Matthee2024, Akins2025, Wang2025_JWSTMIRI}.

In forthcoming updates, we plan to extend the new \lgal{} framework by incorporating advanced dust modelling \citep[e.g.,][]{Parente2023}, improved black hole seeding, growth, and feedback prescriptions \citep[e.g.,][]{Bonoli2025arXiv}, and detailed galactic chemical enrichment models \citep[e.g.,][]{Yates2024}. We also intend to implement the advanced environmental prescriptions of \citet{Ayromlou2019new, Ayromlou2021} within the MTNG framework and to calibrate the model directly against structural observables. Additional processes associated with violent disc instabilities, clump migration, and gas compaction may also be required to improve the structural evolution of low-mass and compact galaxies \citep[e.g.,][]{Zolotov2015}.
At the same time, we will continue to benchmark the semi-analytic prescriptions against state-of-the-art hydrodynamical simulations, including COLIBRE, MillenniumTNG \citep{Pakmor2023}, FLAMINGO \citep{Schaye2023_FLAMINGO}, and THESAN. In parallel, we will explore machine-learning-based approaches to further accelerate parameter inference and large-scale model calibration.

A particularly promising aspect of the updated framework is the introduction of ring-by-ring star formation histories. These preserve the spatial and temporal evolution of star formation within galaxies, enabling direct predictions of resolved observables such as radial sSFR profiles, band-dependent half-light radii, colour gradients, and compactness measures. They also provide a powerful tool for studying quenching pathways and the spatially resolved galaxy population observed with JWST and future facilities. Note also that our physically based galaxy formation model is ideal for producing  realistic mock catalogues throughout cosmologically relevant volumes, which remains computationally challenging for full hydrodynamical simulations. In particular, it will be highly informative to use our model for clustering predictions of different galaxy populations selected at high redshift.

Overall, this work demonstrates that physically motivated modifications to star formation and stellar feedback in dense gas provide a substantial step towards resolving the tension between semi-analytic models and the observed abundance of massive galaxies at cosmic dawn. Without modifying the AGN feedback prescription, the initial mass function or using alternative cosmology, the updated model simultaneously improves the stellar mass function, UV luminosity function, quenched galaxy population, galaxy morphologies, structural evolution, and a more realistic halo-scale baryon conversion efficiencies over a wide range of redshifts. The remaining discrepancies point towards the next generation of physically motivated improvements, particularly in dust, black hole physics, environmental processes, and galaxy compaction. Together with increasingly stringent JWST observations and comparisons to modern hydrodynamical simulations, these developments will enable \lgal{} to evolve from reproducing global galaxy population statistics towards a predictive, physically grounded model of galaxy formation and quenching across cosmic time.

\section*{Acknowledgements}

The analysis presented in this work was carried out on the computing cluster at the Max Planck Institute for Astrophysics (MPA) and on systems at the Max Planck Computing and Data Facility (MPCDF). A part of this work used generative artificial intelligence tools (including Anthropic Claude and Grammarly) to assist with grammar editing and code debugging. AV thanks Rüdiger Pakmor, Simon White, Bo Peng, Qingo Ma, Hitesh Das, Anshuman Acharya, Iker Millan Irigoyen, and Gaoxiang Jin,  for valuable discussions and assistance. MA is supported at the Argelander Institut für Astronomie through the Argelander Fellowship.
A portion of this research was presented and discussed at a workshop of the Munich Institute for Astro-, Particle, and BioPhysics (MIAPbP). The institute is funded by the Deutsche Forschungsgemeinschaft (DFG, German Research Foundation) under Germany's Excellence Strategy – EXC-2094 – 390783311.
This work has made use of the following software and services: \texttt{Astropy} \citep{AstropyCollaboration2018}, \texttt{SciPy} \citep{Virtanen2020_SciPy}, \texttt{Seaborn} and \texttt{Matplotlib} \citep{Hunter2007_Matplotlib, Waskom2021_Seaborn}, and the NASA Astrophysics Data System (ADS).

\section*{Data Availability}
The data underlying this work will be made available from the corresponding author upon reasonable request. The simulation data from the MillenniumTNG project are planned to be made publicly available in the near future.



\bibliographystyle{mnras}
\bibliography{Ref} 




\appendix

\section{MCMC calibration}
\label{app:MCMCcalib}
Calibrating the model using an MCMC approach requires repeated execution of the semi-analytic model, making computational efficiency essential. Following \citet{Henriques2009MCMC}, we reduce the computational cost by selecting a representative subset of merger trees from the MTNG dark-matter-only simulation ($4320^3$ particles in a $(740\,\mathrm{Mpc})^3$ box), rather than running on the full simulation volume.

To do so, we randomly select FoF haloes at $z=0$ within narrow halo-mass bins such that the resulting halo mass function reproduces that of the full volume to within $5\%$. We then verify that this agreement is maintained across redshift, ensuring consistency of the sampled volume.

Formally, let the simulation contain $N=\sum_{i=1}^{I} N_i$ haloes distributed across $I$ halo-mass bins. The luminosity function in bin $j$ can be written as
\begin{equation}
\Phi_j = \sum_{i=1}^{I} N_i \Phi_{ij},
\end{equation}
where $\Phi_{ij}$ denotes the mean number of galaxies in observable bin $j$ for haloes in mass bin $i$. We select $n_i$ haloes per mass bin such that the estimator
\begin{equation}
\label{eq:Treeselection}
\tilde{\Phi}_j =
\sum_{i=1}^{I} \frac{N_i}{n_i}
\sum_{k=1}^{n_i} S_{ijk}
\end{equation}
reproduces the full-simulation result within a tolerance of $\pm5\%$, where $S_{ijk}$ is the number of galaxies in observable bin $j$ for the $k^{\mathrm{th}}$ halo among the $n_i$ selected in mass bin $i$.

In practice, we select complete and random merger trees at $z=0$ such that the halo mass function, stellar mass function, and UV luminosity function match the full-volume results within $5\%$. We then verify that this agreement persists at higher redshifts; if deviations exceed this threshold, additional trees are randomly selected to restore consistency. The full merger histories of the selected trees are retained, and appropriate statistical weights are applied to recover the effective total simulation volume when computing observables. 
To manage compute resources, we additionally avoid selecting extremely large or computationally expensive merger trees.

This procedure selects $\sim0.05\%$ of the total trees, corresponding to an effective representative volume of $(\sim 60\mathrm{Mpc})^3$. The resulting reduction in computational cost is substantial while preserving statistical accuracy across the select redshifts. 

In contrast to the legacy MCMC module, the new implementation is embedded directly within \lgal{} inside the \gadgetfour{} framework. Importantly, it is fully parallelised to efficiently utilise modern high-performance computing resources. This tight integration removes external I/O overhead, improves memory management, and enables distributed parameter exploration across compute nodes. As a result, the updated module is more modular, computationally efficient, and significantly faster, enabling robust multi-parameter calibration against observational constraints across multiple redshifts. In addition, during MCMC calibration we relax the requirements for tracking the positions and velocities of type~2 galaxies (orphan galaxies), reduce the number of time-integration substeps performed between consecutive snapshots, and introduce further run-time optimisations. These changes have a negligible influence on the calibration outcome while speeding up the process significantly. Once the optimal parameter set is identified, the full model is computed for all trees with conservative integration settings.

For a parameter vector $\boldsymbol{\theta}$, the posterior probability is
\begin{equation}
P(\boldsymbol{\theta}\mid \mathcal{D}) \propto
\mathcal{L}(\mathcal{D}\mid \boldsymbol{\theta})\,\Pi(\boldsymbol{\theta}),
\end{equation}
where $\mathcal{D}$ denotes the set of observational constraints, $\mathcal{L}$ is the likelihood, and $\Pi(\boldsymbol{\theta})$ is the prior. In this work, we adopt uniform priors within the allowed parameter ranges, such that the posterior is proportional to the likelihood inside the prior bounds and vanishes outside them.

The MCMC sampling follows a standard Metropolis--Hastings scheme. Given the current parameter vector $\boldsymbol{\theta}$, a new trial point $\boldsymbol{\theta}'$ is generated through a log-normal proposal,
\begin{equation}
\theta_i' = \theta_i \exp\!\left(\sigma_{\rm MCMC}\,G_i\right),
\end{equation}
where $G_i$ is a Gaussian random deviate with zero mean and unit variance, and $\sigma_{\rm MCMC}$ is the proposal step size. This choice ensures positive-definite parameters and approximately symmetric proposals in logarithmic space. The proposed model is accepted with probability
\begin{equation}
A(\boldsymbol{\theta}\rightarrow\boldsymbol{\theta}') =
\min\left[
1,\,
\frac{P(\boldsymbol{\theta}'\mid \mathcal{D})}{P(\boldsymbol{\theta}\mid \mathcal{D})}
\right]
=
\min\left[
1,\,
\frac{\mathcal{L}(\mathcal{D}\mid \boldsymbol{\theta}')\,\Pi(\boldsymbol{\theta}')}{\mathcal{L}(\mathcal{D}\mid \boldsymbol{\theta})\,\Pi(\boldsymbol{\theta})}
\right].
\end{equation}
Since the priors are uniform within their allowed ranges, the prior ratio is unity for proposals inside the bounds.

The total likelihood is constructed as the product of the likelihoods from all active observational constraints and redshifts,
\begin{equation}
\mathcal{L}_{\rm tot} =
\prod_{z}\prod_{m}
\mathcal{L}_{m,z}^{\,w_{m,z}},
\end{equation}
where $\mathcal{L}_{m,z}$ is the likelihood contribution of observable $m$ at redshift $z$, and $w_{m,z}$ is the corresponding weight assigned to that constraint. In practice, the code evaluates the logarithm of the likelihood for numerical stability.

For observables compared through a Gaussian error model, such as stellar mass functions and luminosity functions, we use a chi-square likelihood,
\begin{equation}
\chi^2_{m,z} =
\sum_{j}
\frac{\left[\phi_{j,{\rm model}}-\phi_{j,{\rm obs}}\right]^2}
{\sigma_{j,{\rm obs}}^2},
\end{equation}
such that
\begin{equation}
\mathcal{L}_{m,z} \propto
\exp\left(-\frac{\chi^2_{m,z}}{2}\right).
\end{equation}
For fraction-based or median-based constraints, an analogous Gaussian likelihood is adopted using the corresponding observational uncertainties.

The MCMC chains are evolved for several tens of thousands of steps, including an initial burn-in phase. The bestfits are listed in Table~\ref{tab:MCMC_free_params}.
Overall, the calibrated model achieves a balanced fit across multiple redshifts and observables, demonstrating that the updated physical prescriptions remain consistent with low-redshift constraints while enabling improved agreement with early-Universe galaxy populations (see \secref{Sec:Results}). 

\begin{figure*}
\centering

\setlength{\tabcolsep}{2pt}
\renewcommand{\arraystretch}{0.0}

\begin{tabular}{ccc}

\includegraphics[width=0.30\textwidth]{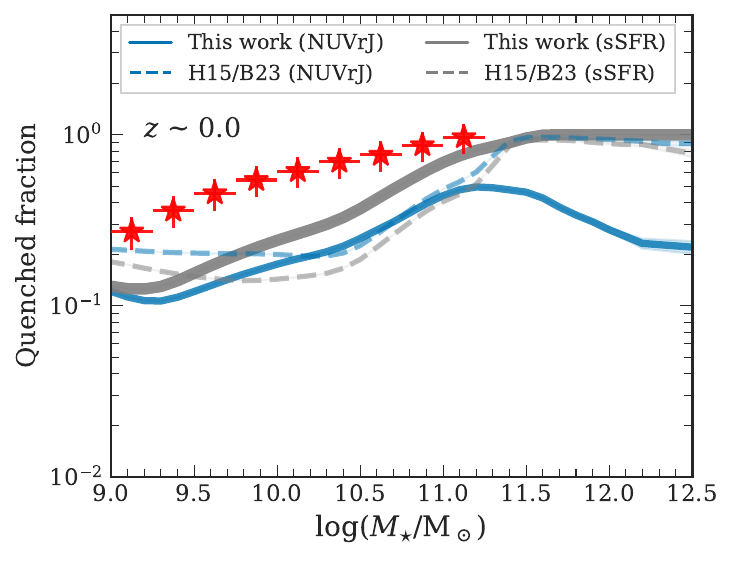} &
\includegraphics[width=0.30\textwidth]{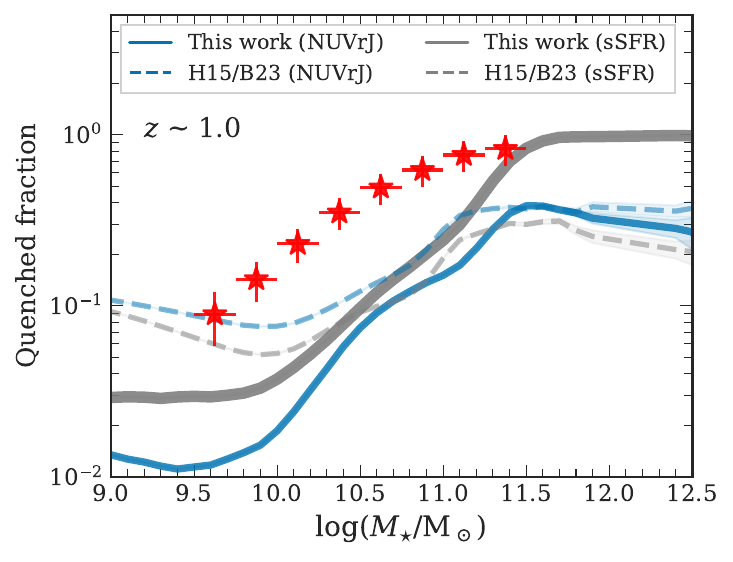}  &
\includegraphics[width=0.30\textwidth]{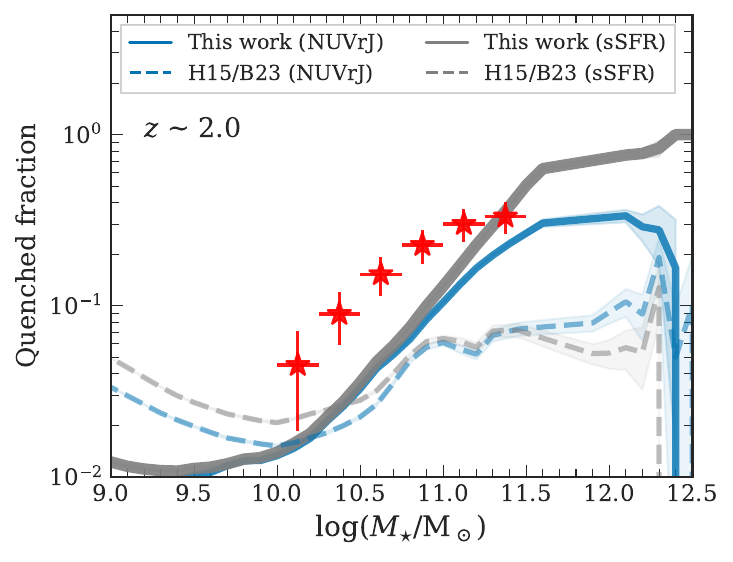}  \\

\includegraphics[width=0.30\textwidth]{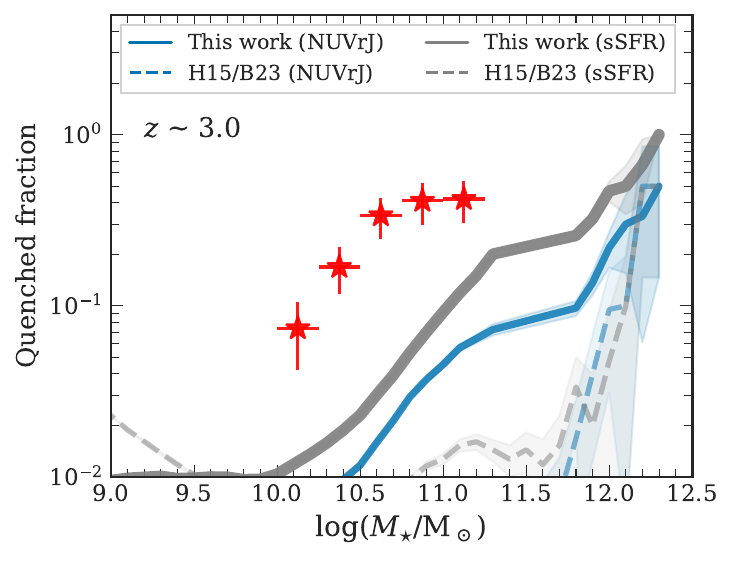}  &
\includegraphics[width=0.30\textwidth]{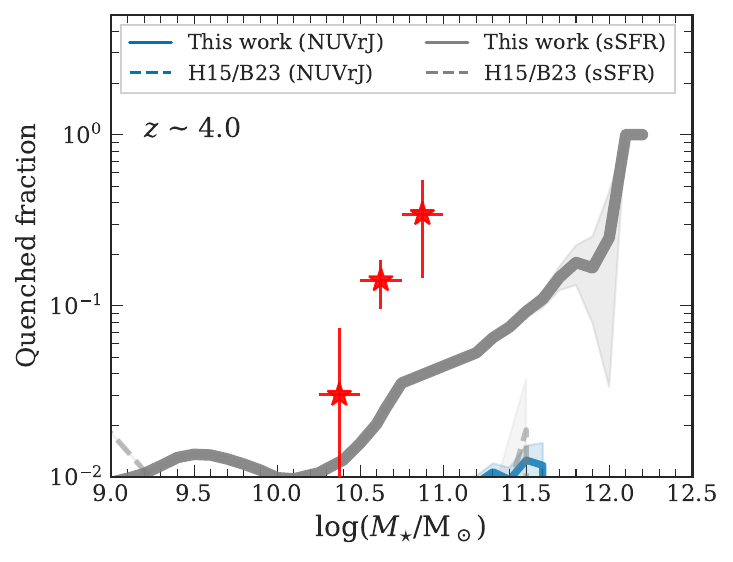}  &
\includegraphics[width=0.30\textwidth]{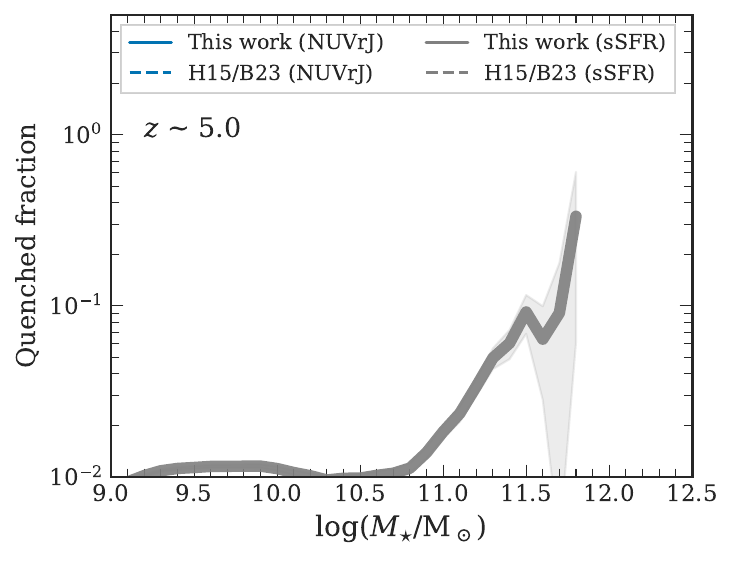}  \\

\end{tabular}

\caption{Evolution of the quenched fraction as a function of stellar mass from $z=0$ to $z\approx 5$. Blue and grey curves show quiescent galaxies selected using the NUVrJ colour-colour criterion and the instantaneous sSFR criterion, respectively. Solid and dashed lines correspond to this work and the \citetalias{Henriques2015}/\citetalias{Barrera2023} model, respectively. Red star symbols indicate the observational constraints included in the MCMC calibration.}
\label{Fig:f_qg}
\end{figure*}

The quenched fraction is one of the observational constraints included in the calibration. Figure~\ref{Fig:f_qg} shows the best-fitting predictions for this work and the legacy models. The updated model produces modest changes at $z\lesssim1$, while substantially increasing the quenched fraction at $z\gtrsim2$.

\bsp	
\label{lastpage}
\end{document}